\documentclass[useAMS,usenatbib]{mnras}

\usepackage{lineno,hyperref}
\usepackage{graphicx}
\usepackage{aas_macros}
\usepackage{caption}

\modulolinenumbers[10]

\def\masyr   {mas\,yr$^{-1}$}
\def\SNT     {{\em squeeze-`n'-tweak }}

\begin{document}


\title [VLBI imaging and analysis pipeline] {SAND: An automated VLBI imaging and analysing pipeline - I. Stripping component trajectories }

\author
[M.~Zhang et al.]
{M.~Zhang,$^{1,2,3}$\thanks{E-mail: ming.zhang@xao.ac.cn (MZ)} A.~Collioud$^3$ and P.~Charlot$^3$ \\
$^1$Xinjiang Astronomical Observatory, Chinese Academy of Sciences, 150 Science 1-Street, Urumqi 830011, China\\
$^2$Key Laboratory for Radio Astronomy, Chinese Academy of Sciences, 2 West Beijing Road, Nanjing 210008, China\\
$^3$Laboratoire d'Astrophysique de Bordeaux, Univ. Bordeaux, CNRS, UMR 5804, B\^at. B18N, All\'ee Geoffroy Saint-Hilaire,\\ 33615 Pessac, France\\
}
\maketitle

\begin{abstract} {We present our implementation of an automated VLBI data
    reduction pipeline dedicated to interferometric data imaging and
    analysis. The pipeline can handle massive VLBI data efficiently
    which makes it an appropriate tool to investigate multi-epoch
    multiband VLBI data. Compared to traditional manual data
    reduction, our pipeline provides more objective results since less
    human interference is involved. Source extraction is done in
    the image plane, while deconvolution and model fitting are
    done in both the image plane and the $uv$ plane for parallel comparison.
    The output from the pipeline includes catalogues of {\sc clean}ed
    images and reconstructed models, polarisation maps, proper motion
    estimates, core light curves and multi-band spectra. We have developed
    a regression {\sc strip} algorithm to automatically detect linear
    or non-linear patterns in the jet component trajectories. This
    algorithm offers an objective method to match jet components at
    different epochs and determine their proper motions.}
\end{abstract}

\begin{keywords}
techniques: image processing - techniques: interferometric - proper motions -
galaxies: active - galaxies: jets - ratio continuum: galaxies -
\end{keywords}



\section{Introduction}\label{intro}

Very Long Baseline Interferometry (VLBI) allows one to get the highest
achievable resolution in astronomy, comparable to an Earth-size
aperture, by coordination of radio telescope arrays around the
world. This technique has been widely used for many studies in
astrophysics, astrometry and geodesy. However, the specific nature of
interferometry hinders a direct acquisition of observables from the
recorded data. The recorded data from all radio telescope elements
must be cross-correlated to form synthetic visibilities and then a
post-processing scheme including calibration and deconvolution must be
accomplished to get images or models of the observed cosmic
sources. At present, VLBI data reduction is still often manually done
using software packages like {\sc aips}\footnote{The {\em Astronomical
    Image Processing System} (AIPS) is distributed by the {\em
    National Radio Astronomy Observatory}
  (NRAO).}~\citep{greisen.98.aspc} and {\sc
  difmap}~\citep{shepherd.97.aspc}. The subtleties of parameter
control and eye guidance make many of those manual processes unlikely
to be repeated with identical outputs by other astronomers and thus
less objective. Moreover, more and more survey and monitoring
observations produce massive amount of data which requires more
efficient reduction methods. {\sc aips} and {\sc difmap} both have
built-in scripting environments permitting batch jobs and automation
to a certain extent. However, their functionalities are too limited to
construct comprehensive data analysis programmes. With recently
matured object-oriented programming language
Python~\citep{rossum.93.nluug} and the encapsulated interface
ObitTalk\footnote{The ObitTalk is part of the Obit package distributed
  by NRAO which offers a set of Python classes interoperable with
  classic {\sc aips}. }~\citep{cotton.08.pasp} to {\sc aips}, it is
now possible to develop an advanced VLBI imaging and analysing
pipeline by utilising the power of both {\sc aips} tasks and a
full-fledged programming language. In parallel to the terminology
`Search \& Destroy' ({\sc sad}), we have named our pipeline `Search \&
Non-Destroy' ({\sc sand}) and released it as an open-source code under
the MIT license~\citep{zhang.16.ascl}.

In addition to component flux models, SAND can extract information on
various axes of the multi-epoch multiband data, including polarisations,
light curves and spectra. For extragalactic radio jets, the structural
patterns, especially for resolved components, can be parameterized.
Additionally, jet kinematics can provide insights into how the jets are
generated and how they interact with the interstellar medium. For a pipeline
designed to mine massive interferometric data, a capability to recognize
kinematic patterns, match components at different epochs, and model their
trajectories is desirable. In this paper, we will concentrate on component
trajectory analysis, as implemented in the {\sc sand} pipeline. Complete
capabilities of {\sc sand} are given in Appendix~\ref{pipeline} for
reference.

\begin{figure}
\centering \includegraphics[width=7.5cm]{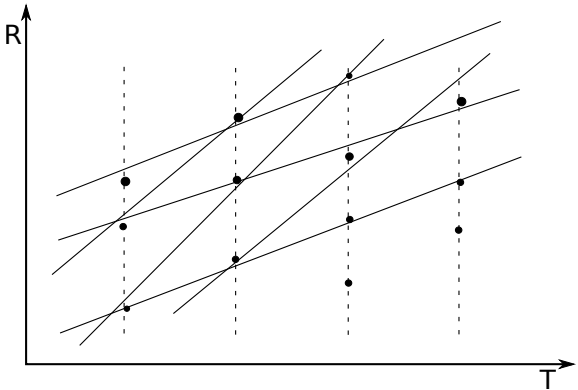}
\caption[Correspondence confusion]{Confusion in the identification of
  jet components at multiple epochs. The black dots indicate jet components.
  The dash lines indicate components observed at the same epochs,
  while the solid lines show the multiple possible paths for those components.}
\label{fig:respconf}
\end{figure}

Apart from uncertainties due to $uv$ sample gridding and
deconvolution, human factors are often responsible for post-processing
errors when VLBI data are reduced manually. A well-known problem is
the confusion in the identification of jet components
at multiple epochs. Unlike stars and quasars, which mostly have
unique overall spectral identities, there is no spectroscopic way to
distinguish jet components individually. Multiple paths may thus be
found when the identification of jet components at successive epochs
is determined visually, as illustrated in Fig.~\ref{fig:respconf}.

In radio interferometry, the images derived from the visibility data
usually show discrete structures. Since many data sets are
undersampled in some dimensions, and there is little prior
information, the application of sophisticated pattern recognition
techniques, such as those used in computer graphics, to resolve the confusion remains
limited. Recently, a novel wavelet decomposition method for recognition
of structural patterns in jet component trajectories has been
developed~\citep{mertens.15.a&a}. This method works well on extended jet
structures, provided a quality {\sc clean}ed image is
available. However, the method only works in the image plane and
requires decomposition into sub-components to cross-correlate the
features. Because of the `{\sc clean} bias' in the image
plane~\citep{condon.98.aj}, the authenticity of such sub-components
is often questionable and additional detection in the
$uv$-plane is required to verify that they are real. As shown
by~\cite{zhang.07.mn}, the image-plane detection through a {\sc
  clean}ed image is guaranteed if the signal-to-noise ratio (SNR) is
at least an order of magnitude higher than the local root mean square
(RMS) noise. In this paper, we present a straightforward iterative
method, based on simple principles, to identify the trajectory patterns
of jet components. The algorithm that we developed is implemented in the {\sc
sand} pipeline. We also discuss the applications and limitations of our method,
along with its complementarity to other methods.

\section{Trajectory representation}

\begin{figure*}
\centering
\includegraphics[height=7cm, angle=-90]{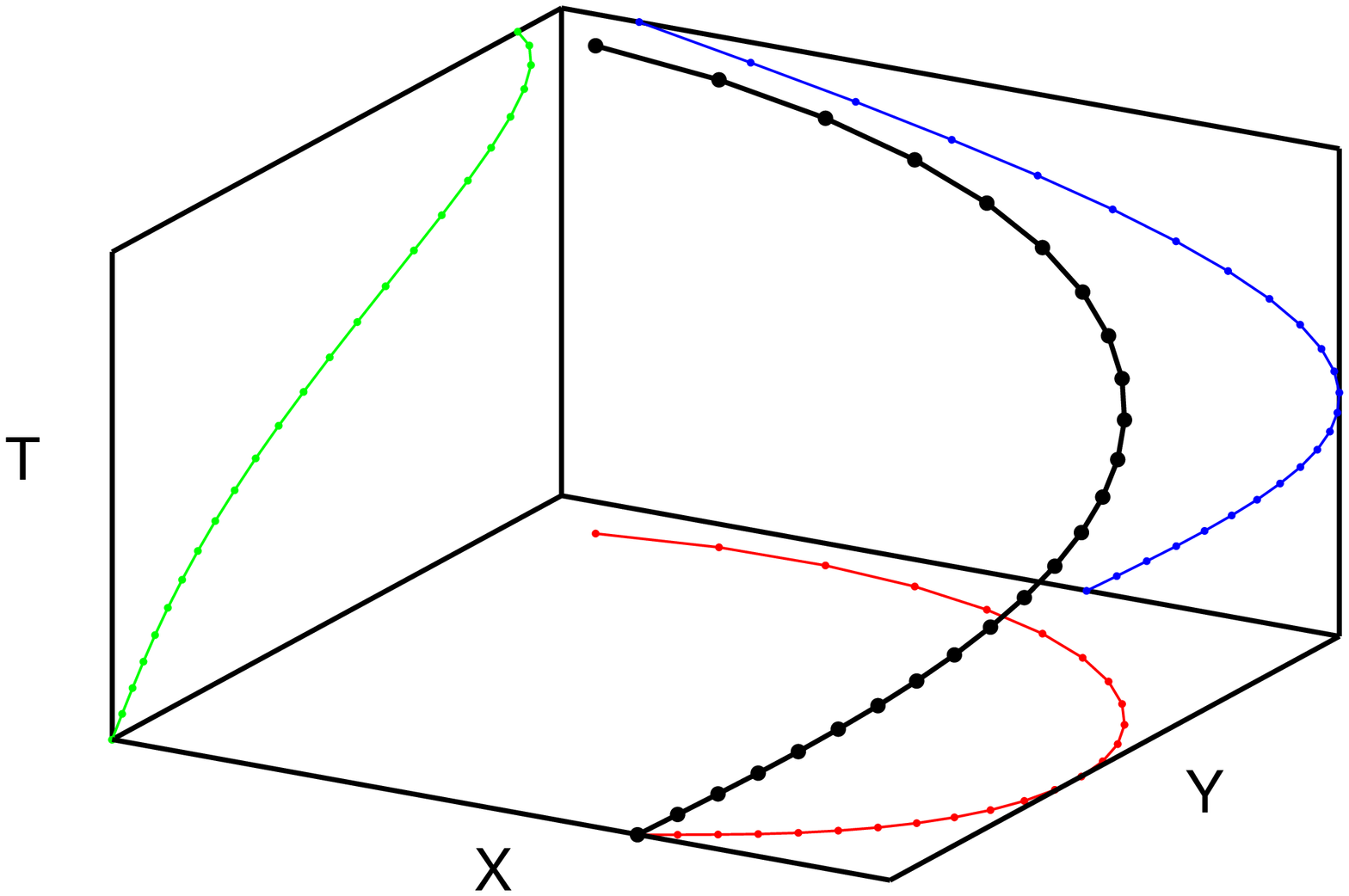}\qquad
\includegraphics[height=7cm, angle=-90]{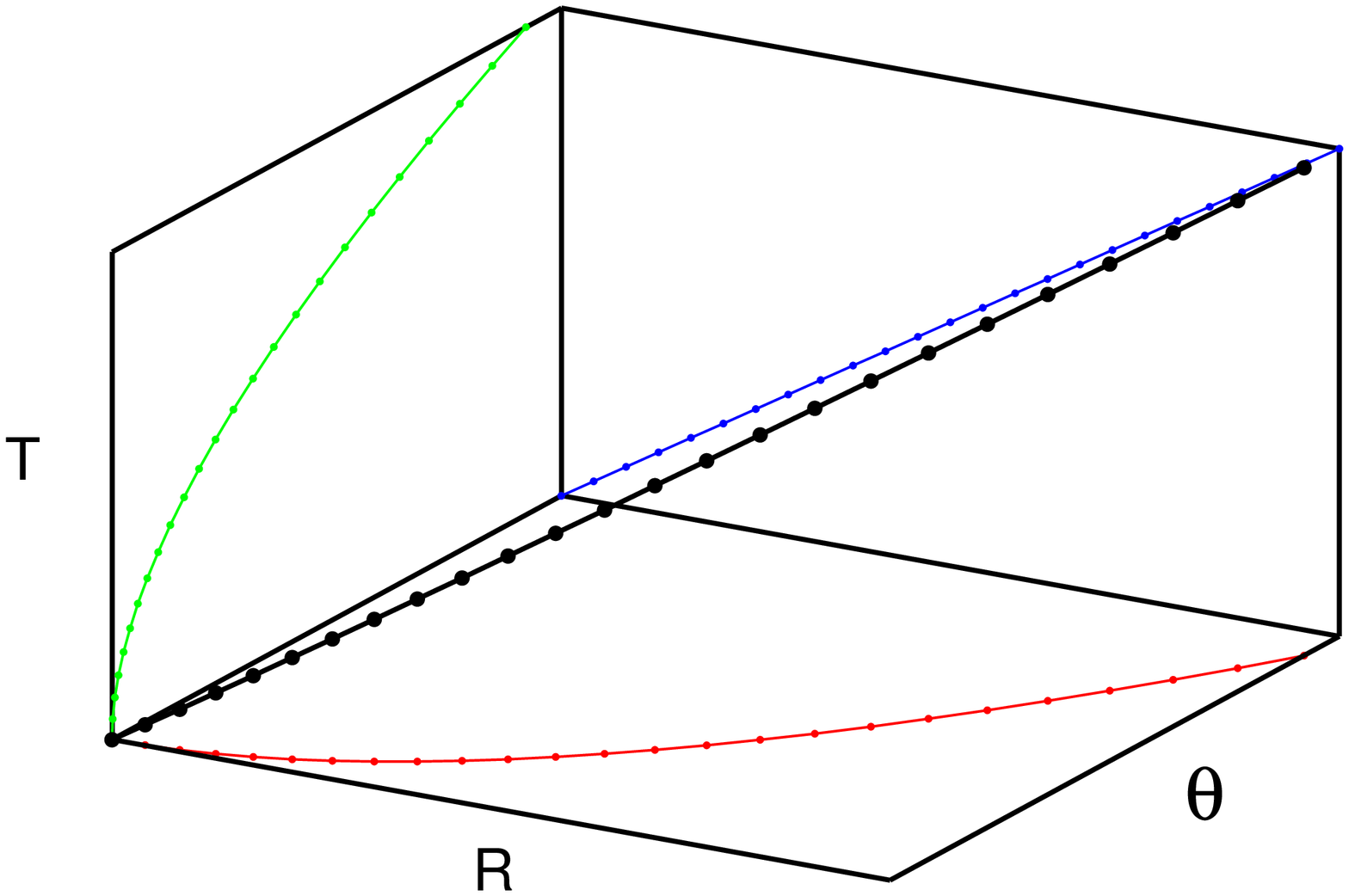}\\
\caption[General trajectory]{Representation of component
  trajectory. In the left panel, the thick black line with dots
  denotes the trajectory in the $(X,Y,T)$ 3-dimensional coordinate
  system, while the thin red, blue and green lines with dots denote
  the trajectories in the $(X,Y)$, $(X,T)$ and $(Y,T)$ projected
  planes; in the right panel, the thick black line with dots denotes
  the trajectory in the $(R,\theta,T)$ 3-dimensional coordinate
  system, while the thin red, blue and green lines with dots denote
  the trajectories in the $(R,\theta)$, $(R,T)$ and $(\theta,T)$
  projected planes.}
\label{fig:gtraj}
\end{figure*}

A trajectory is generally represented as a geometric path which is a
sequence of positions $(X,Y)$ of a given object over
time. Equivalently, the cartesian coordinates $(X,Y)$ may be
substituted by polar coordinates $(R,\theta)$. Introducing the time
parameter $T$, a trajectory may be expressed as a triplet $(X,Y,T)$ or
$(R,\theta,T)$ in a 3-dimensional coordinate system, as represented in
Fig.~\ref{fig:gtraj}. In this framework, any 3D-trajectory pattern may
also be projected onto the $(X,Y)$, $(X,T)$ and $(Y,T)$ planes, with
the projected patterns showing the same data sequence in the three
planes. Examination of all three such projected trajectory patterns
increases the chances of pattern recognition since the complexity of
the patterns differs in the three planes and recognition may be
favoured in one or the other of these planes. Once a trajectory
pattern is identified in one of the planes, the corresponding
3D-trajectory can be easily reconstructed. In the following, we call
{\it trajectory} indistinctly the 3D-trajectory and its
2D-projections.  In practice, it is generally convenient to use polar
coordinates and to examine trajectory patterns in terms of the
evolution with time of the radial distance to the core, or in other
words jet component proper motions.

\section{The regression strip algorithm}\label{regstrip}

To minimize the confusion, increasing the sampling and data quality is
desirable, meanwhile one should resort to an objective method to
figure out the component correspondence at different epochs. To this
end, we have developed an algorithm to iteratively find the most
significant patterns in multi-epoch data, determine component matching
and fit the trajectories over time. Like the `{\sc clean}' deconvolution method
which gradually reaps signal from a residual image, our method progressively
strips out tangled components from a trajectory pattern. By analogy, we call it
`regression {\sc strip}' method.

\begin{figure*}
\centering \includegraphics[width=15cm]{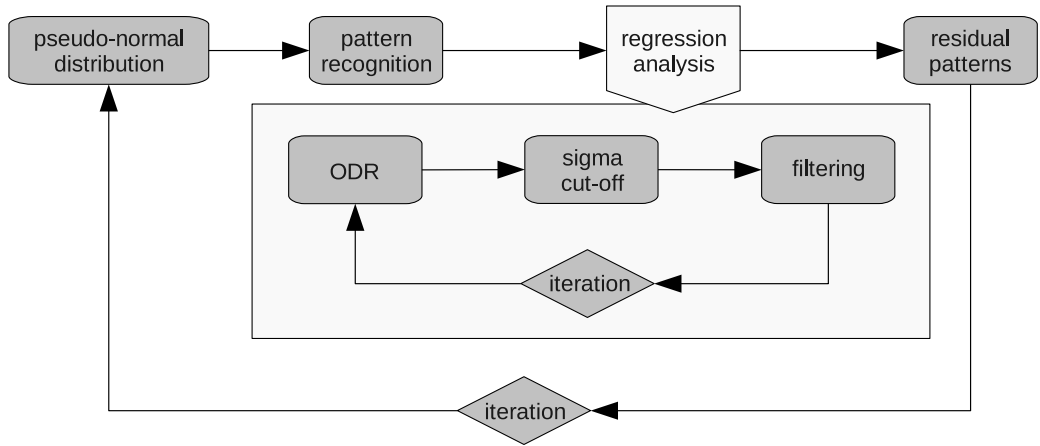}
\caption[Regression {\sc strip} flowchart]{The flowchart of the
  regression {\sc strip} algorithm. The inner light-shaded region
  encompasses the regression analysis accomplished in the {\em minor
    cycle}, while the unshaded outer region represents pattern
  stripping during the {\em major cycle}.}
\label{fig:regflow}
\end{figure*}

The {\sc strip} algorithm consists of two cycles: a {\em major cycle} for
pattern recognition and a {\em minor cycle} for regression analysis. The
flowchart of the algorithm is given in Fig.~\ref{fig:regflow}.

\begin{figure*}
\centering
\includegraphics[width=7.5cm]{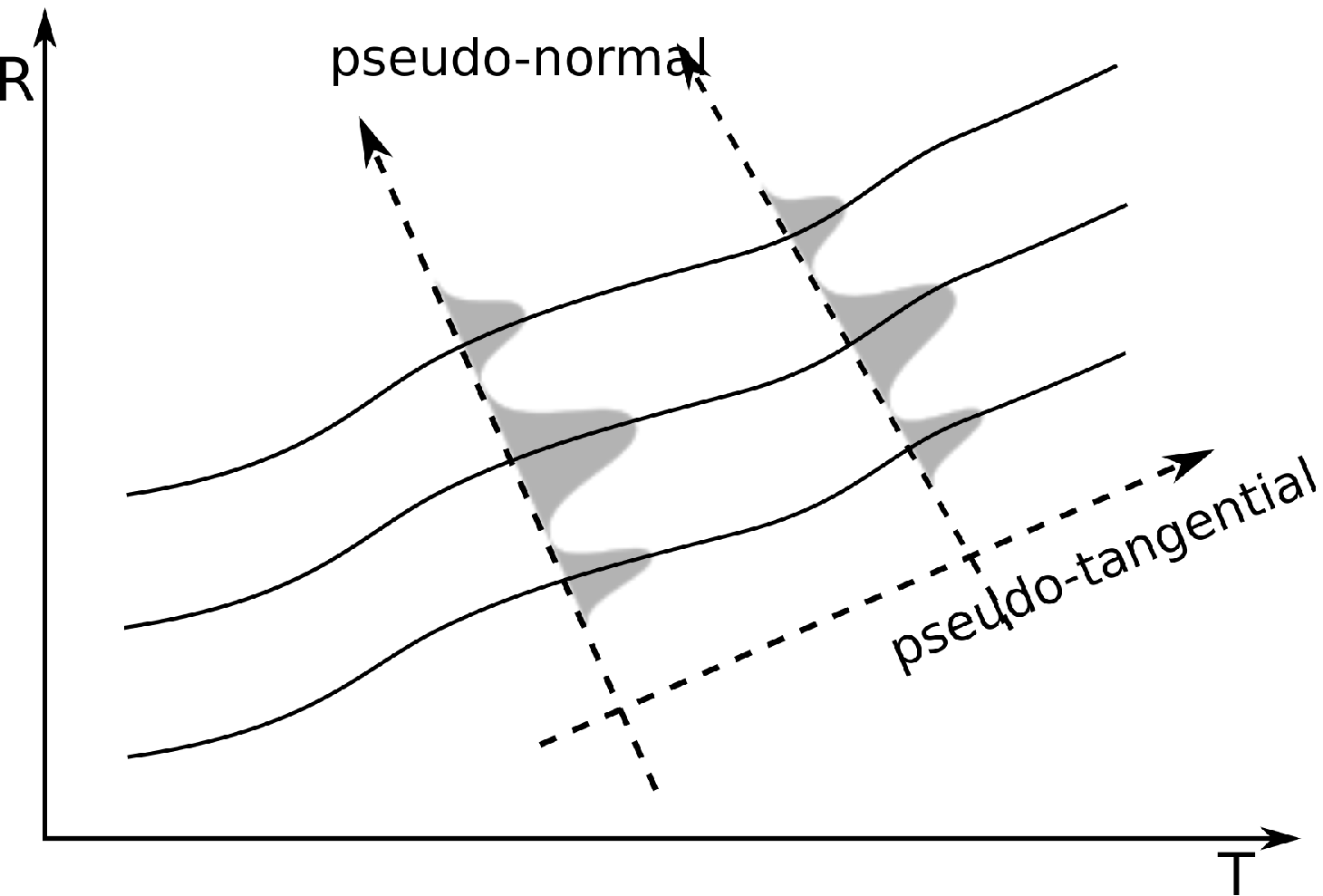}\qquad
\includegraphics[width=7.5cm]{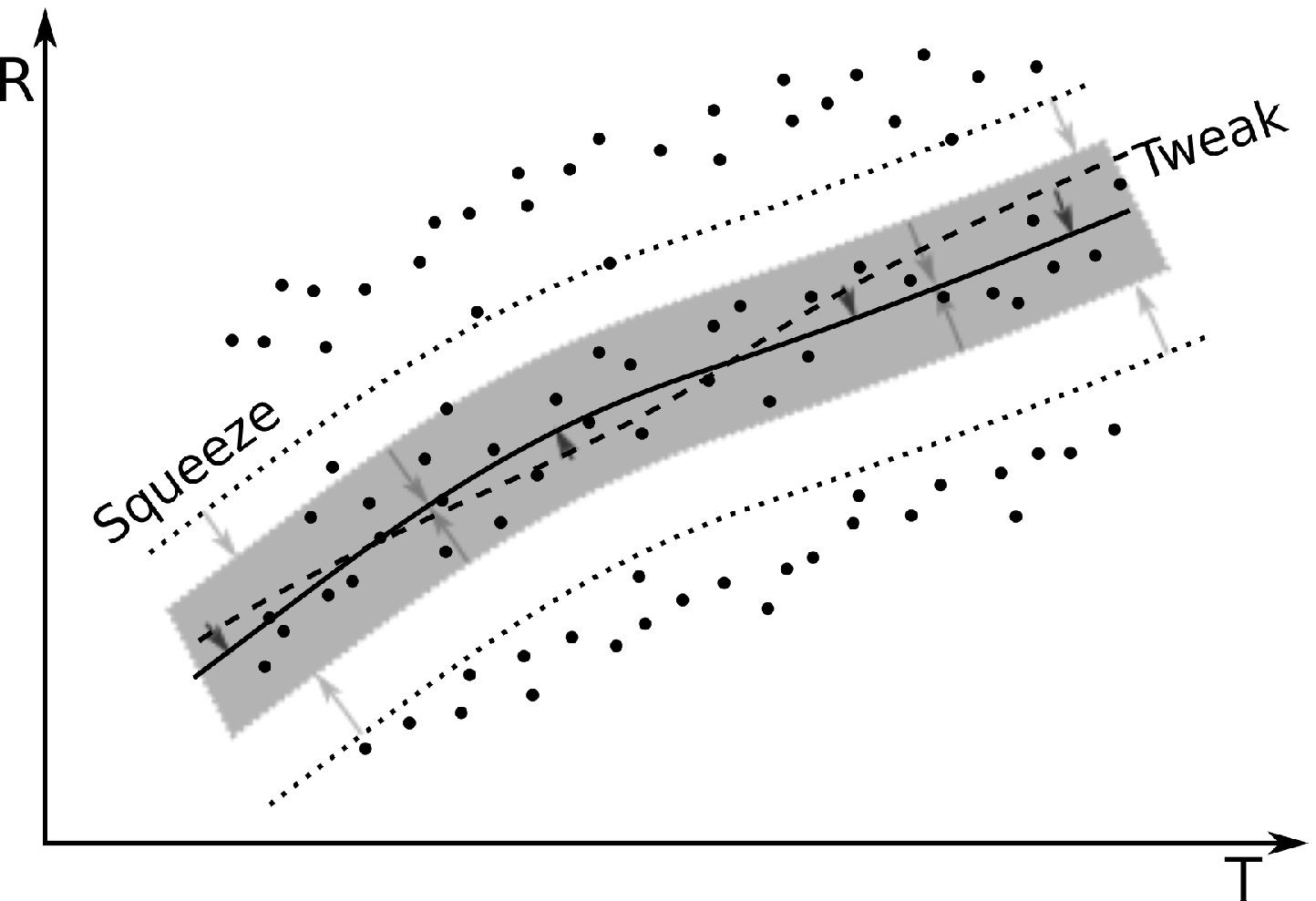}\\
\caption[Pseudo-normal distributions and the \SNT ]{Pseudo-normal
  distributions and regression {\sc strip} algorithm. The left panel
  shows the pseudo-tangential and pseudo-normal directions. The
  pseudo-normal direction is locally orthogonal to the
  pseudo-tangential direction where the pattern flows. The right panel
  shows the \SNT effect of the regression {\sc strip} algorithm. }
\label{fig:regstrip}
\end{figure*}

\begin{itemize}
\item The {\em major cycle} proceeds as follows:

  (i) Calculate the pseudo-normal distribution of the sampled
  component positions. Assuming there are intrinsic linear or
  non-linear patterns in those positions and the patterns for
  different components are similar, we can determine an overall
  direction where the pattern flows. We call it the pseudo-tangential
  direction and its orthogonal direction the pseudo-normal direction,
  as represented in Fig.~\ref{fig:regstrip}. For linear patterns, this
  is a trivial transformation between orthogonal coordinates. For
  non-linear patterns, pseudo-tangential and pseudo-normal directions
  may always be found locally, while an overall adjustment may be made
  in a second stage.

  (ii) Search for significant patterns in the sampled component
  positions. If there are distinguishable patterns in those positions,
  the pattern probability distribution will show maxima in the
  pseudo-normal direction with appropriate binning.The sampled data
  corresponding to those maxima are then drawn and used as input for
  the subsequent regression analysis, representing initial local
  patterns.

  (iii) Carry out regression analyses on the data drawn in
  step~(ii). We can do either linear or non-linear curve fitting. Note
  that the patterns identified at this stage are used only as a
  starting point for the {\em major cycle}. The data are then re-{\sc
    strip}ed in the {\em minor cycle} until the regression process is
  stabilized.

  (iv) Remove the fitted components and get residuals. We assume that
  the regression process can find and fit all components, and that the
  remaining components can be extracted by accomplishing further
  iterations.

  (v) Check results against the iteration condition, if satisfied then go to
  step (i). The iteration condition may be either an iteration number
  or a physical parameter like the residual sample size.
\end{itemize}

\begin{itemize}
\item The {\em minor cycle} proceeds as follows:

  (i) Fit regression curves to the patterns identified during the
  major cycle by minimising orthogonal distances. This fitting is
  accomplished with the {\sc odrpack}
  subroutines~\citep{boggs.89.transmath}.

  (ii) Cut off data points that show departures above a certain
  sigma level. This process not only excludes data belonging to
  adjacent patterns, but also reintroduces those not part of the
  patterns considered in previous iterations. The cut-off level is
  regulated by a loop gain, which is scaled down at every iteration.
  This limits the exclusion of good points as the iteration process
  converges.

  (iii) Filter the components according to their strength or post-fit
  deviations to make sure resolved components are assigned to
  different patterns. As this stage, component correspondence is also
  sorted out.

  (iv) Check results against the iteration condition, go to step (i)
  if required. A similar criterion as that in the {\em major cycle} is
  used to assess convergence.
\end{itemize}

In the {\em minor cycle}, component trajectories are tuned at each
iteration while the deviations are reduced, and the linear or
non-linear shapes are adjusted gradually. This resembles a serial \SNT
effect, as shown in Fig.~\ref{fig:regstrip}.

Our {\sc sand} pipeline only needs sources to be extracted in the
image-plane for the initial input. Component parameters derived in
further stages are obtained from model fitting both in the image-plane
and the $uv$-plane.  This is useful for cross-checking and to
eliminate spurious components. The regression {\sc strip} algorithm
does not require frequent nor even sampling. The basic principles that
make the method work are: (i)~component trajectories show
distinguishable linear or non-linear patterns, as reflected by peaks
above a certain significance level in the pattern probability
distribution along the pseudo-normal direction; (ii)~the {\sc
  stripp}ing algorithm in the regression analysis is progressive, with
the \SNT process not degrading statistical properties of the
pseudo-normal distribution; (iii)~deviations are reduced at every
iteration, ensuring convergence of the algorithm.

\section{Mock data tests}\label{mock}

Our regression analysis does not involve cubic or higher-order
non-linear fitting because the {\sc sand} pipeline serves as an
initial astrometric sifter and quadratic fitting can handle curved
trajectories competently in most cases. In Section~\ref{dsdegen}
below, we demonstrate that more complex trajectories can be treated
locally as a set of superposed scaled quadratic curves. If the trajectory
bends significantly, there should be also a notable rotating pattern
in the position angle (PA) coordinates. Such peculiar sources must be
investigated on a case by case basis. Furthermore, if one really wants
to try the regression {\sc strip} algorithm on highly non-linear data,
one way to proceed is to {\sc strip} the trajectory pattern section by
section.  In this case, one needs to make sure that the binning
provides sampling along the pseudo-tangential direction that is dense
enough in every individual section. In order to demonstrate the
ability of the {\sc sand} pipeline to resolve component trajectories,
we have generated mock data which we have subsequently analysed with
{\sc sand}. The results derived for some typical cases are discussed
below.

\subsection{Affine transformation for linear patterns}

\begin{figure*}
\centering
\includegraphics[width=7.5cm, angle=-90]{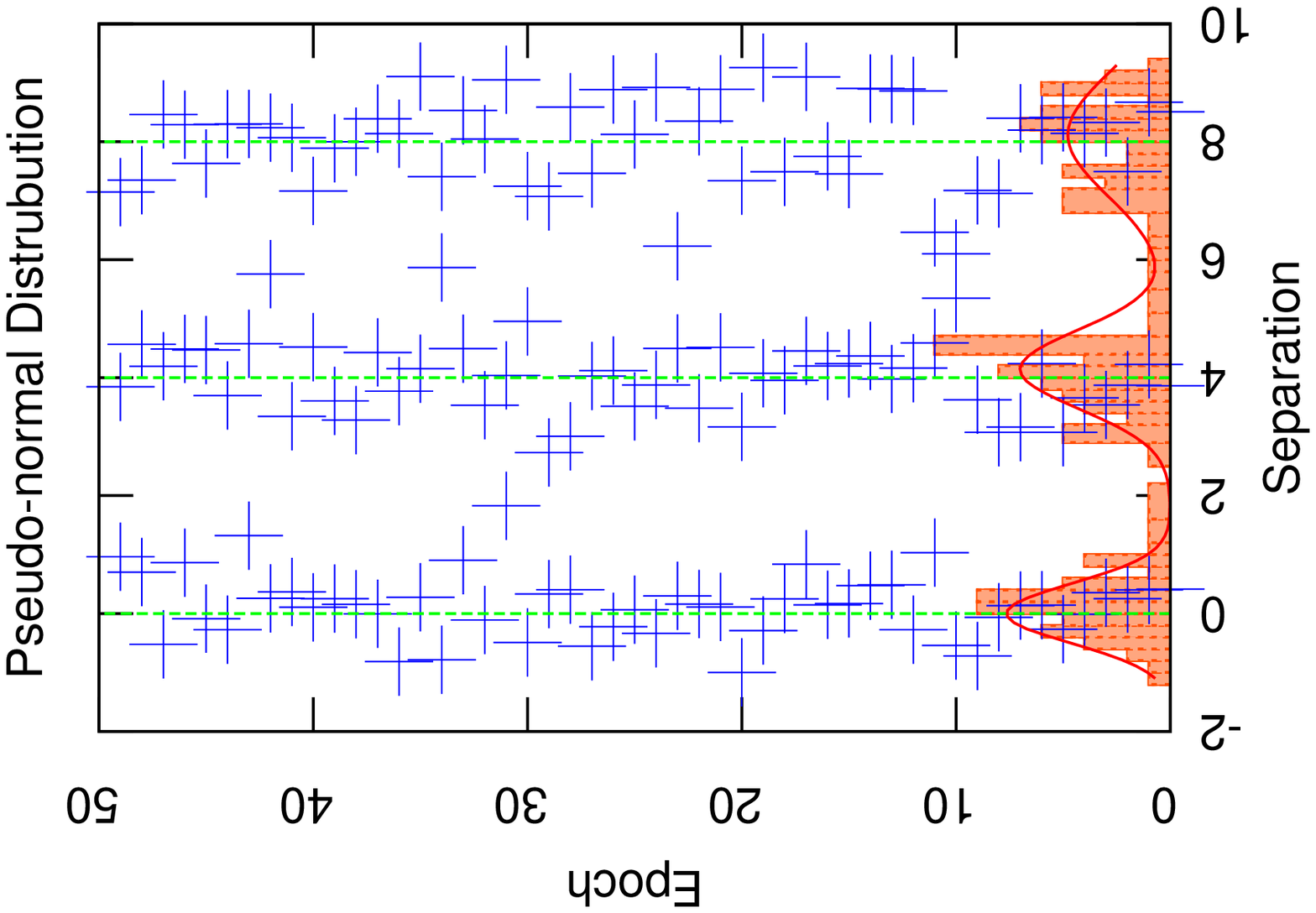}\qquad
\includegraphics[height=10cm, angle=-90]{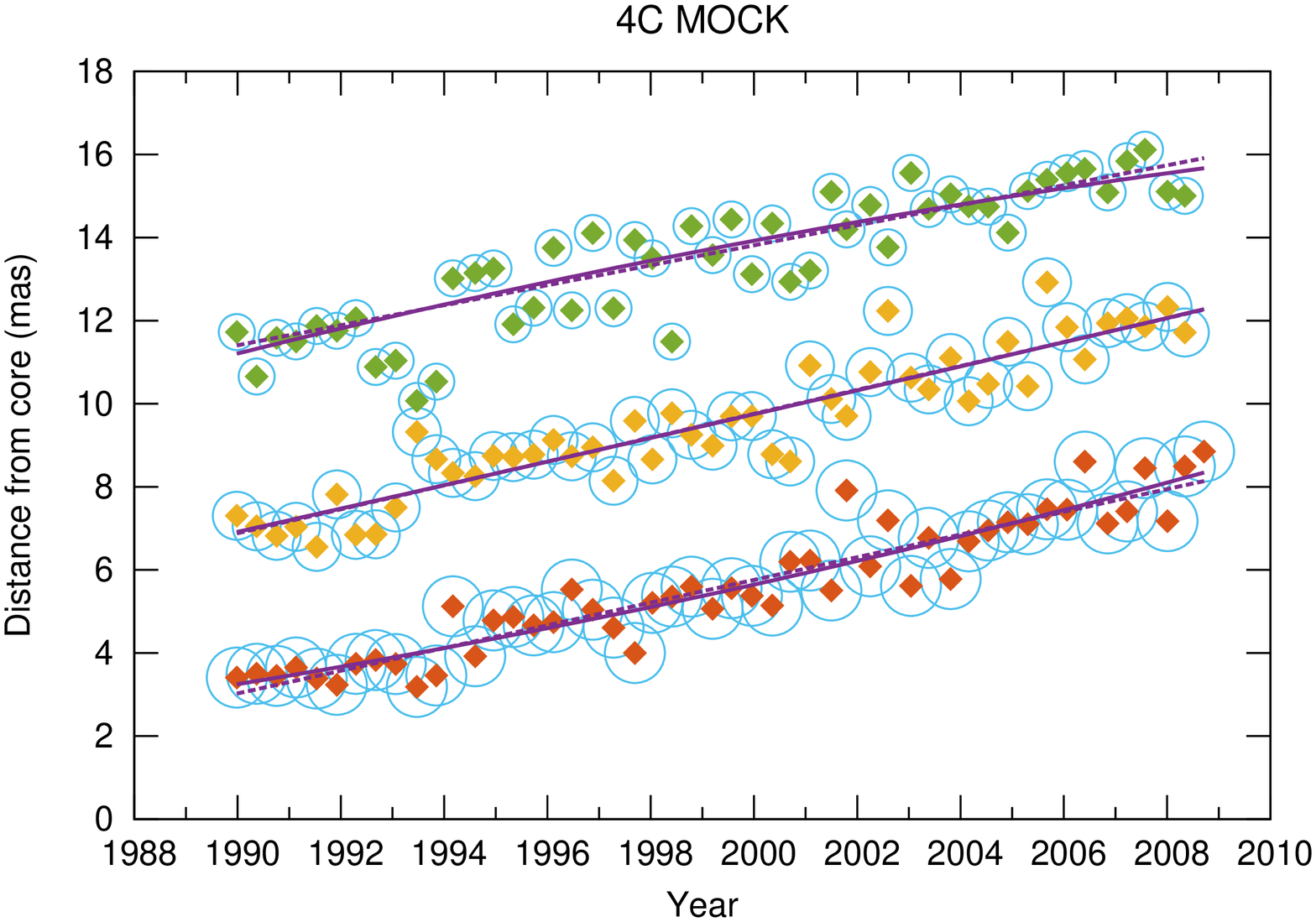}\\
\caption[Pseudo-normal distributions and linear
regression]{Pseudo-normal distributions and regression {\sc strip}
  results with linear mock data. The left panel shows the generated
  mock data in the pseudo-normal dimension. The histograms and the
  superposed multiple Gaussian fits illustrate the significance of
  these pseudo-normal distributions. The right panel illustrates the
  capability of the regression {\sc strip} algorithm to disentangle
  linear proper motions from series of component positions. The purple
  dashed lines indicate linear fits while purple solid lines indicate
  quadratic fits. All components are shown as light-blue circles with
  a size corresponding to the order of extraction, which is flux
  related. They are also plotted as coloured diamonds inside those
  circles, each colour denoting a different pattern (red, yellow and
  green sequentially).}
\label{fig:psulin}
\end{figure*}

If component trajectories are linear, we can demonstrate that there is
an affine transformation between the pseudo-normal coordinates and the
trajectory coordinates, which only involves rotation, translation and
scaling. When patterns are well separated, as in the three-component
test case in Fig.~\ref{fig:psulin}, linear trajectories are easily
identified through the regression {\sc strip} process. When
transformed into the pseudo-normal dimension, those linear trajectory
patterns correspond to distinct pseudo-normal distributions (see left
panel in Fig~\ref{fig:psulin}). The issue of finding component
trajectories then becomes equivalent to searching significant
distributions in the pseudo-normal dimension.

\subsection{The \SNT process for non-linear patterns}

If component trajectories are non-linear, mathematically there is no
such affine transformation to directly convert trajectory coordinates
to pseudo-normal coordinates. However, due to the built-in self-adapting
capability of the regression {\sc strip} algorithm, the \SNT process
is still able to disentangle component trajectories properly, provided
that non-linearity remains moderate.

\begin{figure*}
\centering
\includegraphics[height=8.5cm, angle=-90]{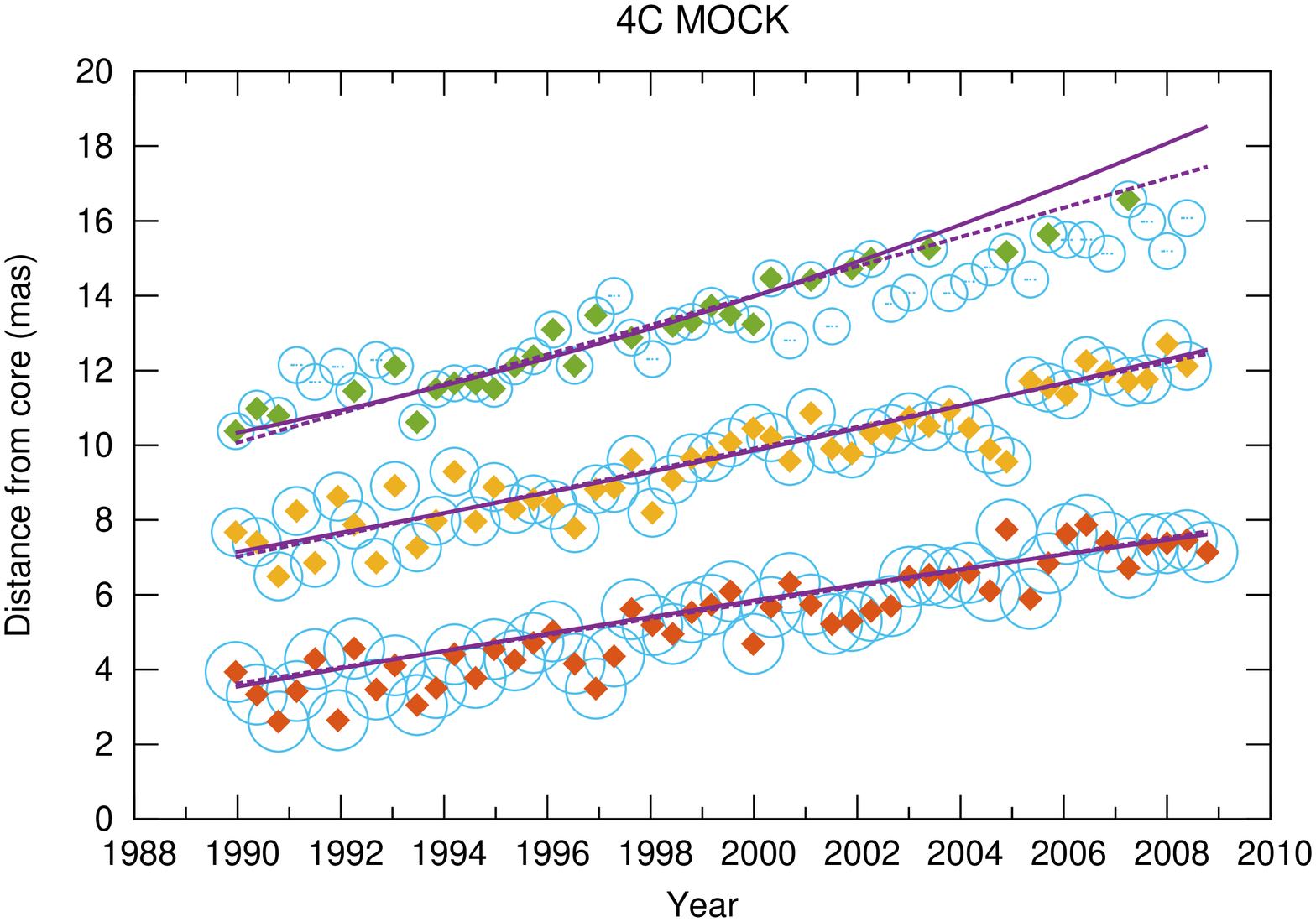}
\includegraphics[height=8.5cm, angle=-90]{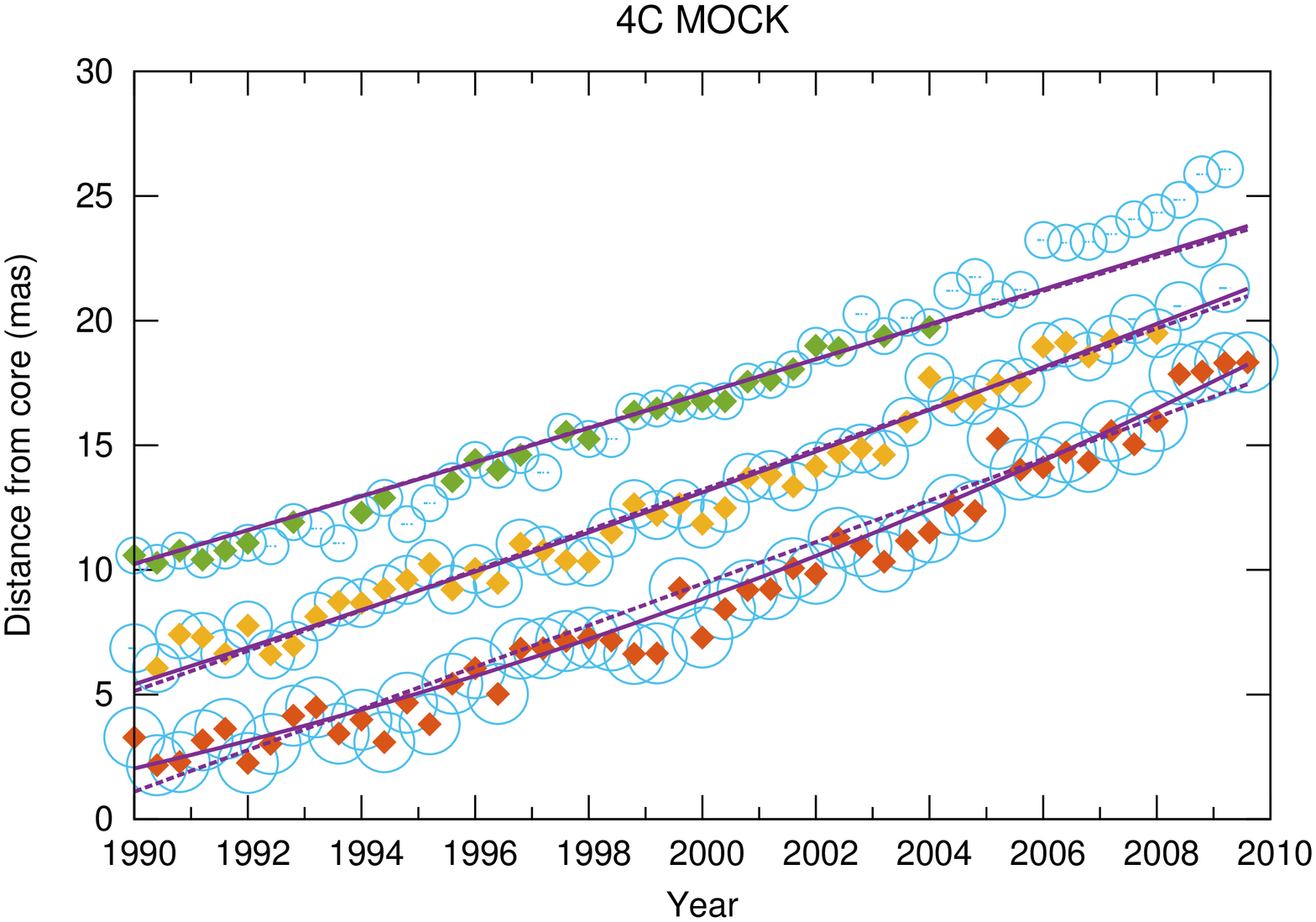}\\
\includegraphics[height=8.5cm, angle=-90]{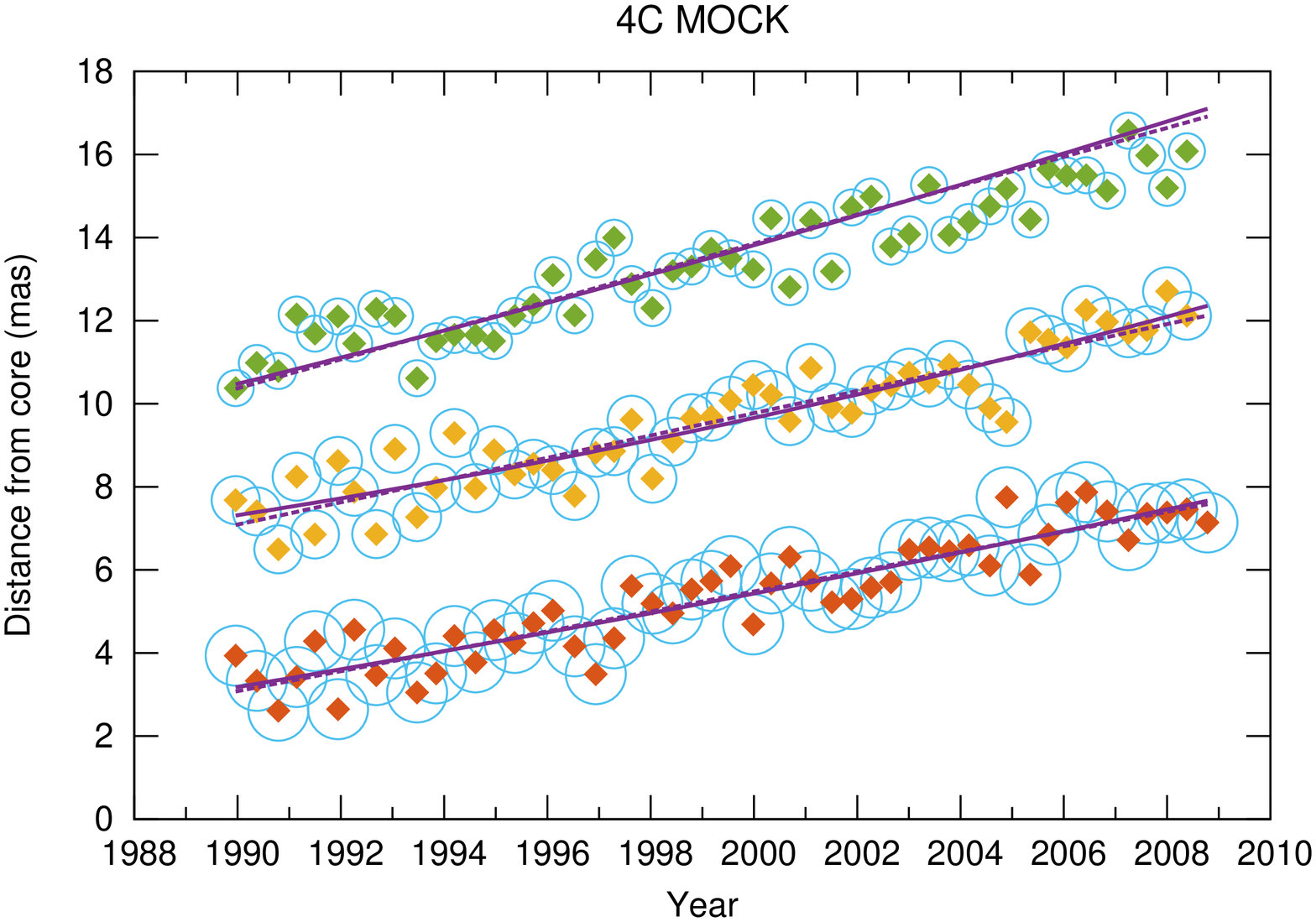}
\includegraphics[height=8.5cm, angle=-90]{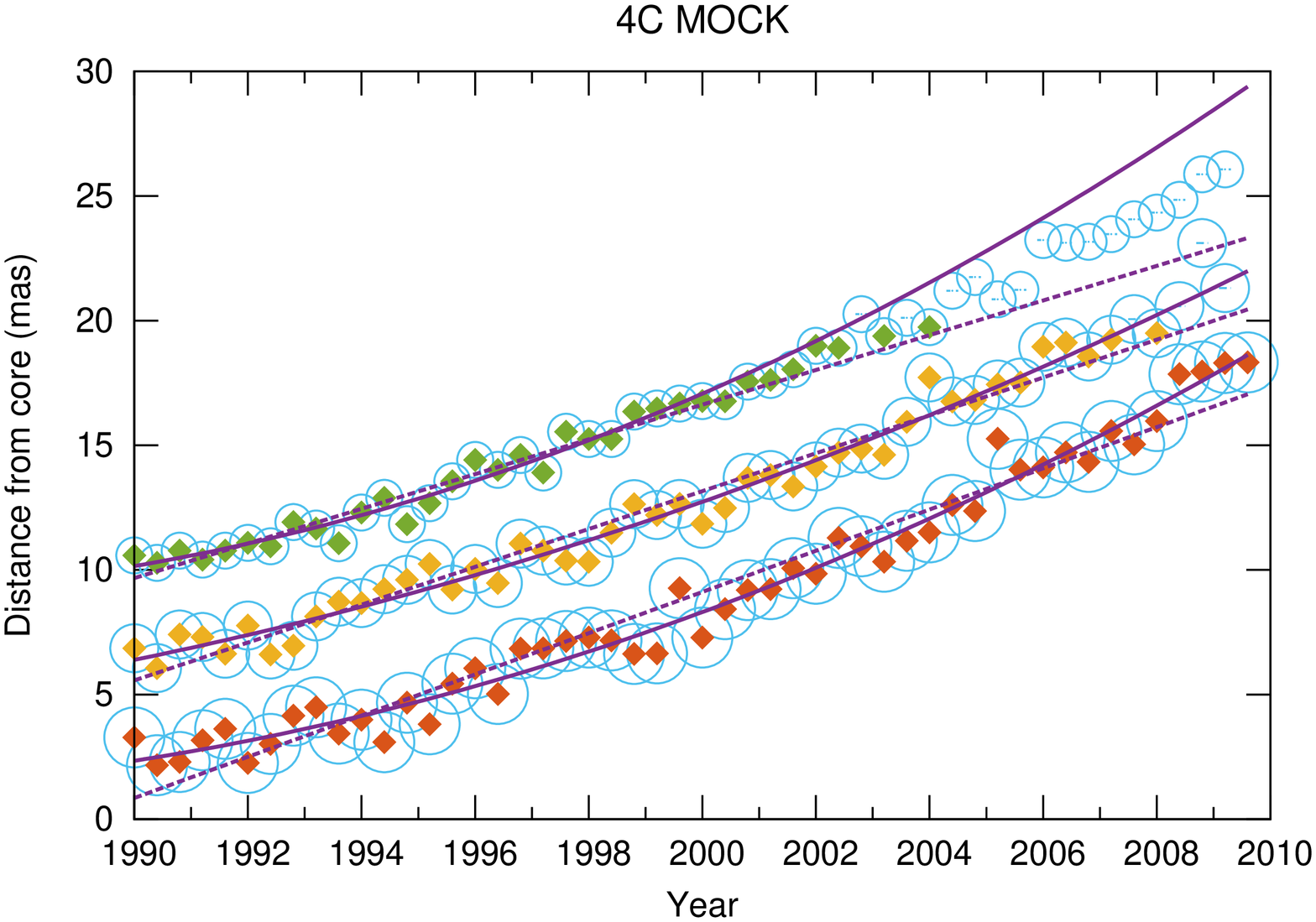}\\
\includegraphics[height=8.5cm, angle=-90]{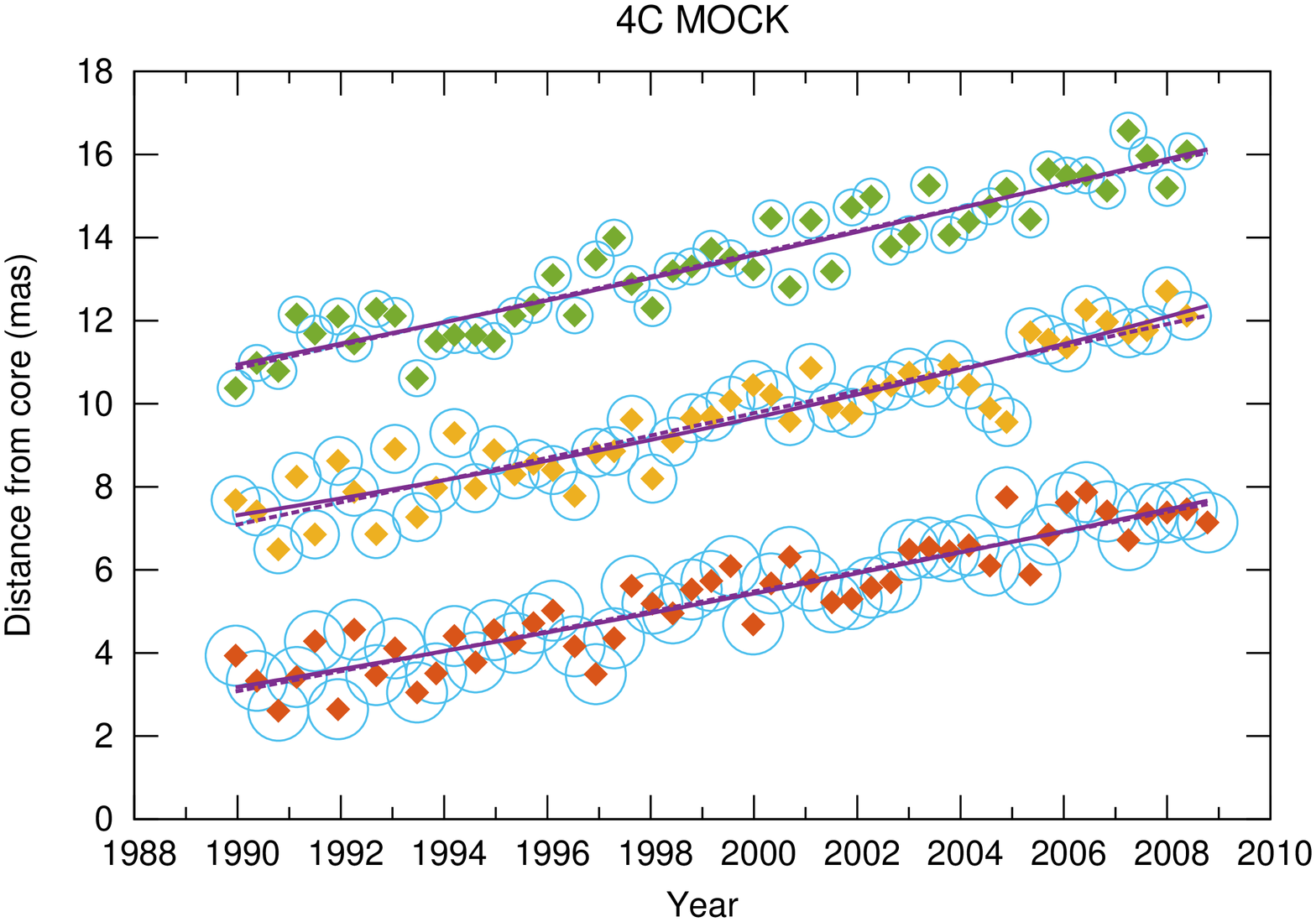}
\includegraphics[height=8.5cm, angle=-90]{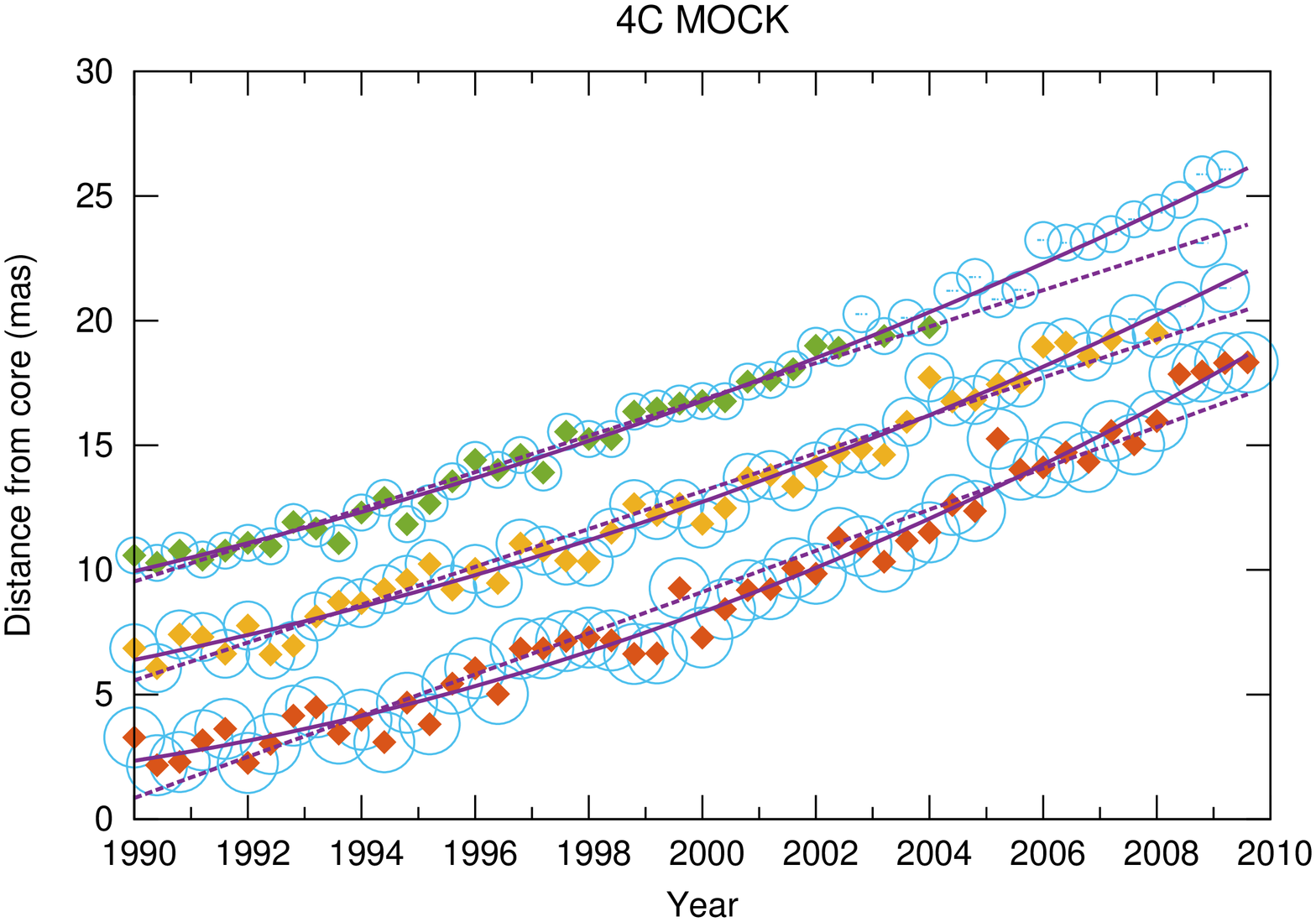}
\caption[The \SNT examples]{Illustration of the \SNT effect with mock
  data. Panels on the left-hand side are for linear data while those
  on the right-hand side are for non-linear data. Each panel provides
  results at a different stage of the iterative process within the
  {\em minor cycle}. From the upper to lower panels, the iteration
  numbers are: 1, 2, and 3 for the linear data (left panels), and 1, 3
  and 5 for the non-linear data (right panels), respectively. Both
  linear and quadratic fits are plotted in every panel for comparison
  purposes. Symbols are the same as those in
  Fig.~\ref{fig:psulin}. The dashed lines inside the blue circles show
  the diminished error bars of the mock data. }
\label{fig:snt}
\end{figure*}

As shown in the left-hand side panels of Fig.~\ref{fig:snt}, for
linear patterns with small dispersion, the component trajectories are
already fitted pretty well after the first iteration, except that in
this case the third component is not picked up at all epochs. This is
because the regression {\sc strip} algorithm first picks up components
in the bin with the highest counts and the resulting deviations from
linear fitting remain small due to the limited sampling. In subsequent
iterations, further components are introduced and the initial tight
sigma cut-off is progressively released. The points scattered from the
trajectory are then gradually recovered and optimal fits are obtained
at the end of the iterative process.

In the case of non-linear patterns, it is more difficult to obtain a
reasonable fit after the first iteration if trajectories shows
significant bending, as shown in the right-hand side panels of
Fig.~\ref{fig:snt}. This is because there is no static
pseudo-tangential direction valid for all data points. However, we
still find that the \SNT process gradually adjusts the quadratic
curves in subsequent iterations to pass through the trajectory
patterns, with convergence obtained after only a few iterations within
the {\em minor cycle}.

We intentionally kept both linear and quadratic fits in
Fig.~\ref{fig:snt} for comparison. As expected, the results of these
are nearly identical for linear trajectories, whereas quadratic fits
are found to reproduce more closely the tail ends of the observed
patterns for non-linear trajectories.

\subsection{Effects of binning and cut-off level}

Most of the examples shown above have visually discernible trajectory
patterns. Additional investigations on how the regression
{\sc strip} method works for less discernible trajectory patterns and
assessment of its limits in those cases remains of interest for a
wider use of the algorithm.

Considering that a distribution is always scalable, we may introduce a dimensionless quantity, the {\em degree of
 separation} of two normal distributions, to investigate such cases. This quantity is defined as:

\begin{equation}
  DS := \frac{\Delta\mu}{\sigma_{\rm max}} ,
\label{eq:ds}
\end{equation}
where $\Delta\mu$ is the separation between the means and $\sigma_{\rm
 max}$ is the larger of the standard deviations of the two distributions. A
large $DS$ value indicates that the two adjacent populations of points
are well-separated, which implies a higher statistical significance
for the pseudo-normal distributions. For example, the $DS$ value for
the mock data in Fig.~\ref{fig:psulin} is equal to 6.6, which should
be viewed as a `well-resolved' case.

The implementation of the regression {\sc strip} algorithm in the {\sc
 sand} pipeline incorporates parameters to control the number of
bins, the sigma cut-off, the number of iterations and the loop
gain. The number of iterations and loop gain mainly control the
convergence speed of the regression {\sc strip} process. On the other
hand, the number of bins and sigma cut-off play important roles in
trajectory pattern recognition. It appears that the statistics
involved in such recognition are indeed very sensitive to how the data
are binned when the sampling size is limited, especially where
neighbouring trajectory patterns start to amalgamate.

It is worth pointing out that the PA information has not been
considered in our mock data tests since our intent was to characterize
proper motions in trajectory coordinates. However, the {\sc sand}
pipeline can deal with patterns in both trajectory and PA coordinates
at the same time. Normally, a linear proper motion pattern in
trajectory coordinates corresponds to a collimated pattern in PA
coordinates. In the regression {\sc strip} process, a sigma cut-off on
both distance and PA deviations may hence be considered. In order
to assess the pattern recognition limits of the algorithm, several
challenging cases are considered in the sub-sections below.

\subsubsection{Linear patterns with small $DS$}

For `well-resolved' trajectory patterns, it is found that linear
patterns can be extracted correctly if the number of bins is set to a
couple of tens and the cut-off level is set to a couple of
sigmas. However, for smaller $DS$ values (see below), the data points
gradually tangle in trajectory coordinates, which requires
fine tuning on the number of bins and sigma cut-off.

\begin{figure*}
\centering
\includegraphics[height=8.5cm, angle=-90]{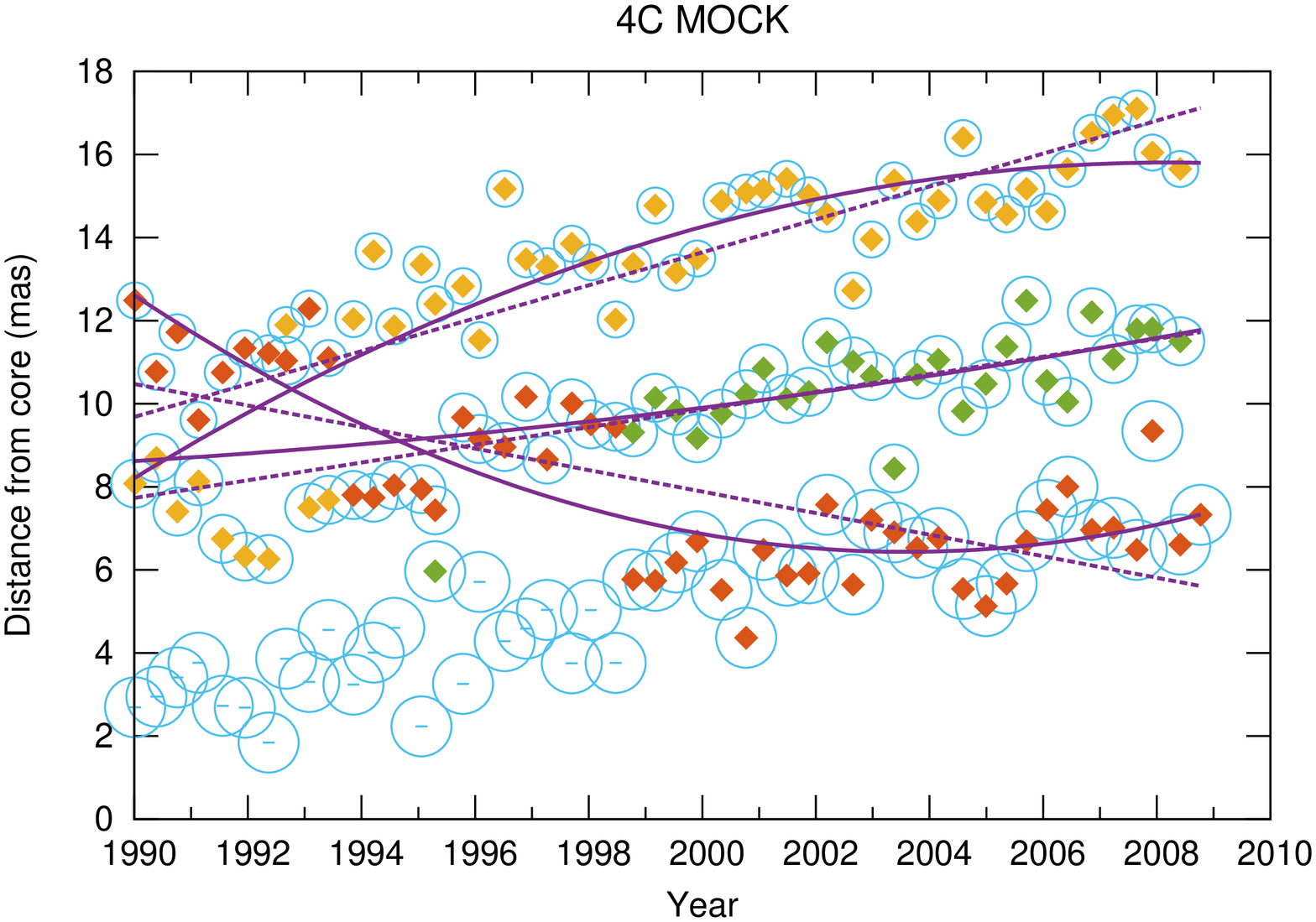}
\includegraphics[height=8.5cm, angle=-90]{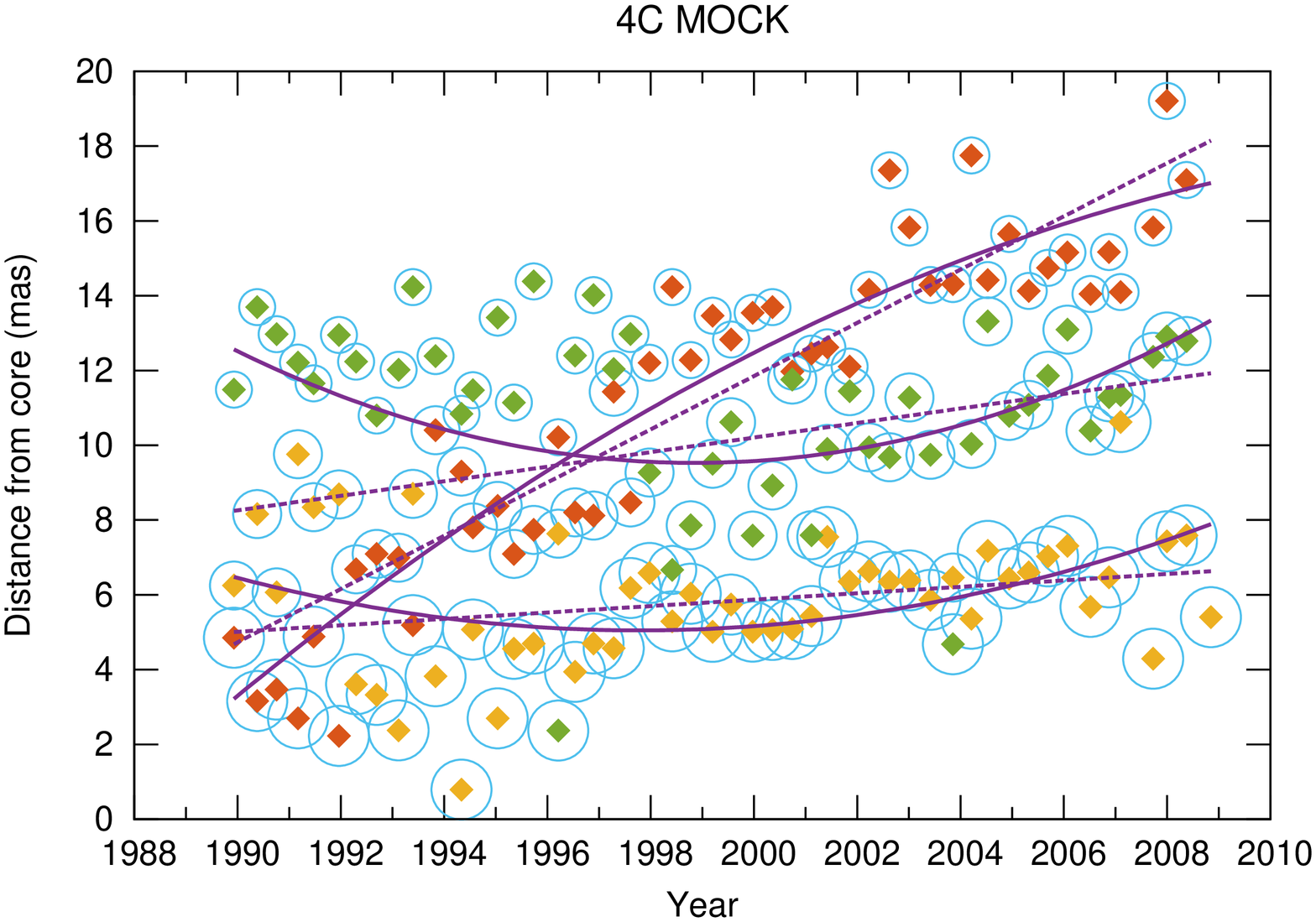}\\
\includegraphics[height=8.5cm, angle=-90]{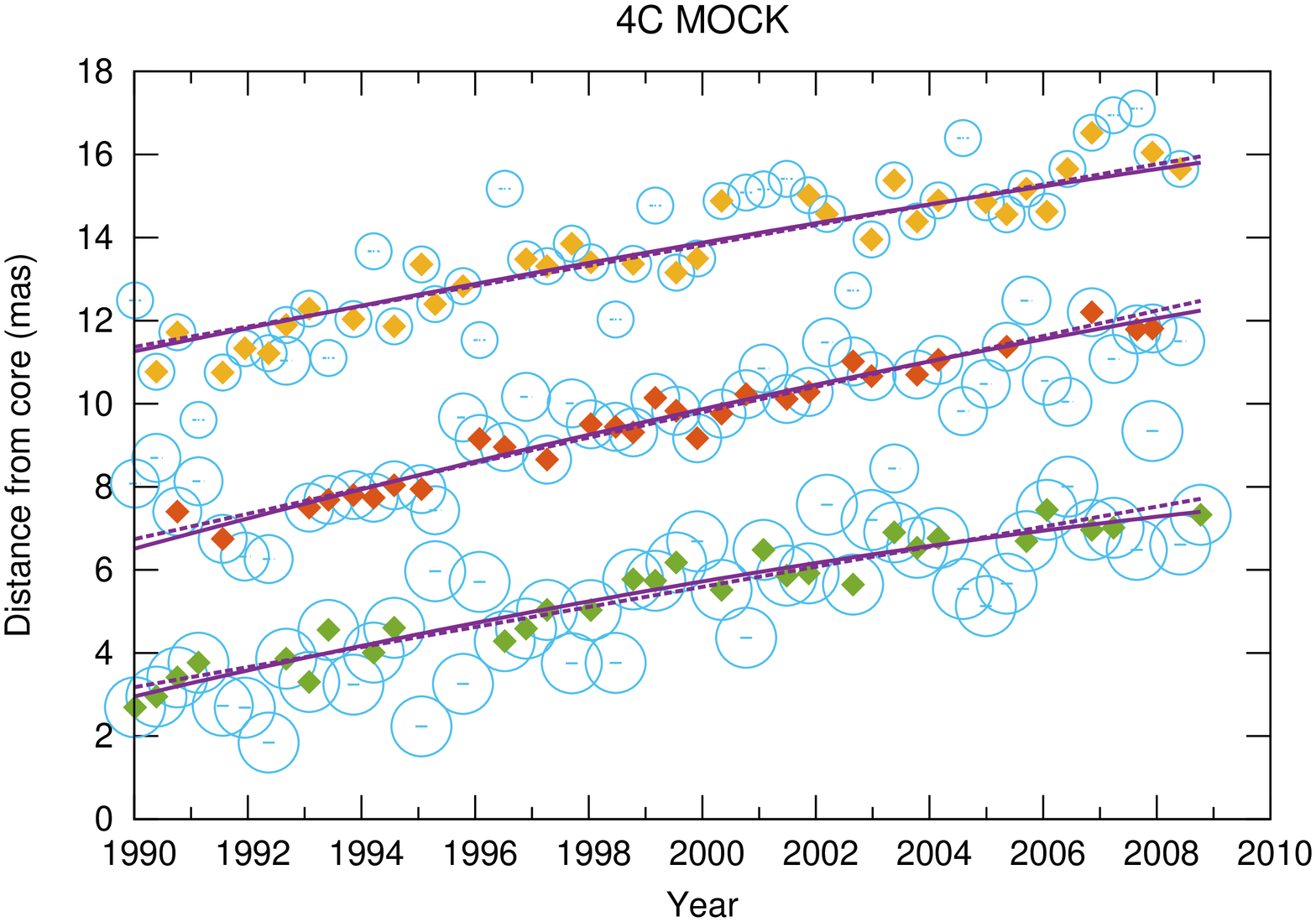}
\includegraphics[height=8.5cm, angle=-90]{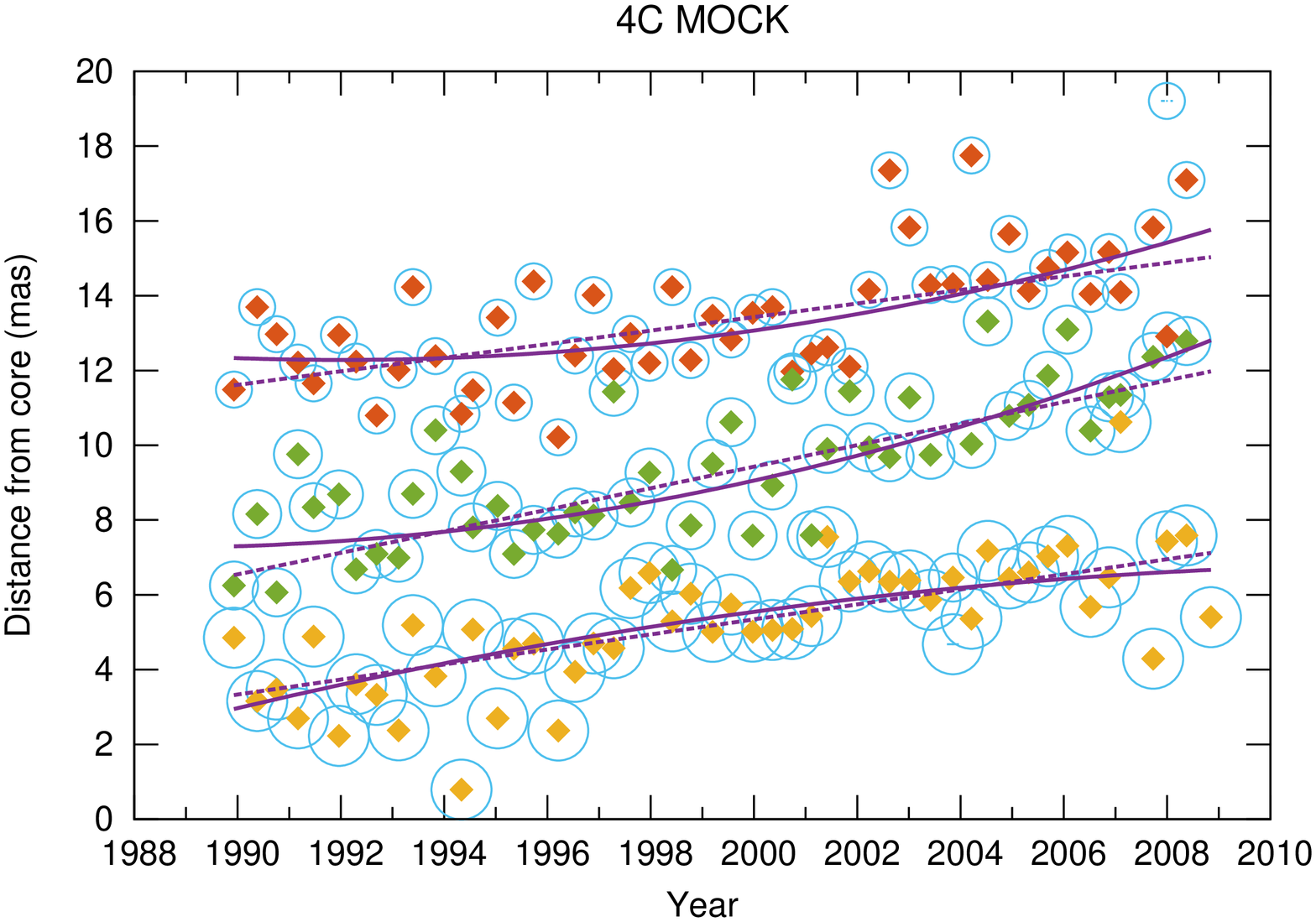}

\caption[The linear bin-cut examples]{Effects of binning and cut-off
  level on linear mock data. Panels on the left-hand side are for a
  $DS$ value of 4, while \textcolor[rgb]{1.00,0.00,0.00}{those on the
    right-hand side} are for a $DS$ value of 3.1. The number of bins
  and sigma cut-off are (10, 6) and (20, 3) for the upper and lower
  left-hand side panels, and (10, 6) and (30, 6) for the upper and
  lower right-hand side panels, respectively. Symbols are the same as
  those in Fig.~\ref{fig:psulin}.}
\label{fig:bincut1}
\end{figure*}

The two panels in the left-hand side of Fig.~\ref{fig:bincut1} show
results of the {\sc strip} algorithm in a case where $DS=4$. When the
data are gridded with 10~bins and the cut-off level is set to
6$\sigma$ (upper panel), one sees that the algorithm is not able to
properly disentangle the trajectory patterns and leads to implausible
results. This is because a small number of bins and a large sigma
cut-off encompass too many points from adjacent patterns when
neighbouring trajectory patterns are entangled, a situation that
confuses the fit. Additionally, the data points left out from {\sc
  stripp}ing can further degrade the situation in subsequent
iterations. On the other hand, when the number of bins is increased to
20 and the cut-off level is reduced to 3$\sigma$ (lower panel), the
fit becomes more reasonable. The trade-off, however, is that some data
points remain left out from the fitted trajectories due to a smaller
tolerance in the data selection.

The two panels in the right-hand side of Fig.~\ref{fig:bincut1}
provide results for a test case where $DS=3.1$. Visually, it is
already difficult to discern any obvious pattern if plotting all data
with the same symbol. Clearly, the increased scatter in the sampled
data (as implied by the smaller $DS$ value) is causing confusion in
the recognition of adjacent patterns. As before, with 10 bins and a
6$\sigma$ cut-off level (upper panel), the fitted trajectories make no
much sense. When increasing the number of bins to 30 (lower panel),
the situation improves and reasonable results are then derived. This
shows that one is less likely to get confusion at early {\sc
  stripp}ing stages if starting from a smaller bin width. Evidently,
the use of such a reduced width should not compromise the data
sampling.

\subsubsection{Non-linear patterns with small $DS$}

Because non-linear trajectory patterns do not flow in a straight
pseudo-tangential direction, increasing the number of bins will
degrade statistics at some point. Though a larger sigma cut-off could
compensate the effect to some extent, it may also compromise the fits
since a higher fraction of data from adjacent patterns are then also
likely to be picked up, hence adding further confusion in the pattern
recognition.

\begin{figure*}
\centering
\includegraphics[height=8.5cm, angle=-90]{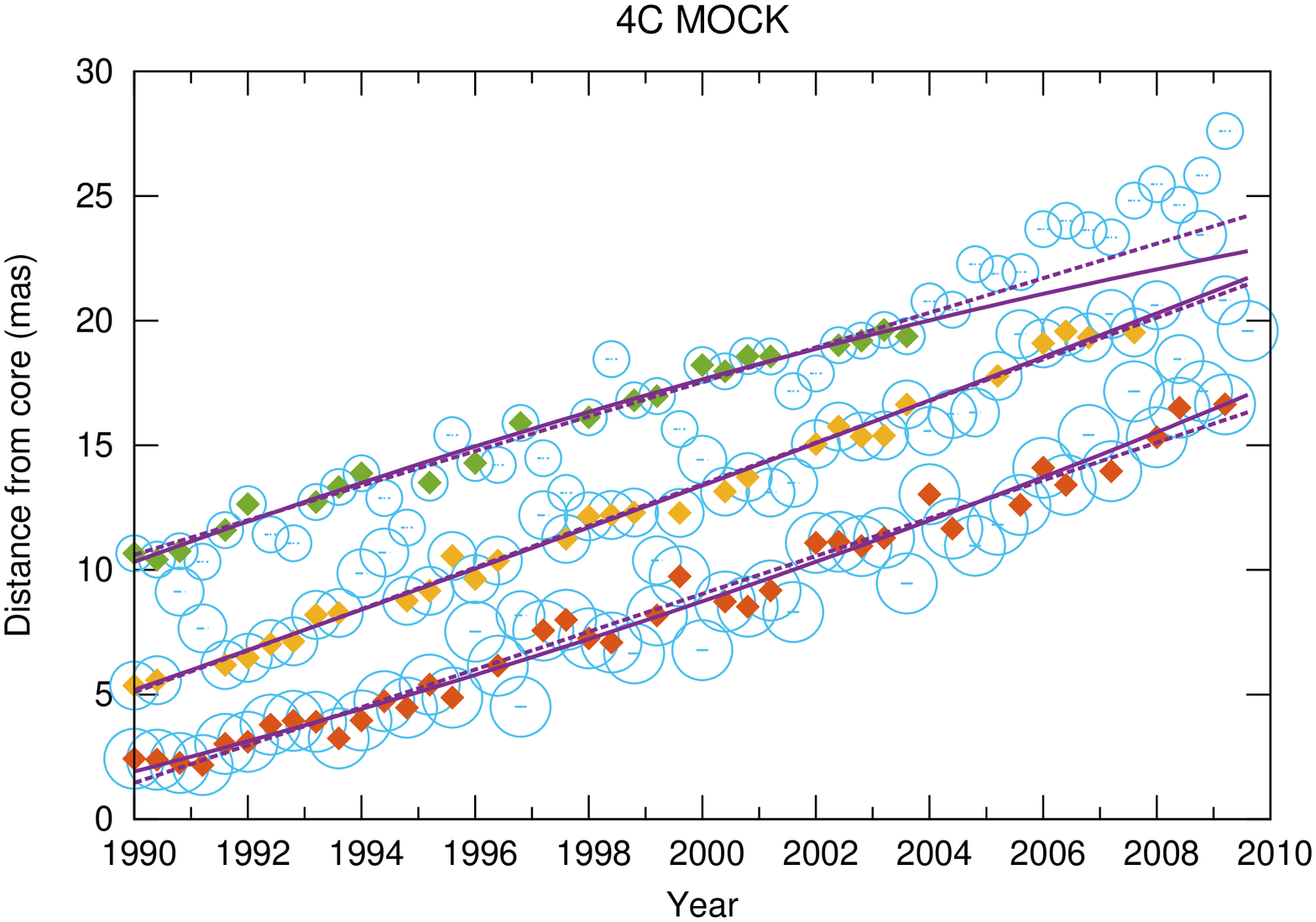}
\includegraphics[height=8.5cm, angle=-90]{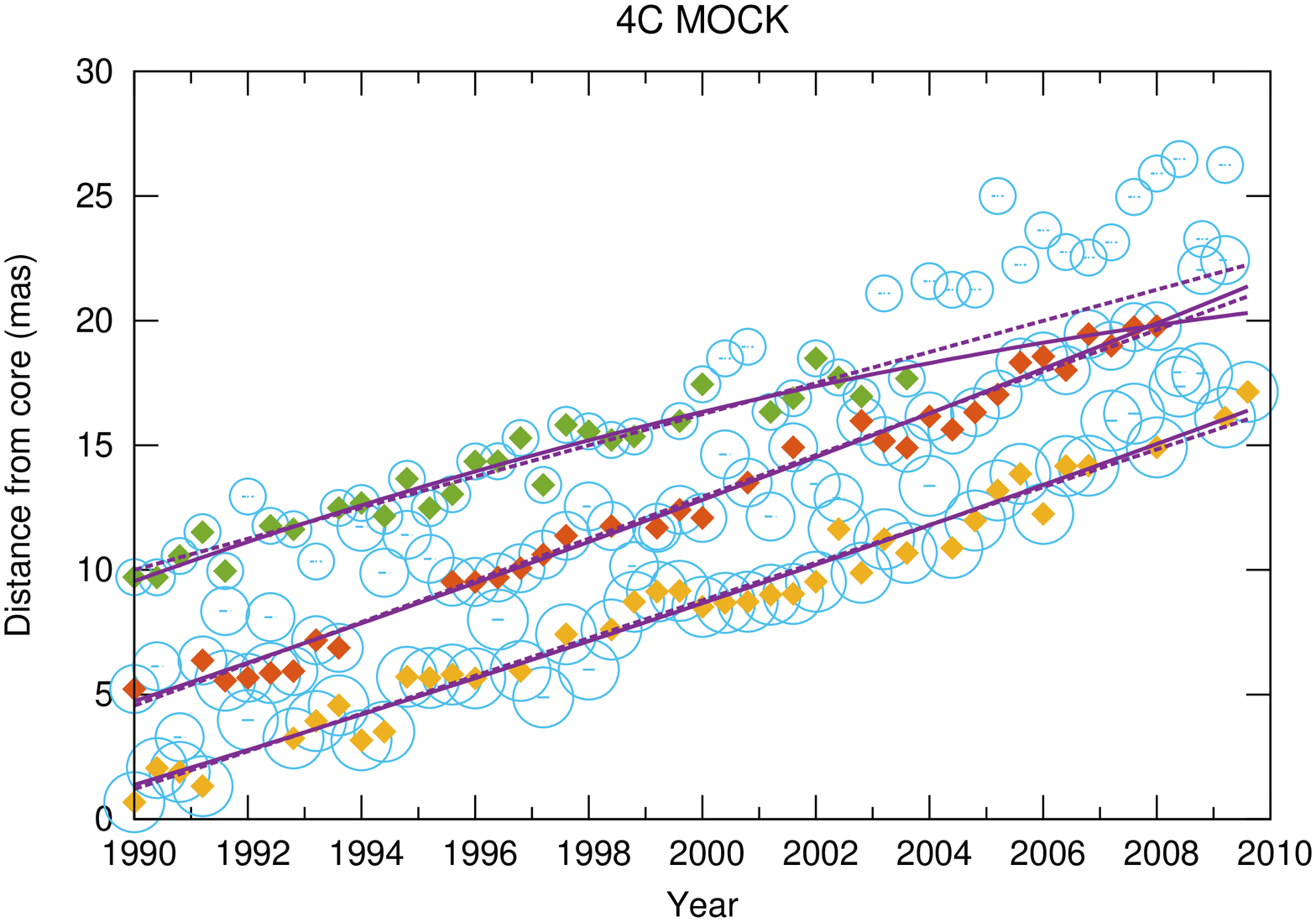}\\
\includegraphics[height=8.5cm, angle=-90]{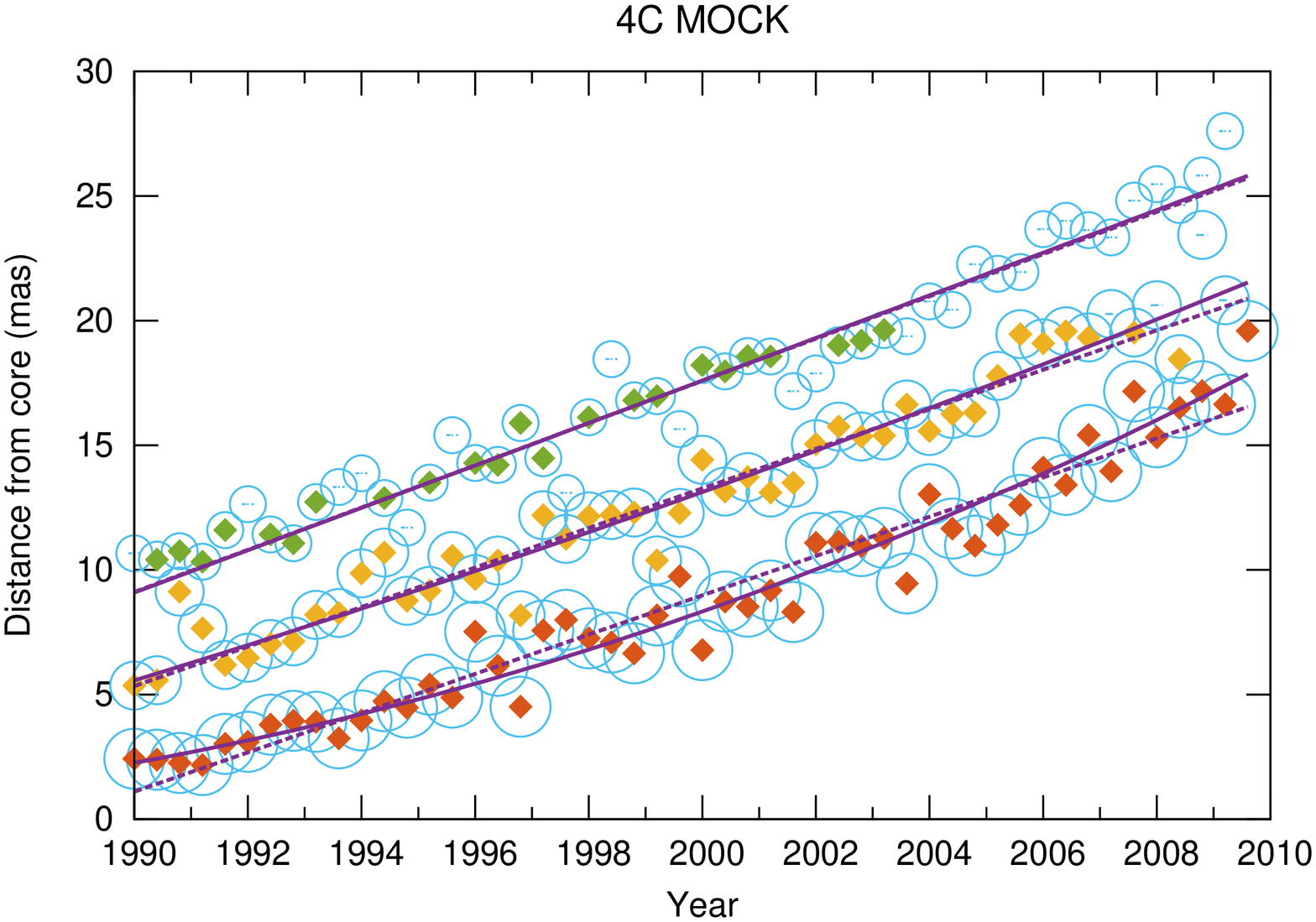}
\includegraphics[height=8.5cm, angle=-90]{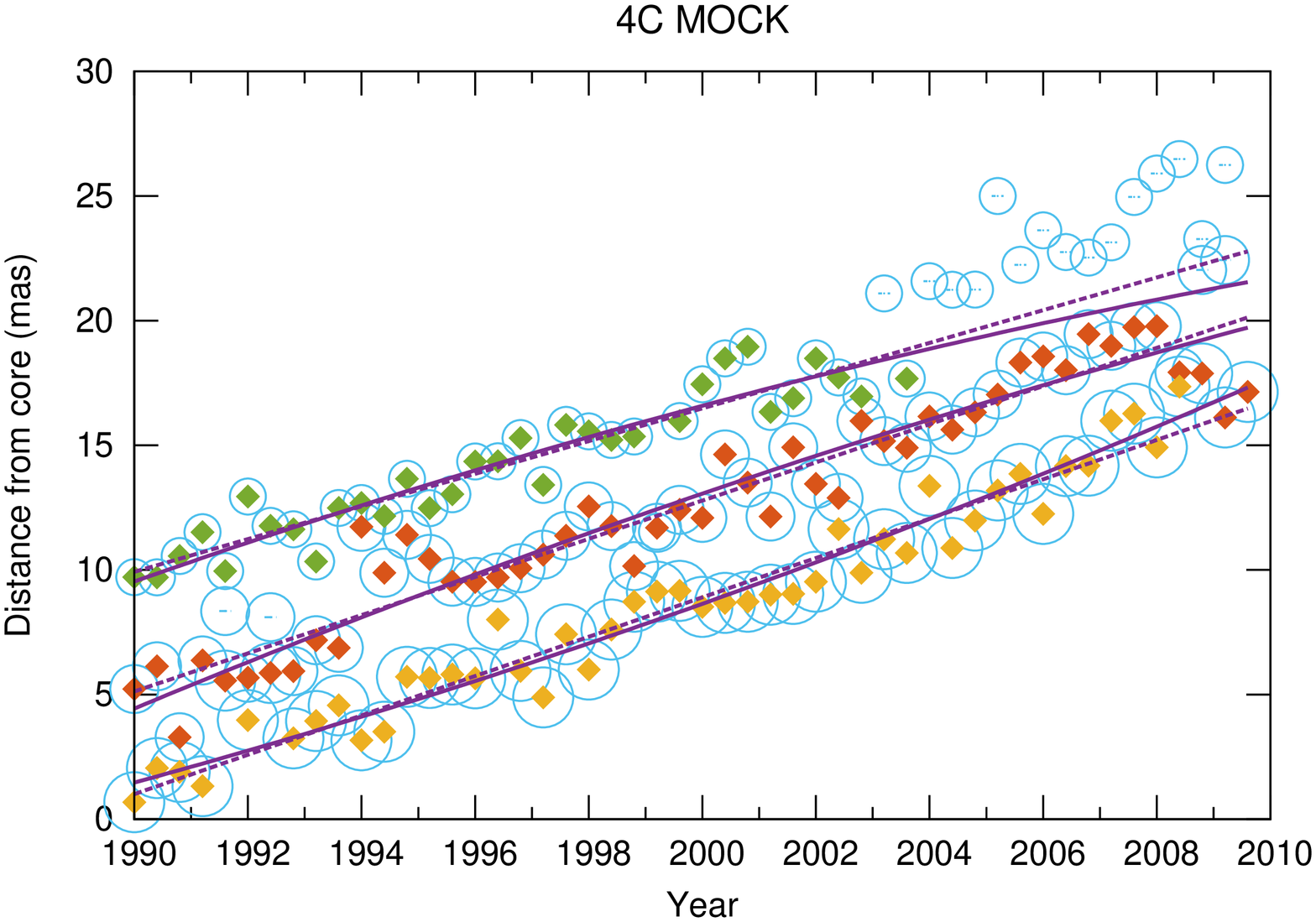}\\
\includegraphics[height=8.5cm, angle=-90]{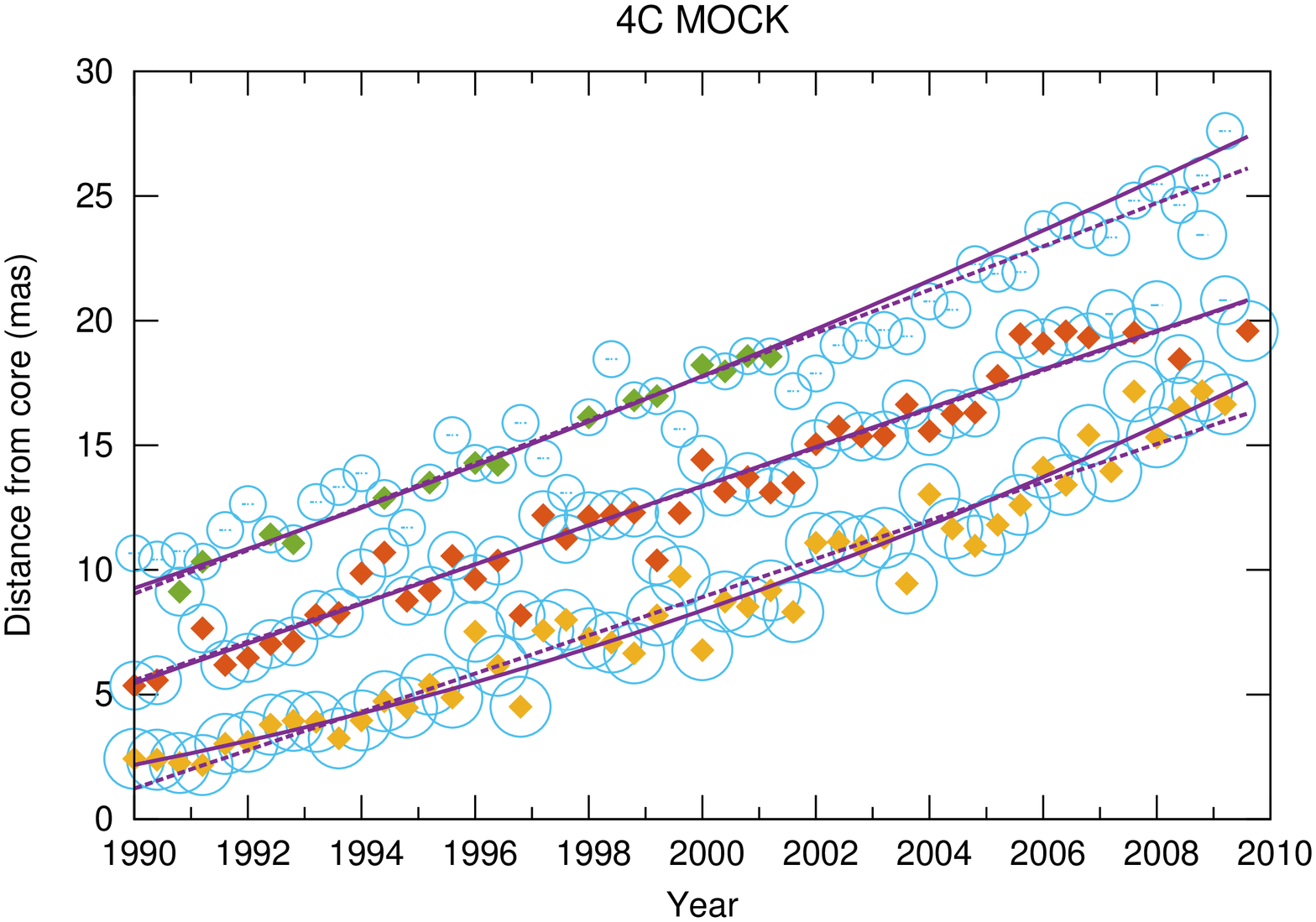}
\includegraphics[height=8.5cm, angle=-90]{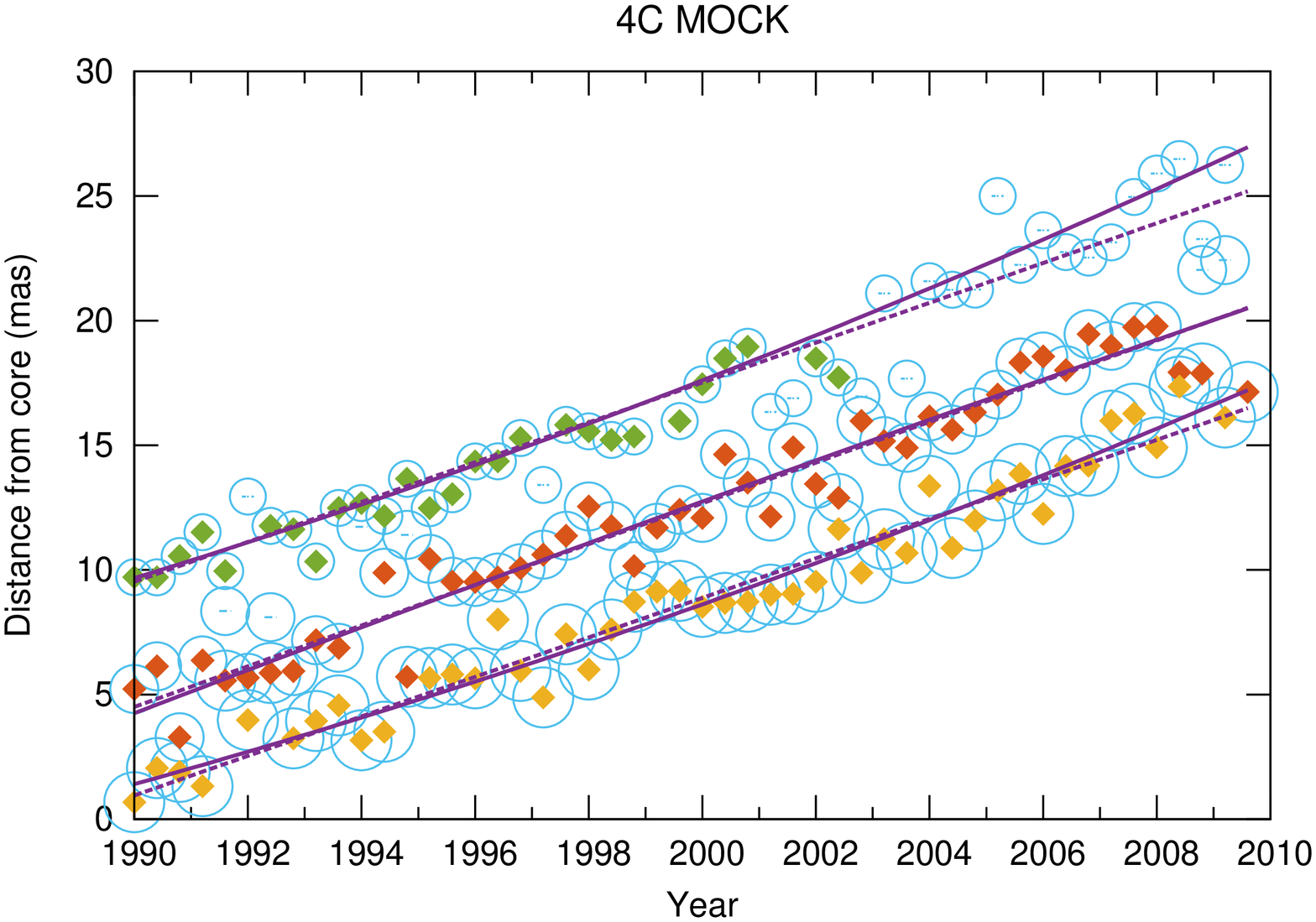}
\caption[The non-linear bin-cut examples]{ Effect of binning and
  cut-off level on non-linear mock data. Panels on the left-hand side
  are for a $DS$ value of 4, while those on the right-hand side are
  for a $DS$ value of 3.1. The number of bins and sigma cut-off are
  (10, 3), (10, 6) and (20, 6) for the upper, middle and lower
  left-hand side panels, and (10, 3), (10, 6) and (15, 6) for the
  upper, middle and lower right-hand side panels,
  respectively. Symbols are the same as those in
  Fig.~\ref{fig:psulin}.}
\label{fig:bincut2}
\end{figure*}

The three panels in the left column of Fig.~\ref{fig:bincut2} show
results of such non-linear fits for $DS=4$. Going from the upper panel
to the middle panel, the sigma cut-off level is increased from 3 to 6
while keeping the number of bins equal to 10. As noticeable when
comparing the plots in the two panels, a notable effect of this change
is that a number of data points previously left out are
recovered. When the number of bins is increased to 20 (lower panel),
confusion appears to be reduced, but the data points picked-up for the
linear fit during initial {\sc stripp}ing are also thinned out.

The panels in the right column of Fig.~\ref{fig:bincut2} show similar
tests but with a value of $DS$ reduced to 3.1, i.e. in a case with
more entangled trajectory patterns. Since quadratic patterns are more
capable to bend the direction of the flow than linear patterns do, the
data points initially picked-up for the \SNT process need to be
carefully inspected , otherwise the fitted quadratic patterns can go
astray.  Comparing results in the upper and middle panels, which were
derived with 10 bins but with a cut-off level at 3-$\sigma$ and
6-$\sigma$, respectively, one can see that the fitted trajectories
mostly differ towards their tail ends. This is because setting a
certain sigma cut-off can either include or exclude some pivotal
points which affect the curvature of the fitted patterns. Decelerating
patterns may always be rejected since our mock patterns were set to
accelerate. An increase of the number of bins from 10 to 15, as in the
case reported in the lower right panel of Fig.~\ref{fig:bincut2},
allows one to recover the curved trajectory of the third component.

\subsubsection{Extreme cases with $DS<3$}

\begin{figure*}
\centering
\includegraphics[height=8.5cm, angle=-90]{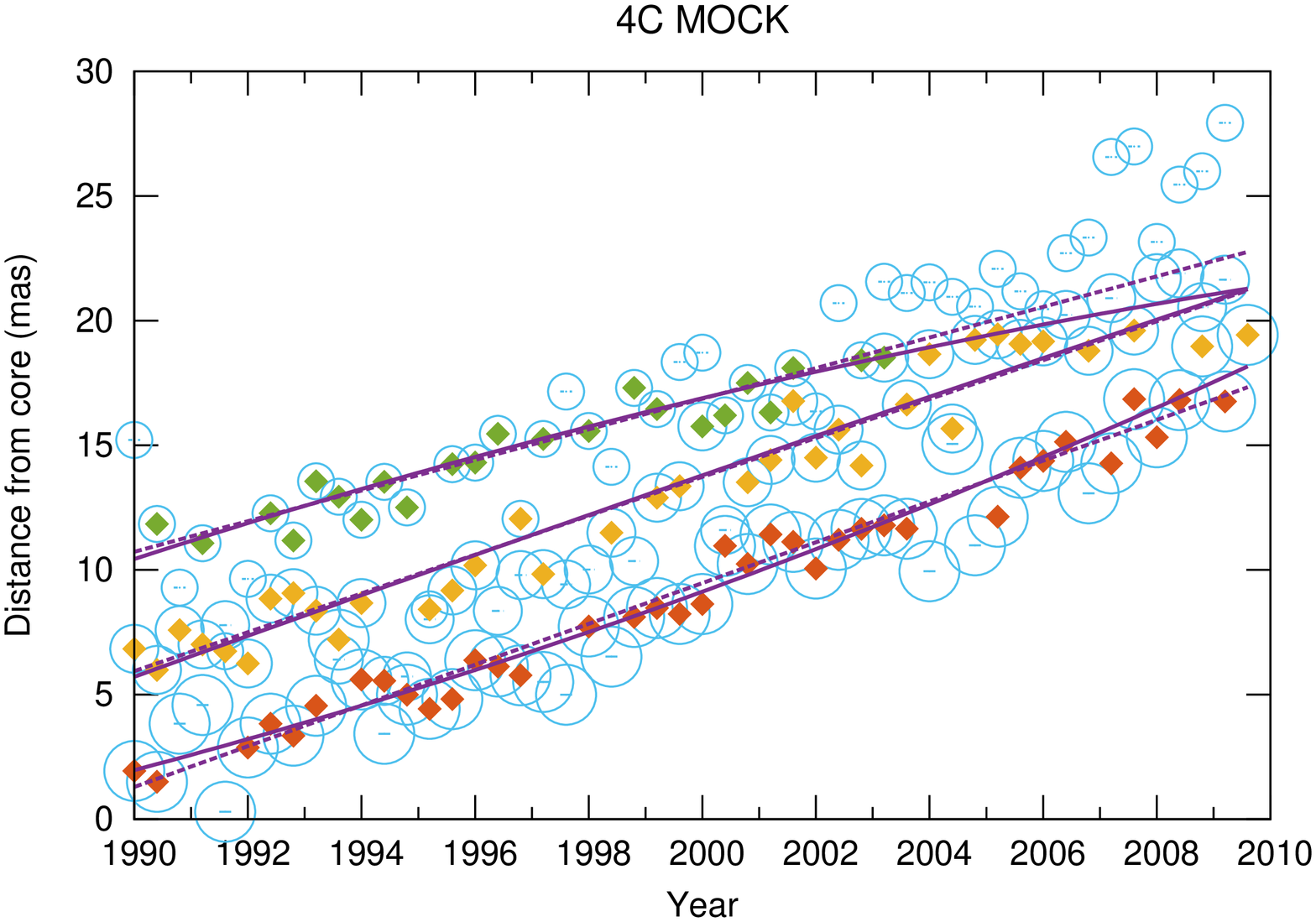}
\includegraphics[height=8.5cm, angle=-90]{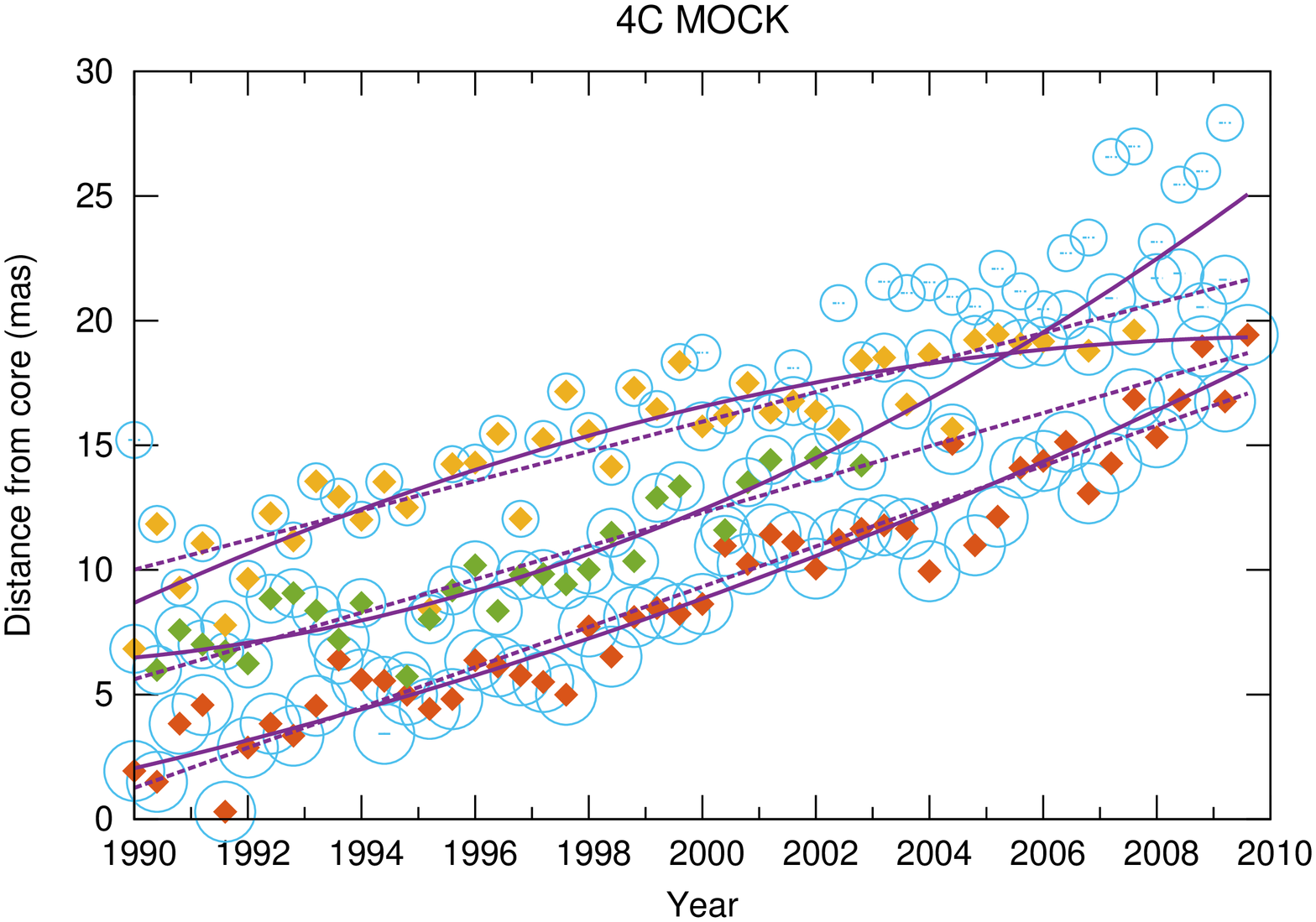}\\
\includegraphics[height=8.5cm, angle=-90]{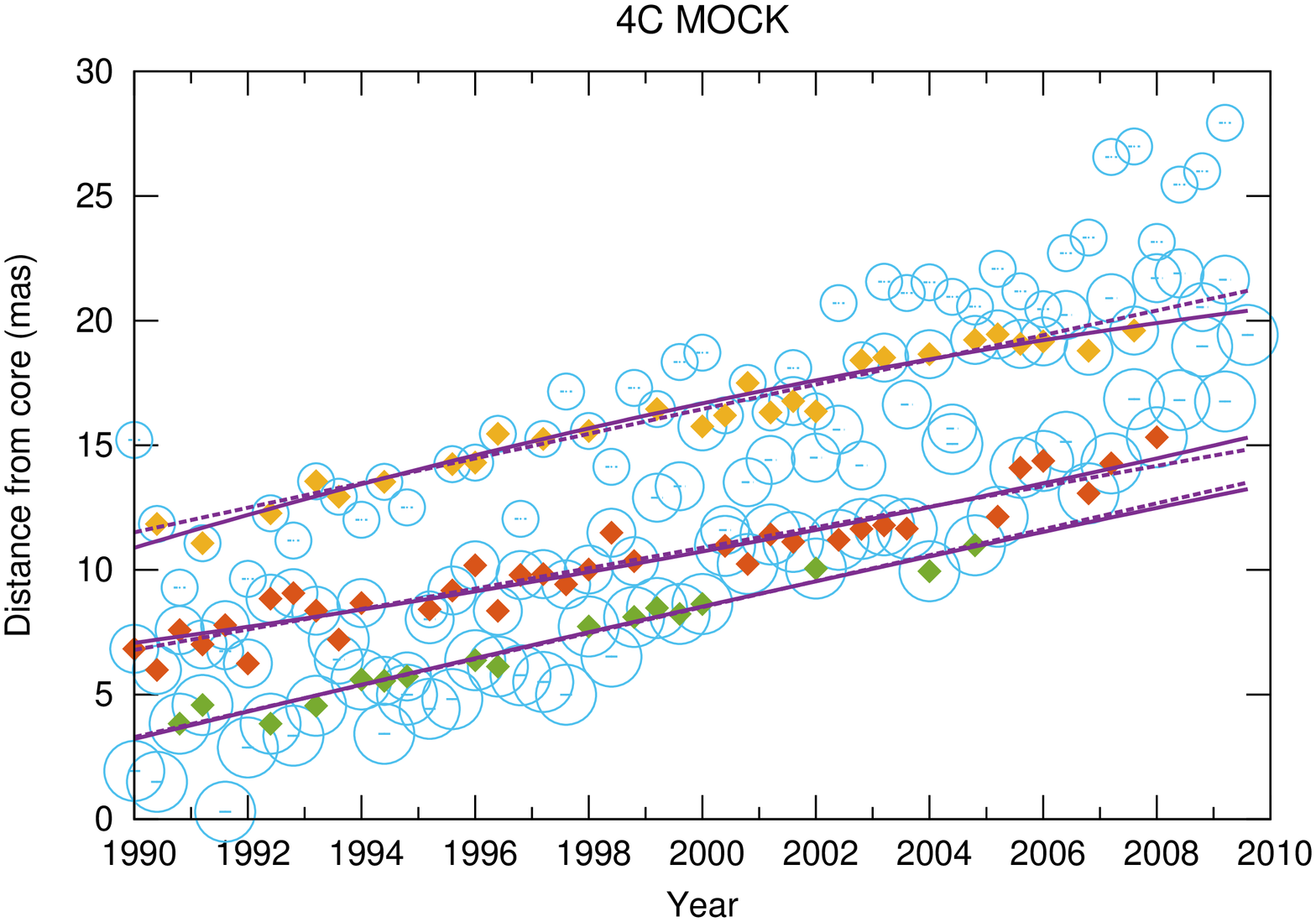}
\includegraphics[height=8.5cm, angle=-90]{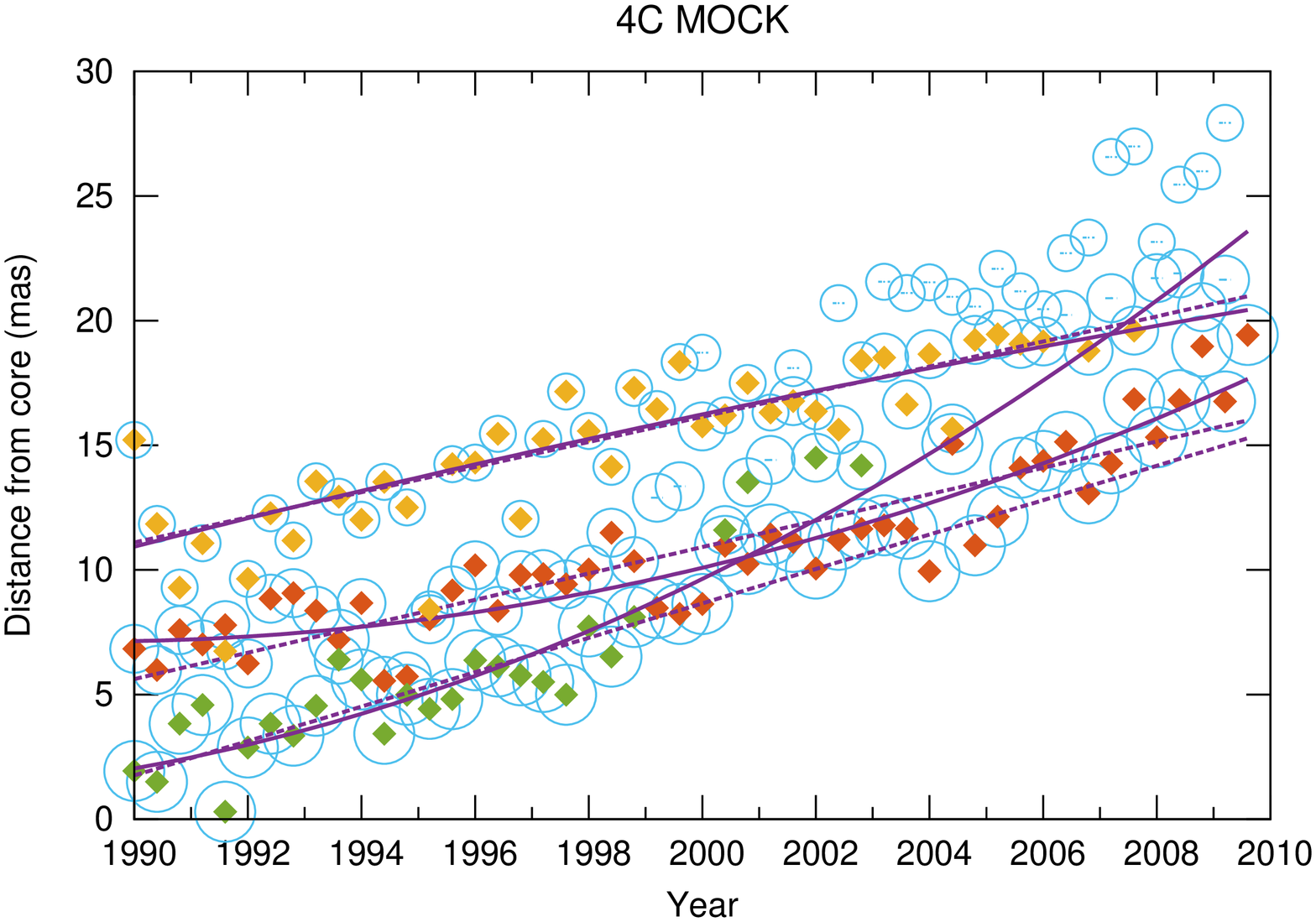}
\caption[The DS2.6 examples]{Effects of binning and cut-off level on
  non-linear mock data with $DS=2.6$. From left to right and for the
  upper to lower panels, the number of bins and sigma cut-off are (10,
  3), (10, 6), (15, 3) and (15, 6), respectively. Symbols are the same
  as those in Fig.~\ref{fig:psulin}. }

\label{fig:bincut3}
\end{figure*}

Results for a non-linear case with $DS=2.6$ are reported in
Fig.~\ref{fig:bincut3}. Comparing with the known patterns of the mock
data indicates that the regression {\sc strip} algorithm cannot
properly disentangle the trajectory patterns in such a case no matter
how the number of bins and sigma cut-off are tuned. Some results may
look plausible but they either miss out many data points or produce
wrong accelerations.

From the examples above, we find that the regression {\sc strip}
algorithm can resolve trajectory patterns easily when the $DS$ value
is larger than 3. However, when the $DS$ value is smaller than 3, the
algorithm approaches its limit. This is analogous to the situation for
a Gaussian beam where two convolved point sources need to be at least
one beam width apart ({\sc fwhm} $\simeq 2.4\sigma$) to be
resolved. Based on the statistics for our pseudo-normal distribution,
it appears that the threshold for $DS$ should be even raised further
due to the discrete and limited sampling of the data sets.

\subsection{DS-curvature degeneracy}\label{dsdegen}

For non-linear quadratic trajectories, we find that there is
a degeneracy between $DS$ and the local curvature. The local
curvature of a univariate quadratic function expressed in the
form of $y(x)=ax^2+bx+c$ is given by
\begin{equation}
k=\frac{y''}{(1+y'^2)^{3/2}} = \frac{2a}{[1+(2ax+b)^2]^{3/2}} \leq 2a .
\label{eq:curvt}
\end{equation}
At the extreme curvature points, $k_{\rm max}=2a$ which is a
constant. For different sets of quadratic trajectories, the local
pattern configurations only differ by a scaling factor,
as illustrated by tests using our regression {\sc strip} algorithm.

\begin{figure}
\centering
\includegraphics[height=8.5cm, angle=-90]{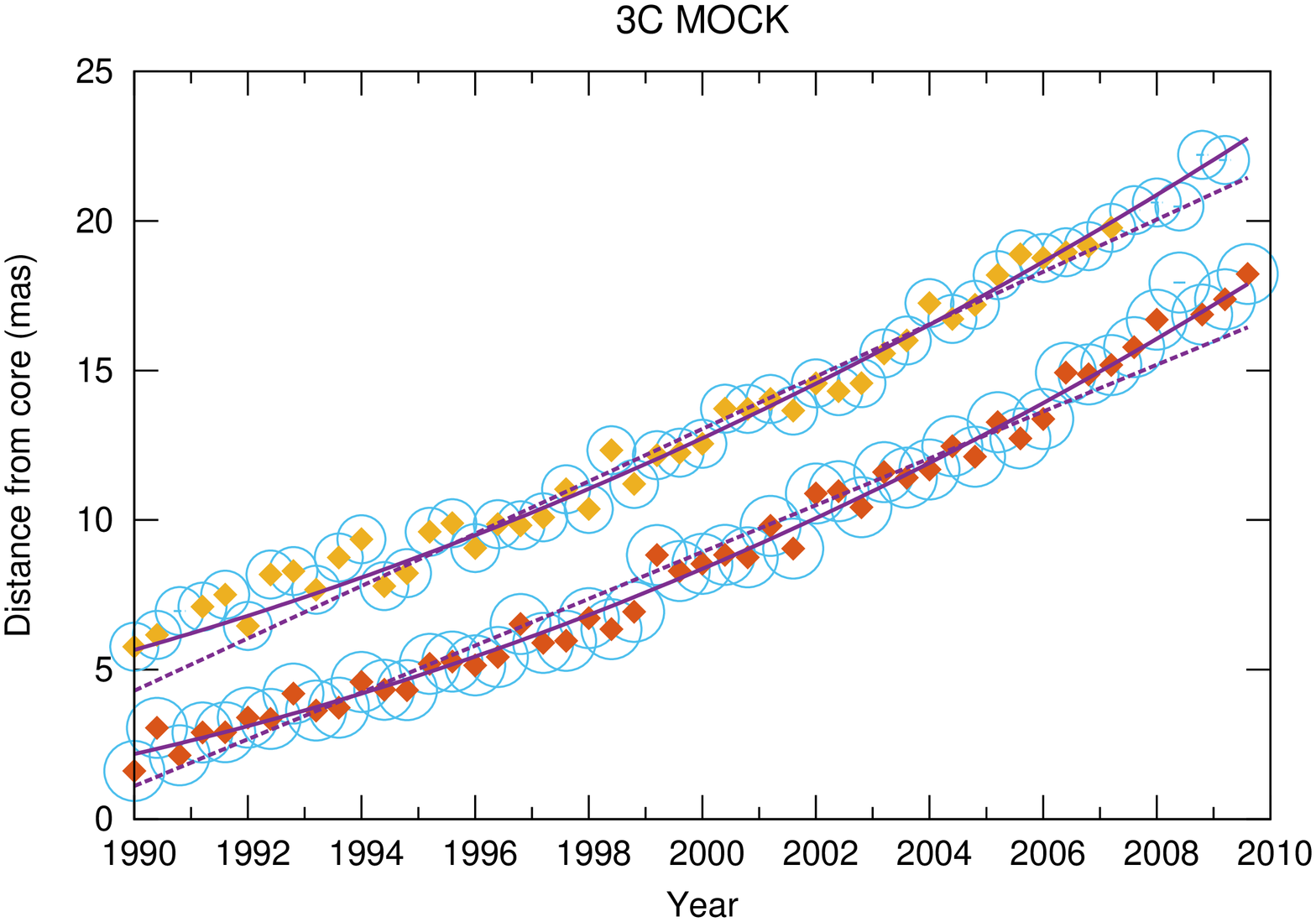}
\includegraphics[height=8.5cm, angle=-90]{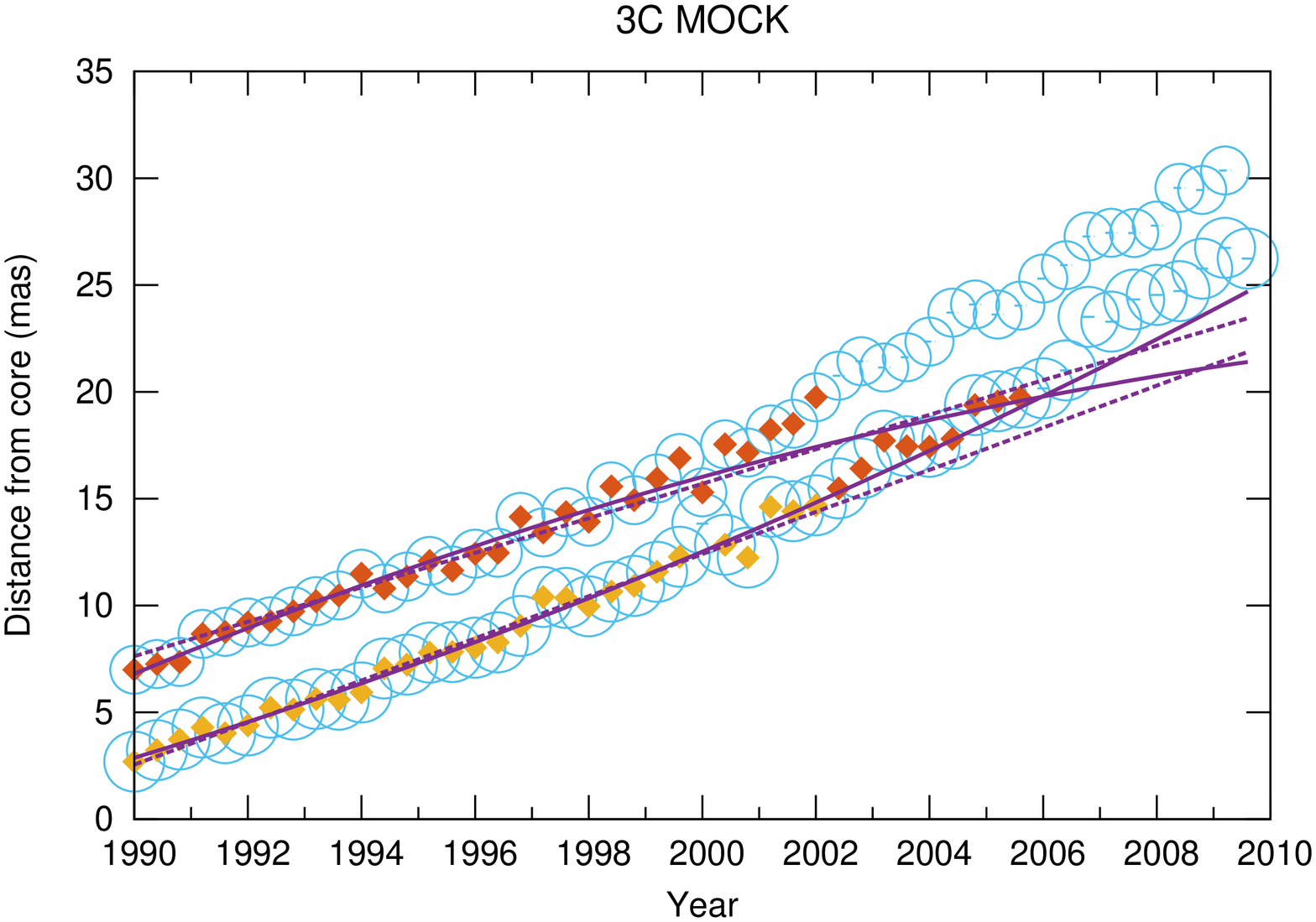}
\includegraphics[height=8.5cm, angle=-90]{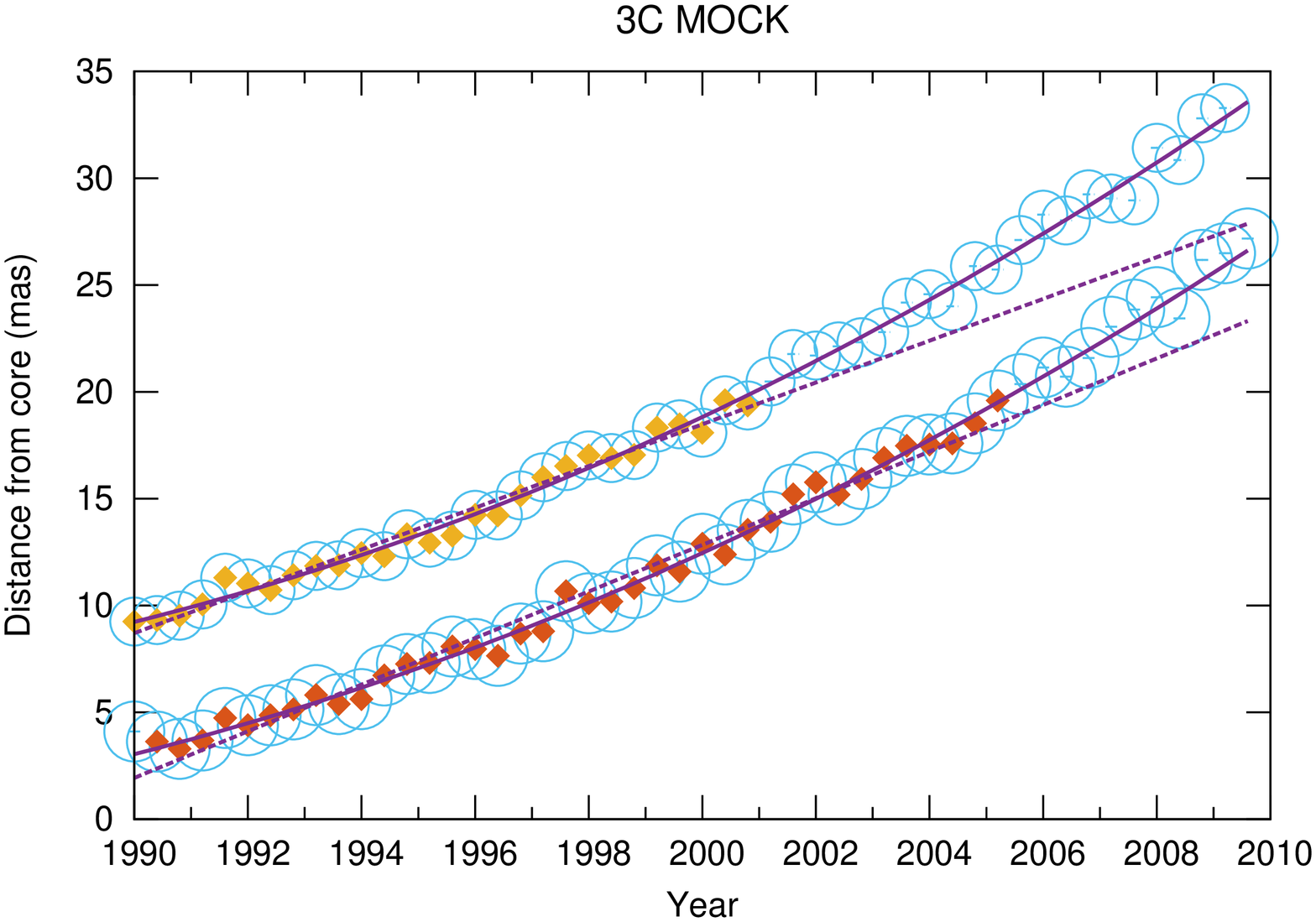}
\caption[DS-Curvature degeneracy]{The DS-curvature degeneracy in the
  case of quadratic trajectories. In these plots, the $DS$ and $k_{\rm
    max}$ values are (8, 0.02), (8, 0.03) and (12, 0.03) for the
  upper, middle and lower panels, respectively. Symbols are the same
  as those in Fig.~\ref{fig:psulin}. }
\label{fig:dsdegen}
\end{figure}

\begin{figure}
\centering
\includegraphics[width=7.5cm]{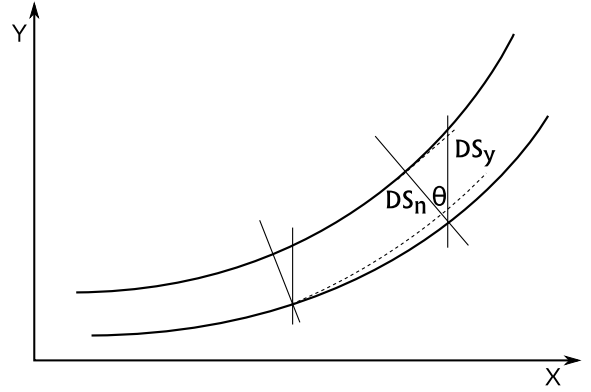}
\caption[DSn-DSy]{Illustration of the $DS$ variation for non-intersecting
quadratic curves. $DS_{n}$ is the component of $DS$ in the pseudo-normal
direction, $DS_{y}$ is the scaling of $DS$ in the Y direction, and
$\theta$ is the angle between $DS_{n}$ and $DS_{y}$.}
\label{fig:dsndsy}
\end{figure}

The plots in Fig.~\ref{fig:dsdegen} show that for two well stripped
trajectory patterns, the regression {\sc strip} algorithm gets
confused if the curvature is scaled to one and a half times the
original value. However, the confusion
goes away if the $DS$ value is also
scaled by the same amount. If noting $DS_{n}$ the component of $DS$ in
the pseudo-normal direction and $DS_{y}$ the scaled value of $DS$ in
the Y direction (see Fig.~\ref{fig:dsndsy}), it can be proved that

\[
DS_{n} \leq DS_{y} \cdot \cos\theta, \quad\textrm{and}\quad
x\uparrow\; \theta\downarrow\; DS_{n}\uparrow\; DS_{y}\uparrow .
\]
In this case, the trajectory patterns are so well separated near their
tail ends where $x\gg1$ that the algorithm has no problem picking them
up even with a larger sigma cut-off.

\section{Trials with real observations}\label{trials}

\subsection{Description of data}

VLBI observations are generally categorized into two broad classes,
those dedicated to geodesy and astrometry and those dedicated to
astrophysics. They are both based on the same technique but have
different scopes. Astrometric VLBI is targeted to finding compact and
stable sources over the entire sky to construct highly-accurate
celestial reference frames, while astrophysical VLBI is more focused
on extended and variable sources which are best to study internal
physical processes in the sources. Since the radio source count drops
with frequency, astrometric VLBI has mostly concentrated on observing
at centimetre wavelengths. On the other hand, astrophysical VLBI may
be conducted up to millimetre wavelengths for higher
resolution. Nevertheless, there are a number of common sources between
these programmes since all-sky surveys have been conducted on each
side. In this light, our pipeline is of high interest since it can
efficiently cross-examine multi-epoch multi-band VLBI data from
various programmes for better modelling.

For demonstrating the capabilities of our pipeline, we have
tested it on data from several VLBI monitoring programmes, available
either publicly or through our collaboration. These include the
RDV\footnote{ The RDV ({\em Research and Development VLBA}) programme
  is a joint geodetic and astrometric research programme of NASA, NRAO
  and USNO carried out at S/X dual-band (2/8~GHz).},
MOJAVE\footnote{The MOJAVE ({\em Monitoring Of Jets in Active galactic
    nuclei with VLBA Experiments}) programme is a VLBA programme
  carried out at K band (15~GHz) to monitor radio brightness and
  polarization variations in QSO jets.} and
VLBA-BU-Blazar\footnote{{\em Boston University Gamma-ray Blazar
    monitoring program with the VLBA at 43~GHz} } programmes. The RDV
observations have a primary geodetic-astrometric scope, while the MOJAVE and
VLBA-BU-Blazar observations have astrophysical scopes.

\subsection{The source of interest: 1308+326}

The radio source 1308+326 is a well-known quasi-stellar object (QSO)
that has been a target of the RDV, MOJAVE and VLBA-BU-Blazar
monitoring programmes. VLBI images of 1308+326 (see
Fig.~\ref{fig:chtX}) reveal that its morphology became more complex
and more extended starting from around 2000.  For this reason, the RDV
observations were strongly reduced afterwards and there is only sparse
data beyond 2004. On the other hand, the MOJAVE data of 1308+326 span
over two decades (since 1995). The more recent VLBA-BU-Blazar
observations were initiated in 2009. In this paper, we use 1308+326
only for demonstration purposes, rather than for a thorough study of
the kinematics of its jet.

\subsection{Detection of proper motions}\label{promo}

As shown above, trajectory pattern recognition takes place at the same
time as the regression analysis in the regression {\sc strip}
algorithm implemented in the {\sc sand} pipeline. When a trajectory
pattern is identified, proper motions are thus determined
concurrently. Such proper motions may be further used to study jet
kinematics and constrain active galactic nucleus (AGN) models. Tests
using mock data have shown that the algorithm can cope with a variety
of situations. In reality, one has to bear in mind that the results of
the fits depend at some level on the data sampling, hence a favourable
situation may deteriorate if the data sampling is inadequate. In the
following, we discuss the proper motions determined with our algorithm
at the S~band, X~band, K~band and Q~band frequencies.

\begin{figure*}
\centering
\includegraphics[width=6cm, angle=-90]{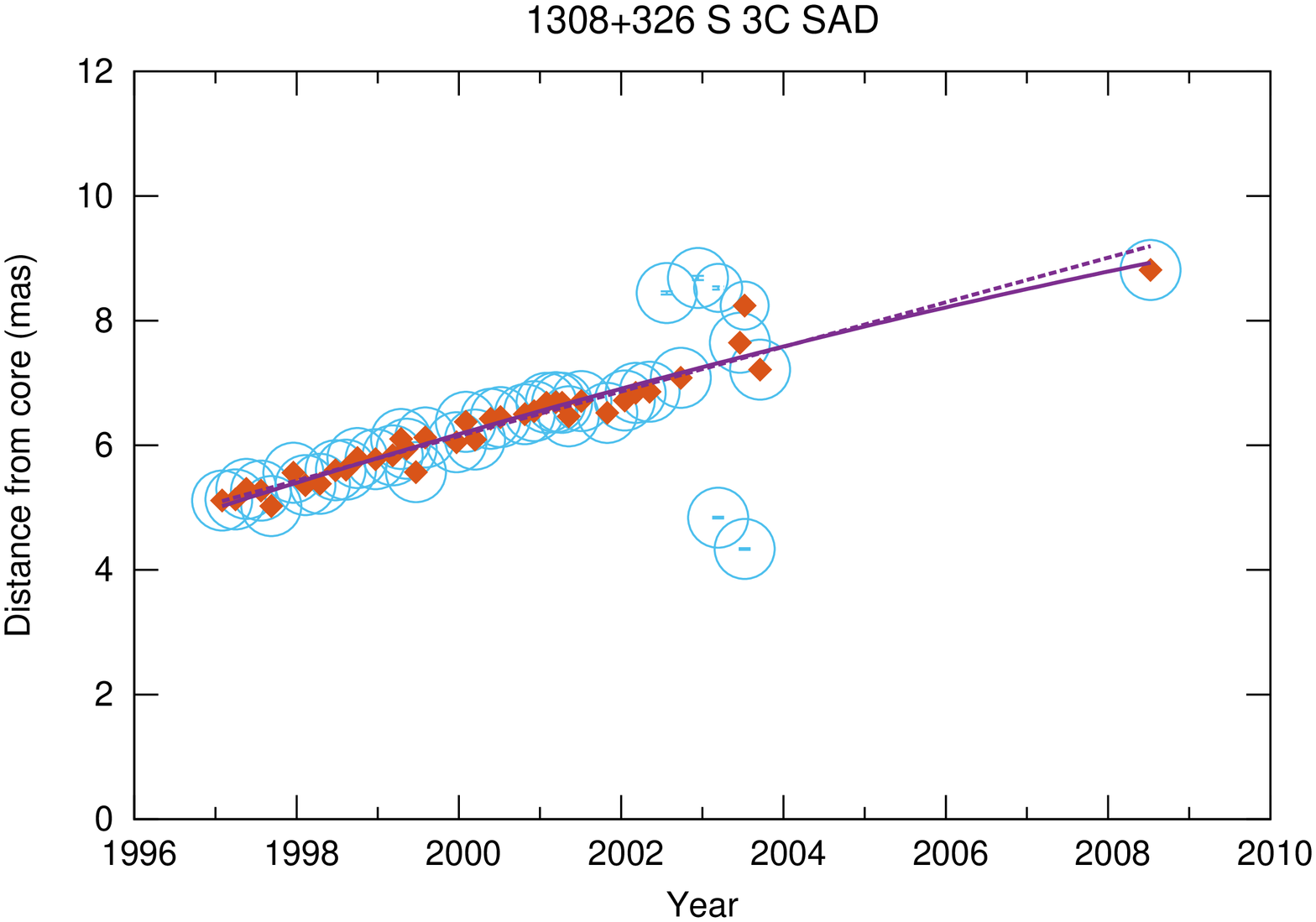}
\includegraphics[width=6cm, angle=-90]{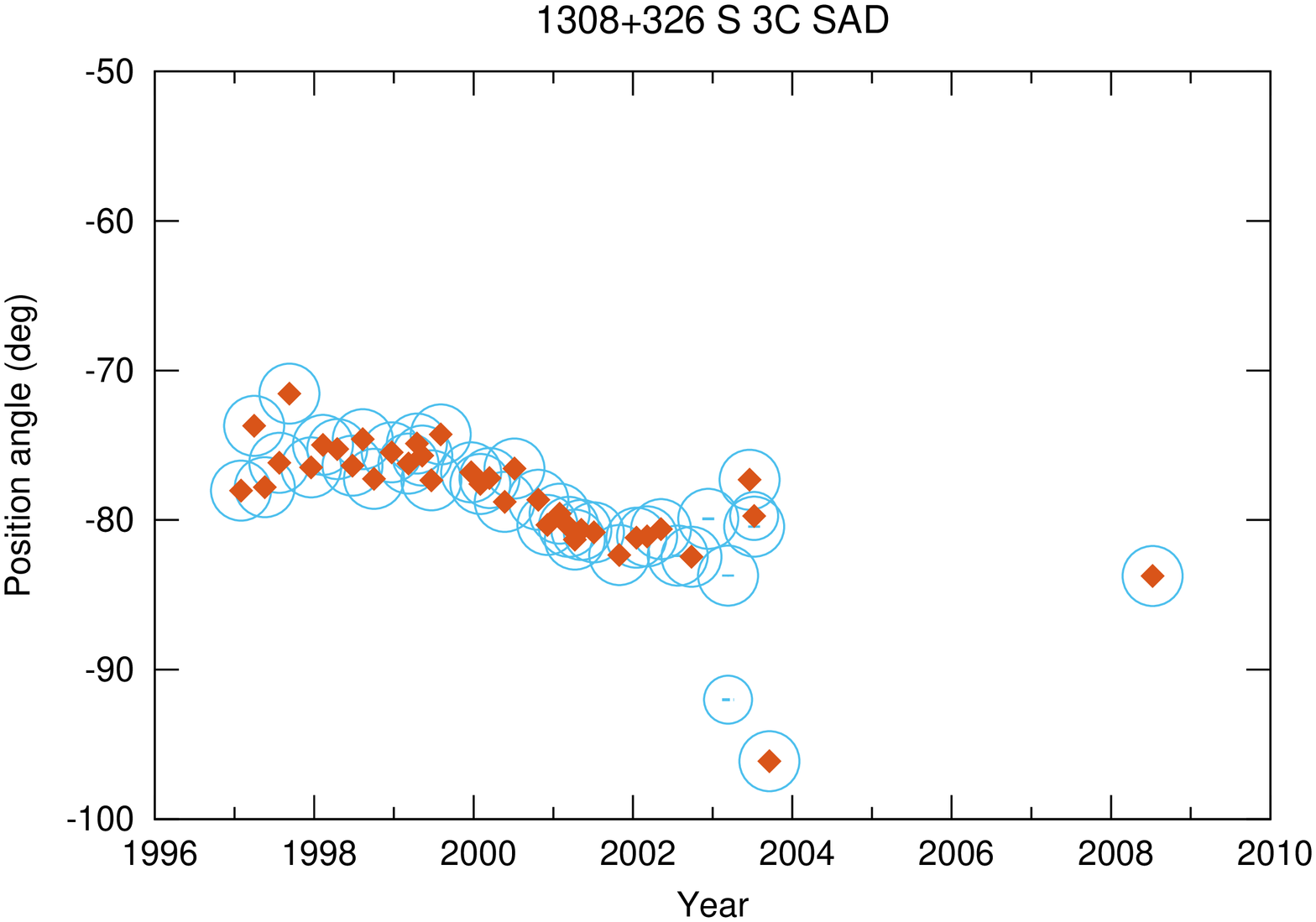}\\
\includegraphics[width=6cm, angle=-90]{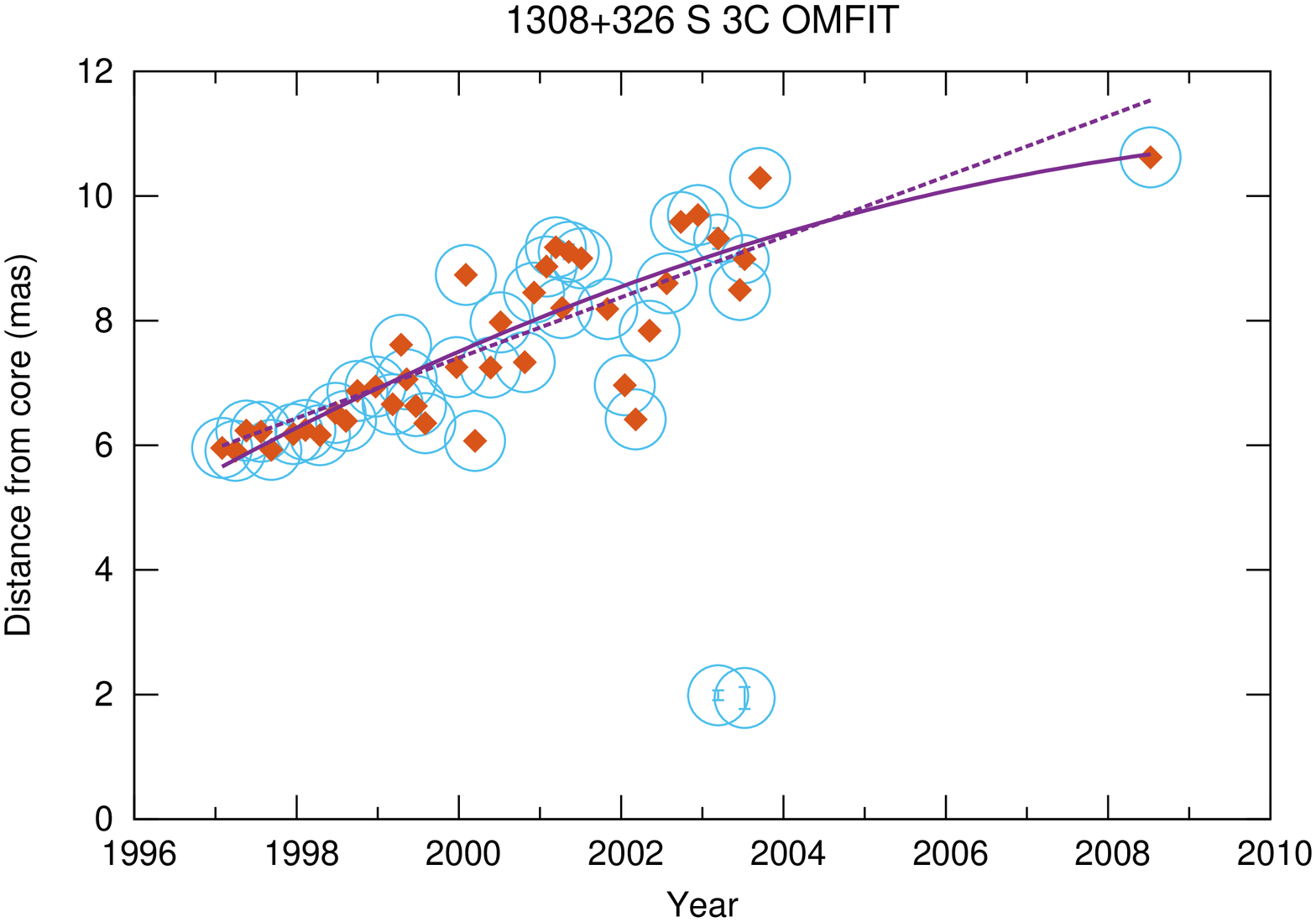}
\includegraphics[width=6cm, angle=-90]{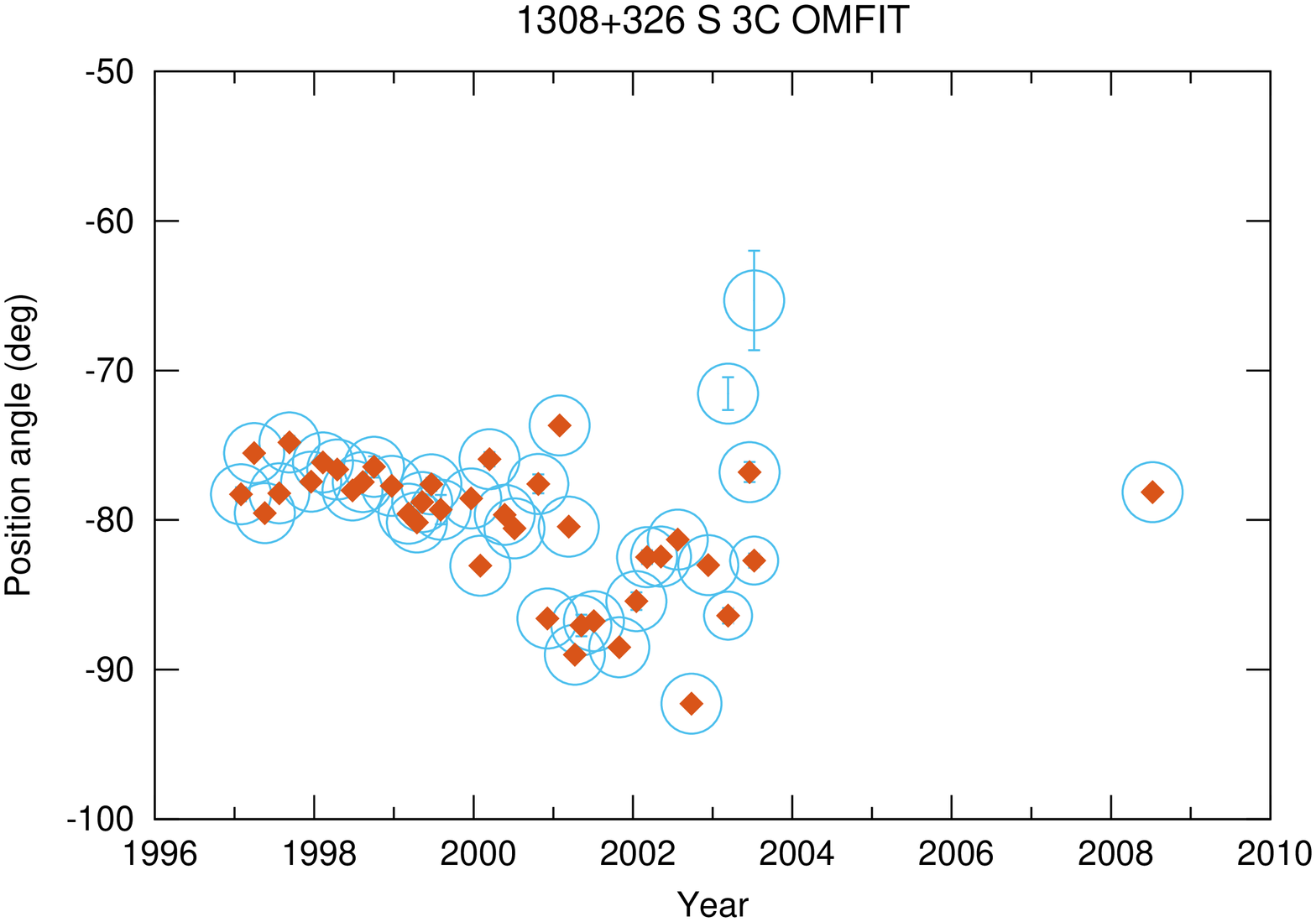}
\caption[S-band proper motion patterns]{Proper motion patterns for
  1308+326 at the S band frequency, as inferred from model-fitting in
  the image-plane (upper panels) and $uv$-plane (lower panels). The
  left panels show distances to the core while the right panels show
  radial position angles. Symbols are the same as those in
  Fig.~\ref{fig:psulin}.}
\label{fig:promS}
\end{figure*}

At the S band frequency, the detected proper motion pattern is
significant and one sees that it can be characterized simply with a
single component (Fig.~\ref{fig:promS}). This is because the
resolution at S band is lower. While a straight line fits the proper
motion of this unique component fairly well, a number of outliers are
found around epoch 2003. Such a scatter may be caused by a newly-born
component emerging from the core by that time. However, no firm
conclusion can be drawn in the absence of regular observations beyond
2004. From the PA plots in Fig.~\ref{fig:promS} (right panels), it is
found that the detected jet component roughly moves along a straight
line, though with a slight turning over time. Comparing the fits in
the image plane (upper panels) and $uv$-plane (lower panels) shows
that the results derived with the two approaches are consistent within
half a milliarcsecond. A linear fit of the component proper motion
then leads to an angular speed of 0.358$\pm$0.015~\masyr. In a final
stage, our pipeline automatically checks the NED\footnote{The
  NASA/IPAC Extragalactic Database (NED) is operated by the Jet
  Propulsion Laboratory, California Institute of Technology.}
database to retrieve the source redshift and derive its angular
diameter distance with an embedded cosmological calculator. At a
redshift of 0.996, the angular proper motion for 1308+326 corresponds
to an apparent superluminal velocity of~18.1~c.\footnote{Here and
  elsewhere in this paper we use a flat $\Lambda$CDM cosmological
  model with $\Omega_m=0.3$, $\Omega_\Lambda=0.7$ and $H_0=72$ km\
  s$^{-1}$Mpc$^{-1}$.}

\begin{figure*}
\centering
\includegraphics[width=6cm, angle=-90]{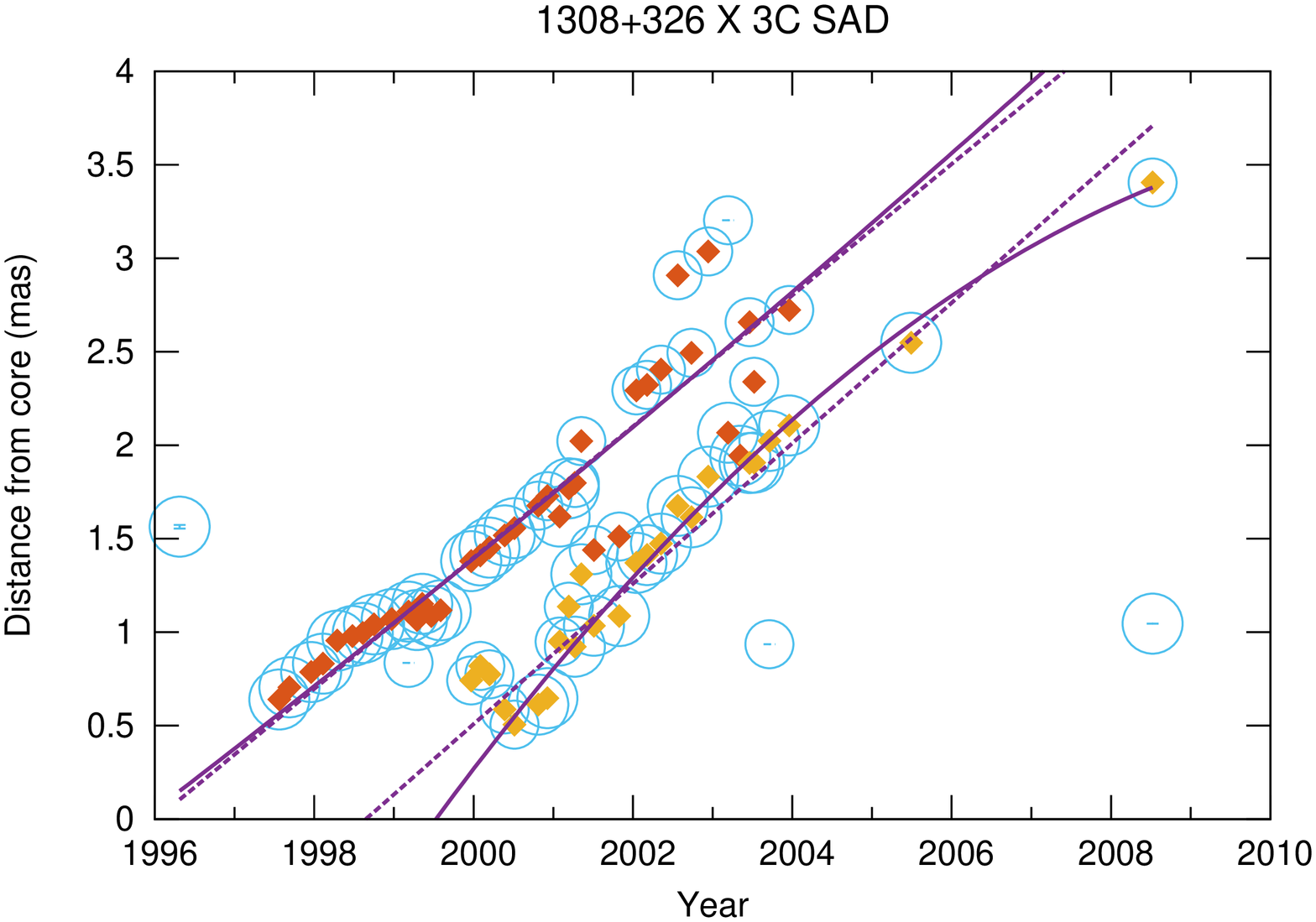}
\includegraphics[width=6cm, angle=-90]{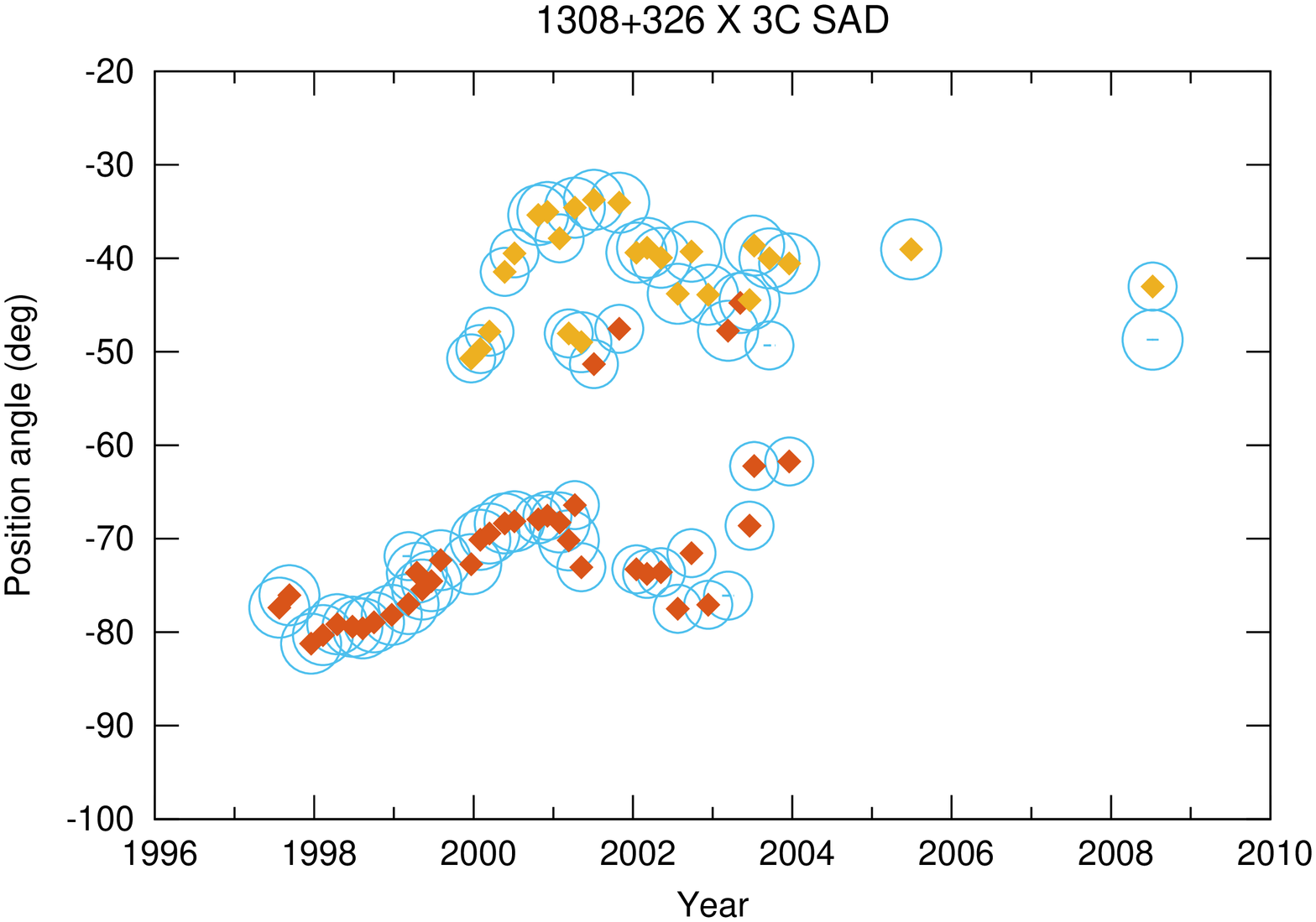}\\
\includegraphics[width=6cm, angle=-90]{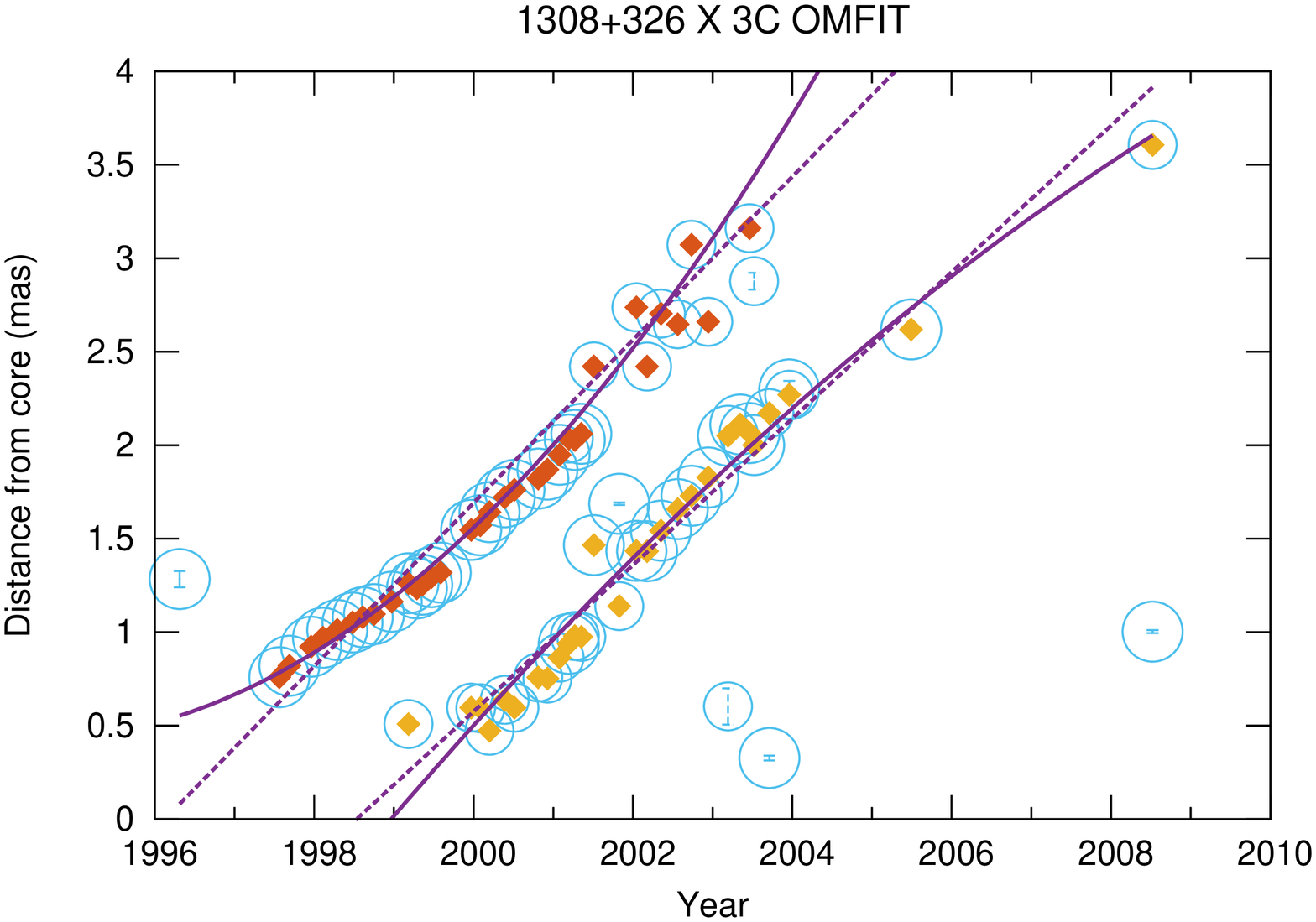}
\includegraphics[width=6cm, angle=-90]{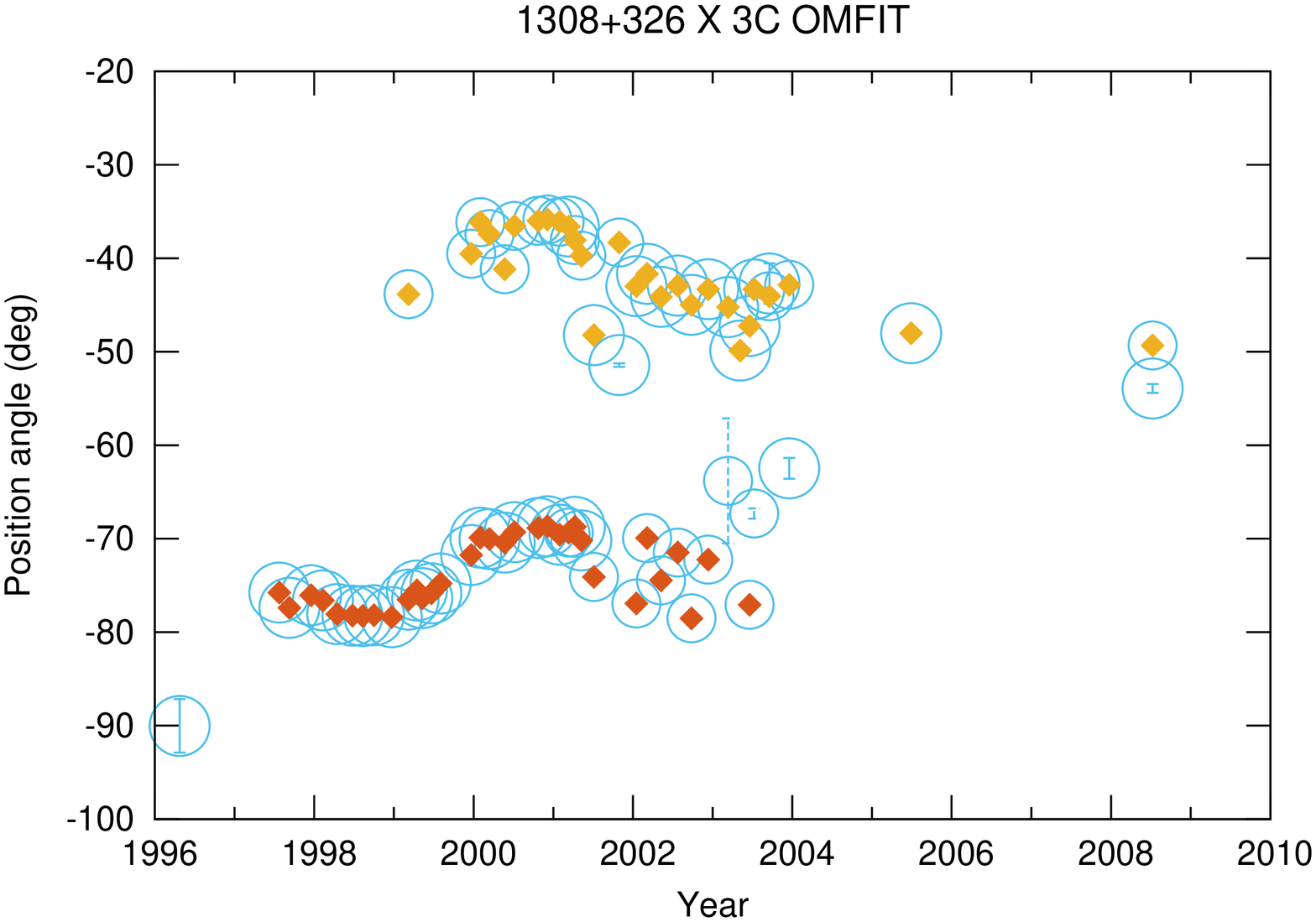}
\caption[X-band proper motion patterns]{Proper motion patterns for
  1308+326 at the X band frequency, as inferred from model-fitting in
  the image-plane (upper panels) and $uv$-plane (lower panels). The
  left panels show distances to the core while the right panels show
  radial position angles. Symbols are the same as those in
  Fig.~\ref{fig:psulin}.}
\label{fig:promX}
\end{figure*}

At the X band frequency, two significant proper motion patterns are
identified, both of which can be easily fitted with a straight line or
a quadratic line (Fig.~\ref{fig:promX}). The $DS$ value for these
patterns is about 3.8. In Fig.~\ref{fig:promX}, the different colours
assigned to components in the same pattern indicate that all such
components were not extracted at the same iteration. The components
are mostly ordered according to their flux when extracted, but this
does not correspond to reality because component brightness varies
with time. The correct component assignment is determined when the fit
is tuned during the \SNT process. Just as at the S~band frequency,
components belonging to the same pattern roughly line up but the
direction of motion is changing with time (see plots in the right
panels of Fig.~\ref{fig:promX}). The scatter around 2003, when a new
component is suspected to emerge, is found at X band too. The fitted
angular speeds of the two detected components are
0.351$\pm$0.015~\masyr and 0.376$\pm$0.018~\masyr, corresponding to
apparent superluminal velocities of 17.8~c and 19.0~c, respectively.

\begin{figure*}
\centering
\includegraphics[width=6cm, angle=-90]{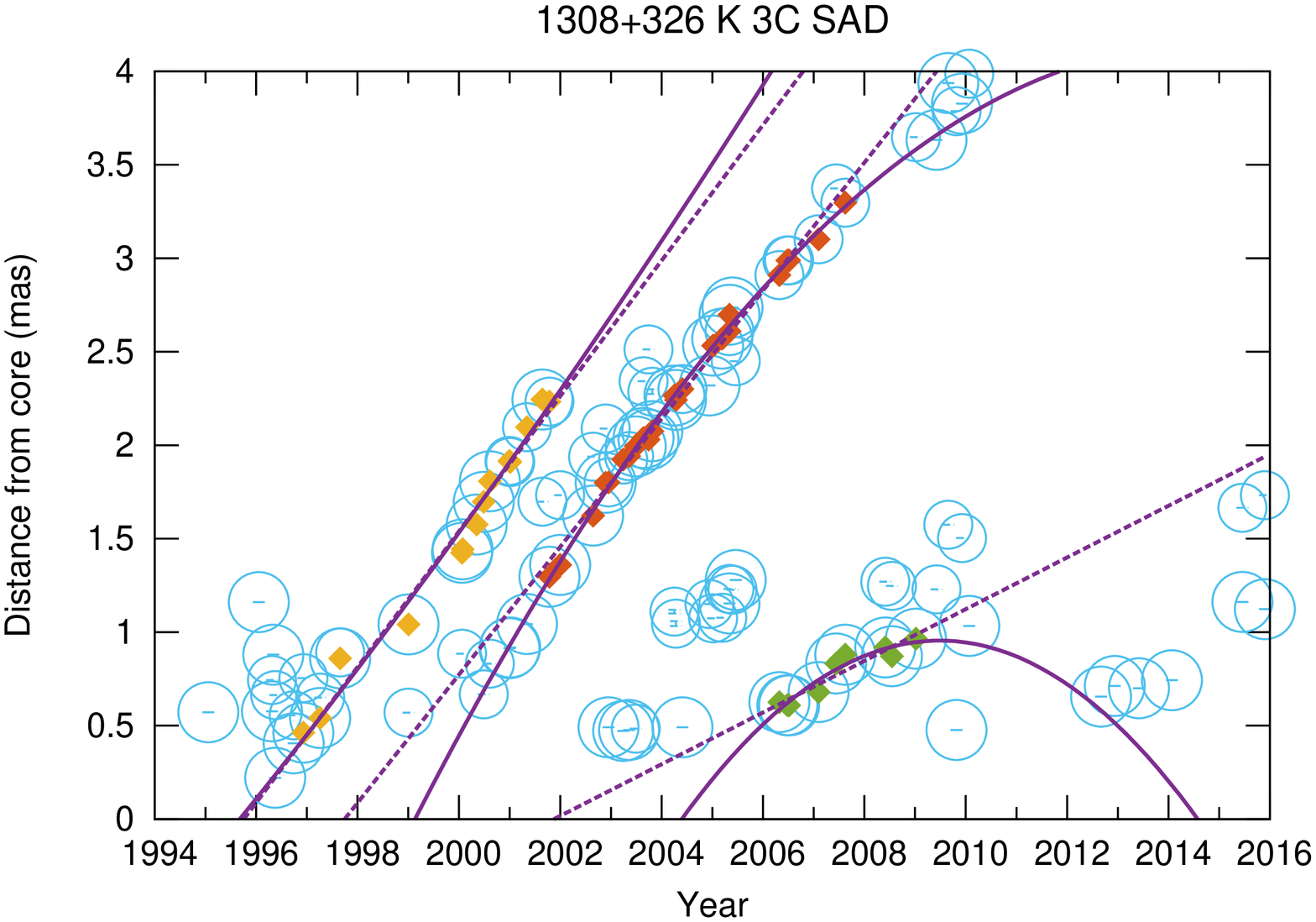}
\includegraphics[width=6cm, angle=-90]{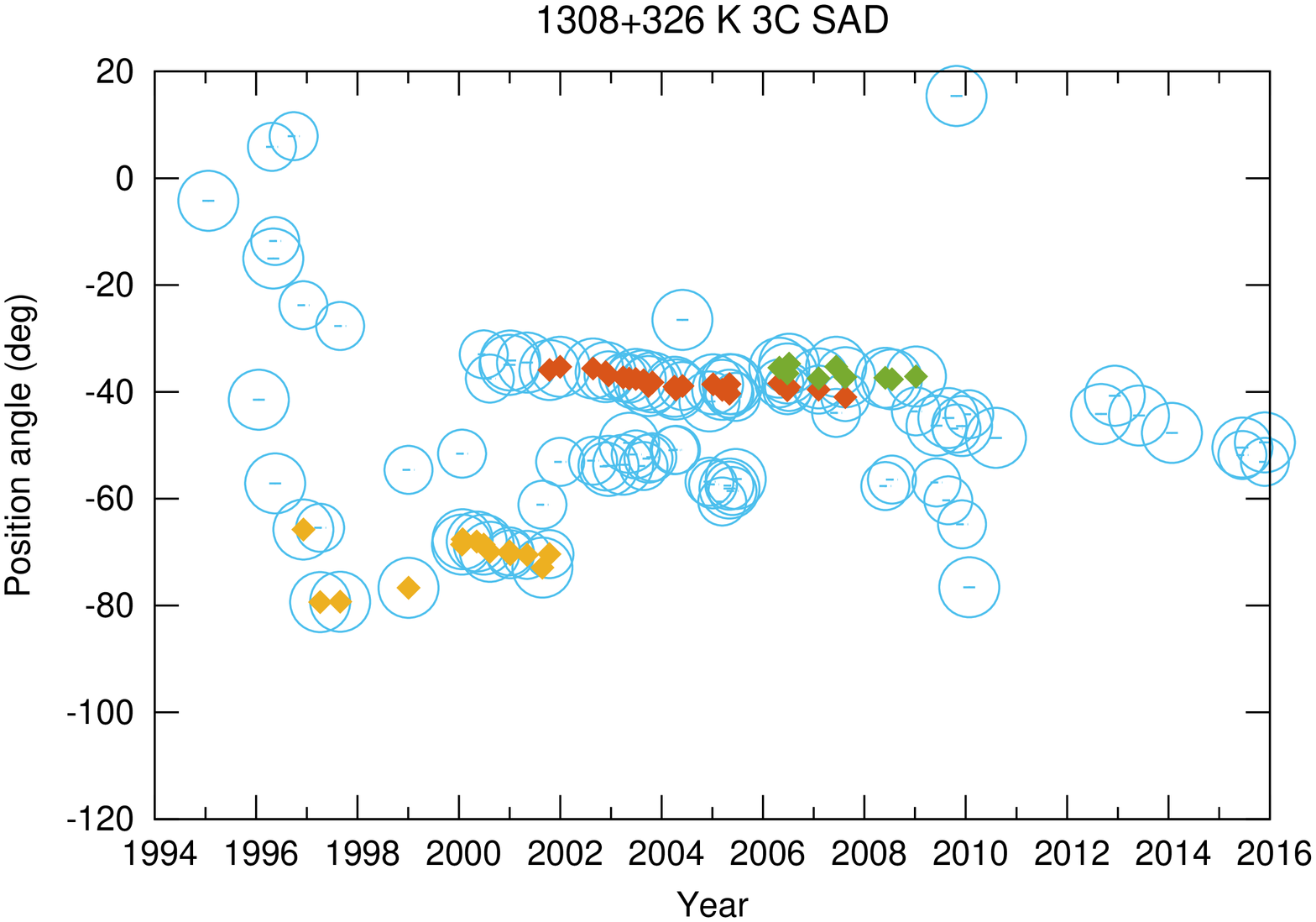}\\
\includegraphics[width=6cm, angle=-90]{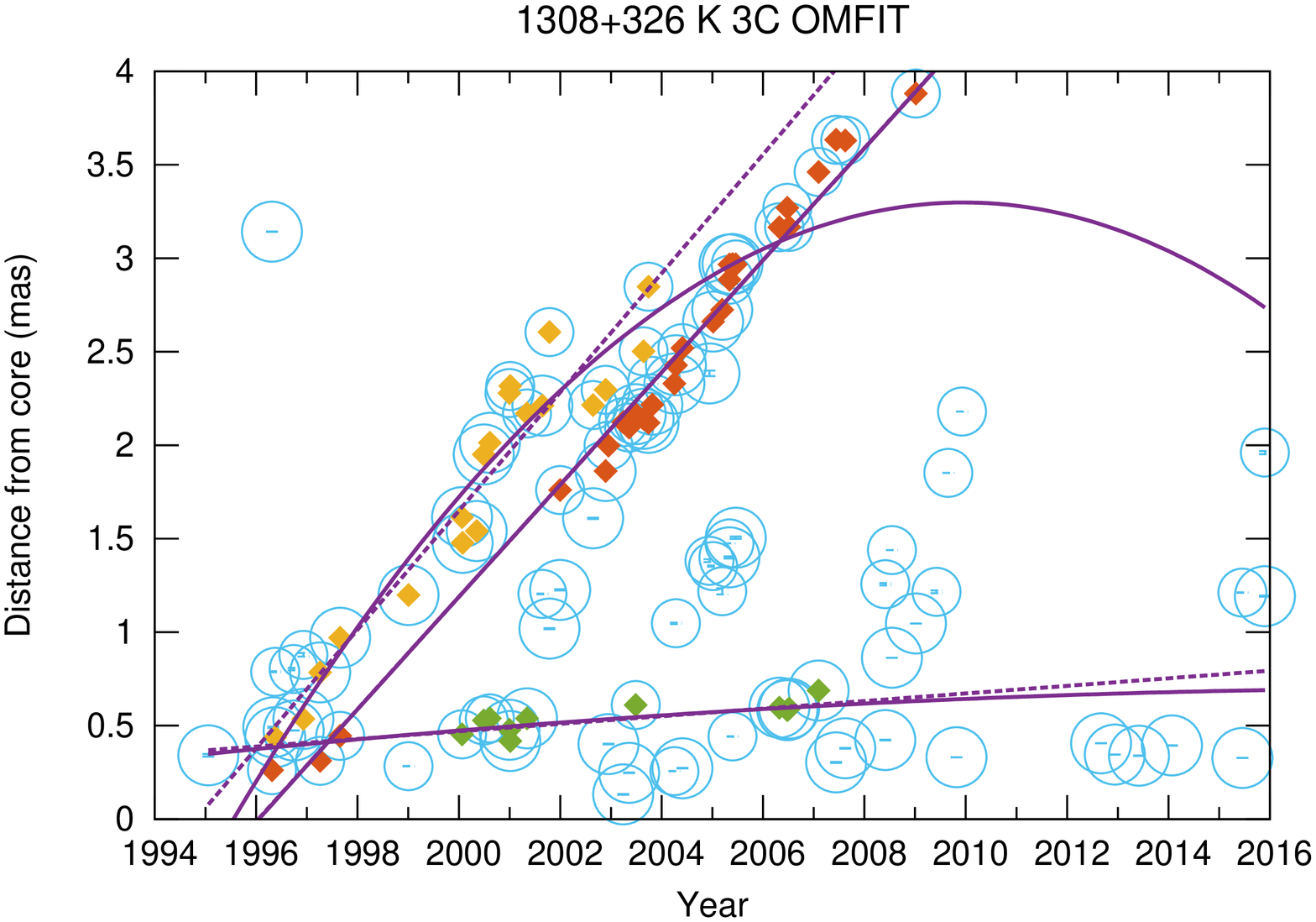}
\includegraphics[width=6cm, angle=-90]{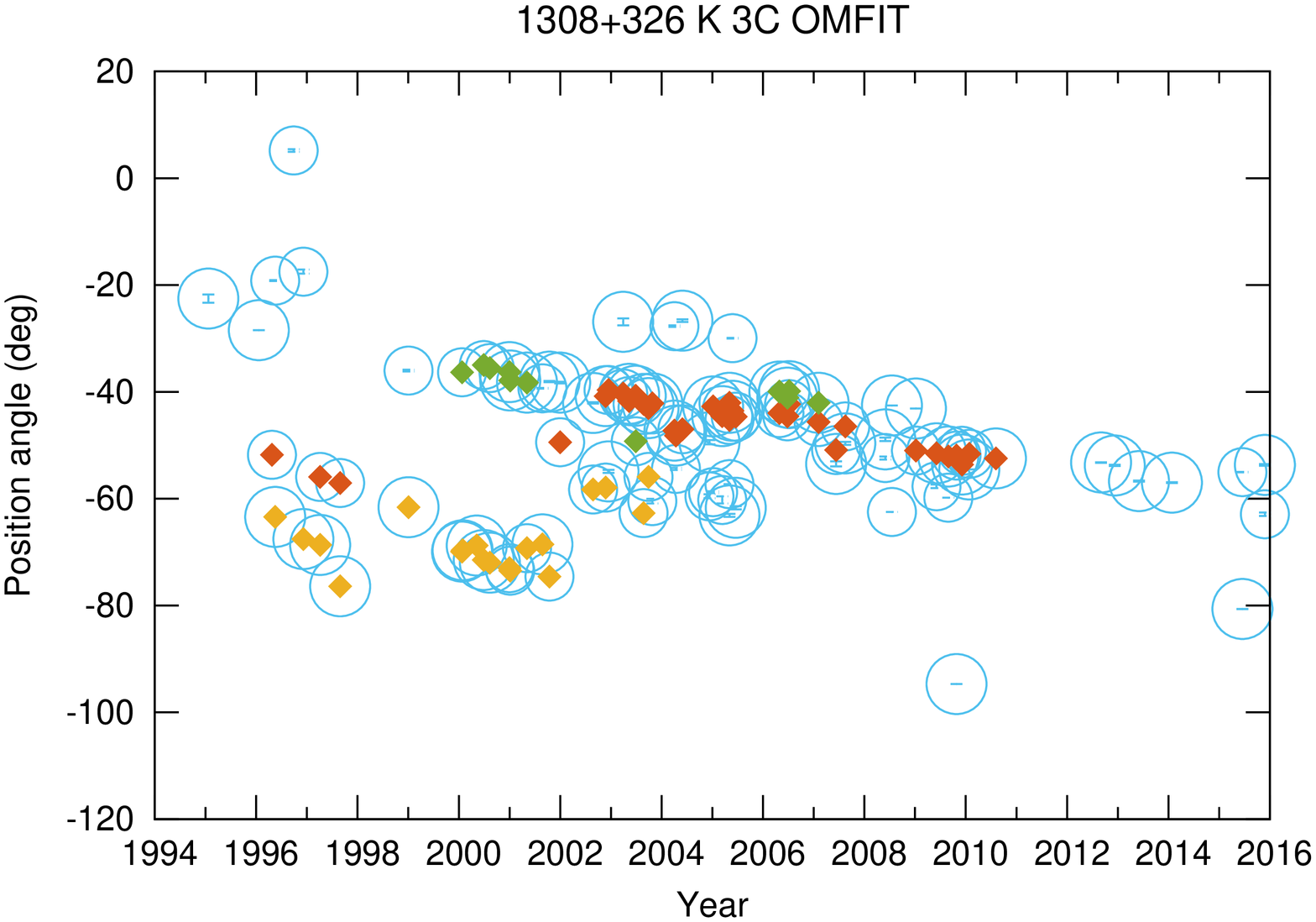}
\caption[K-band proper motion patterns]{Proper motion patterns for
  1308+326 at the K band frequency, as inferred from model-fitting in
  the image-plane (upper panels) and $uv$-plane (lower panels). The
  left panels show distances to the core while the right panels show
  radial position angles. Symbols are the same as those in
  Fig.~\ref{fig:psulin}.}
\label{fig:promK}
\end{figure*}

Based on the series of images generated by the {\sc sand} pipeline,
1308+326 was found to have more complex structure at higher
frequencies, namely K band and Q band. This is not only because the
extended radio emission is resolved into discrete components at these
higher frequencies, as a result of higher resolving power, but also
because the higher-frequency images reveal intense kinematics of the
jet close to the AGN core.  The complexity of the brightness
distribution is also reflected in the difficulty of proper motion
pattern recognition, as shown in Fig.~\ref{fig:promK}. To illustrate
how observing frequency affects such recognition, we have
conservatively set the control parameters for our tests. Although the
MOJAVE data cover a longer time span, the sampling of these data is
neither frequent nor even. As a result, regression {\sc stripp}ing
stopped after fitting three proper motion patterns, with the residual
data points not having enough significance for further fitting. The
$DS$ value of the first two fitted patterns is about 2.4. Of course,
we could visually fit at least two additional patterns but these would
not be reliable because of the lack of data points. Furthermore, we
see from the plots in Fig.~\ref{fig:promK} that our pipeline even
tried to do a quadratic fit for the third proper motion pattern when
model-fitting the data in the image plane. Since the residual data
points from epoch 2012 could belong to the emerging fourth component
and such an erratic behaviour is not verified with $uv$-plane fitting,
it appears difficult to trust those fits. In practice, we can simply
restrain the initial statistical criteria and let the regression {\sc
  strip} algorithm exclude such situations.

\begin{figure*}
\centering
\includegraphics[width=6cm, angle=-90]{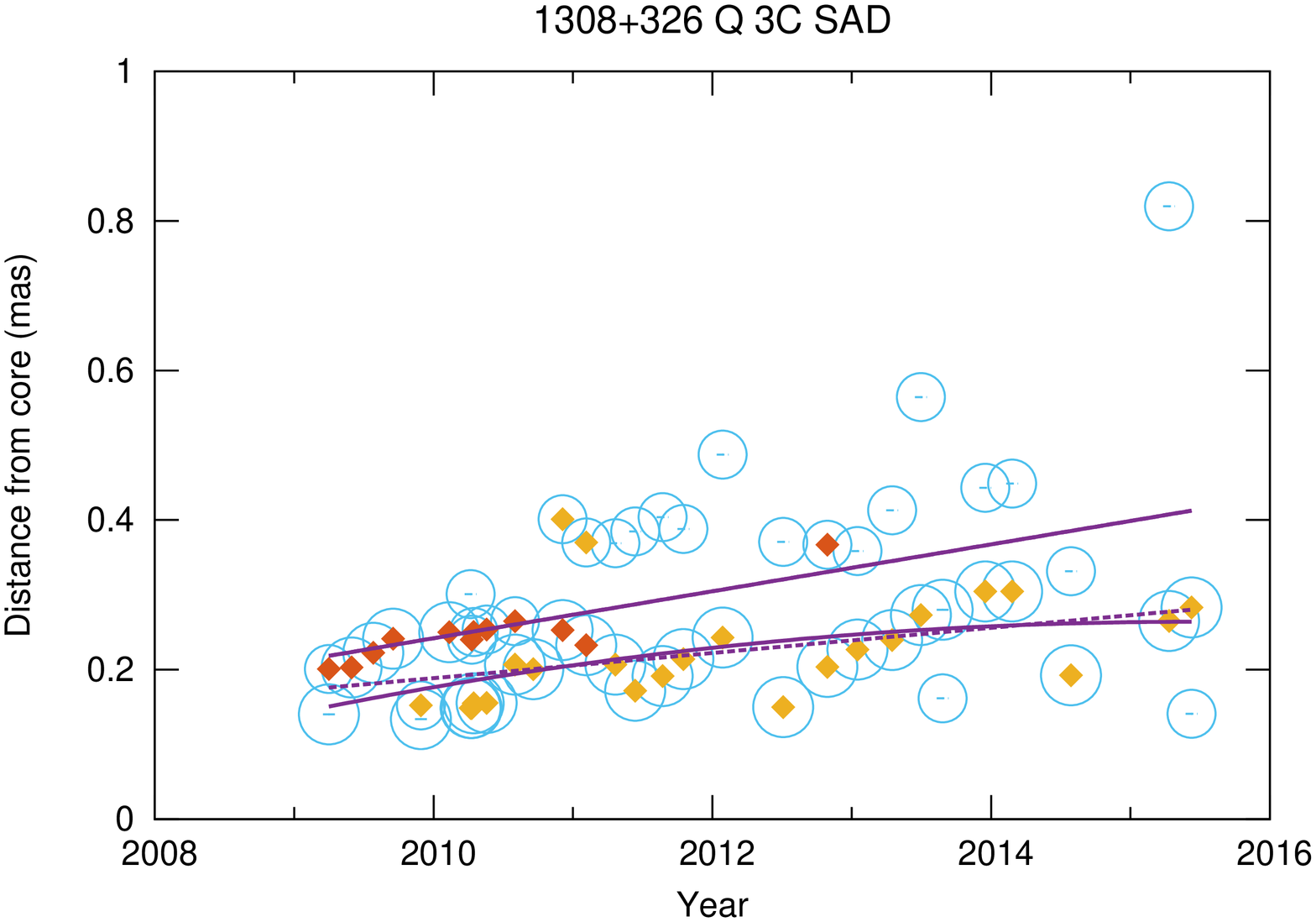}
\includegraphics[width=6cm, angle=-90]{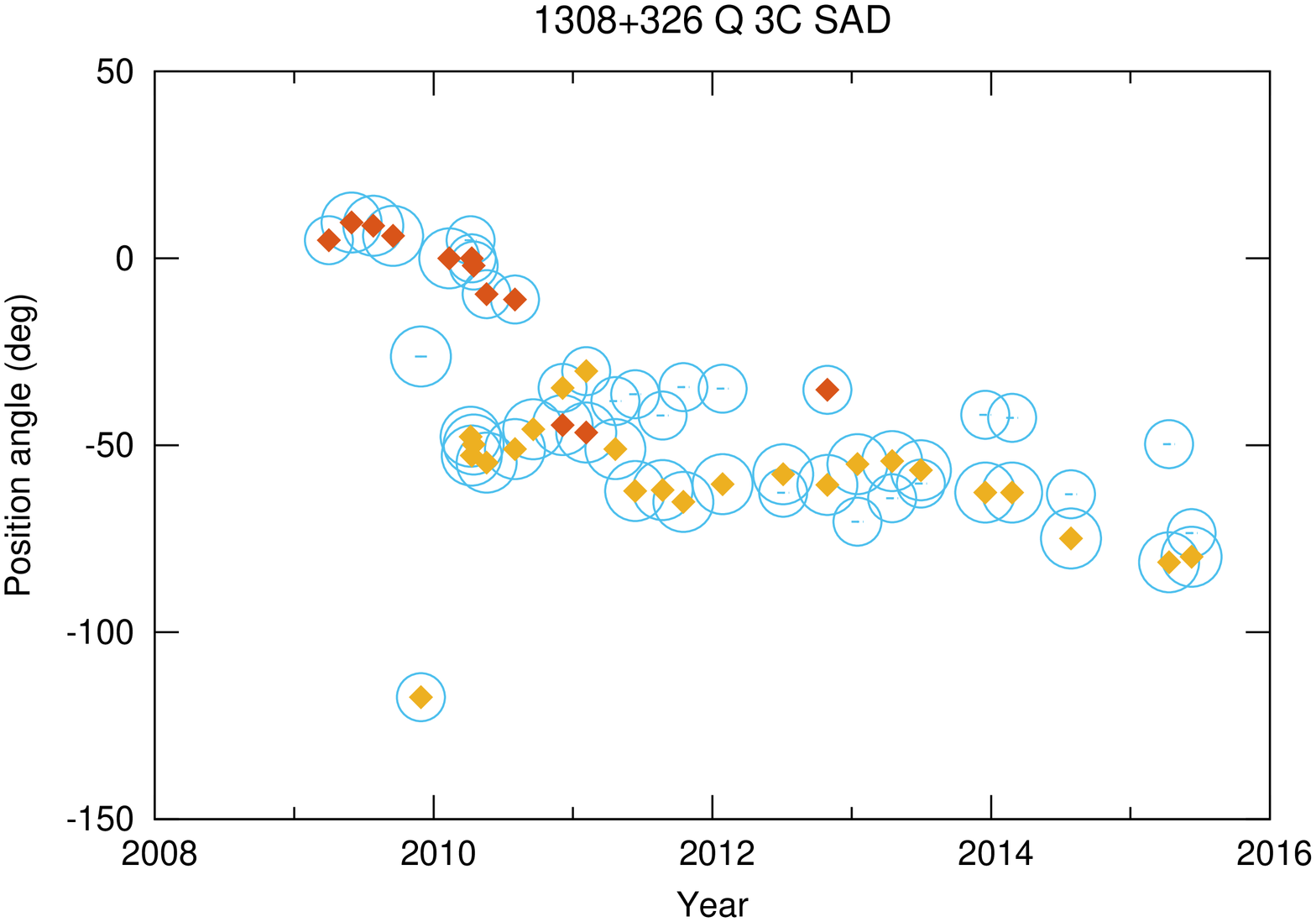}\\
\includegraphics[width=6cm, angle=-90]{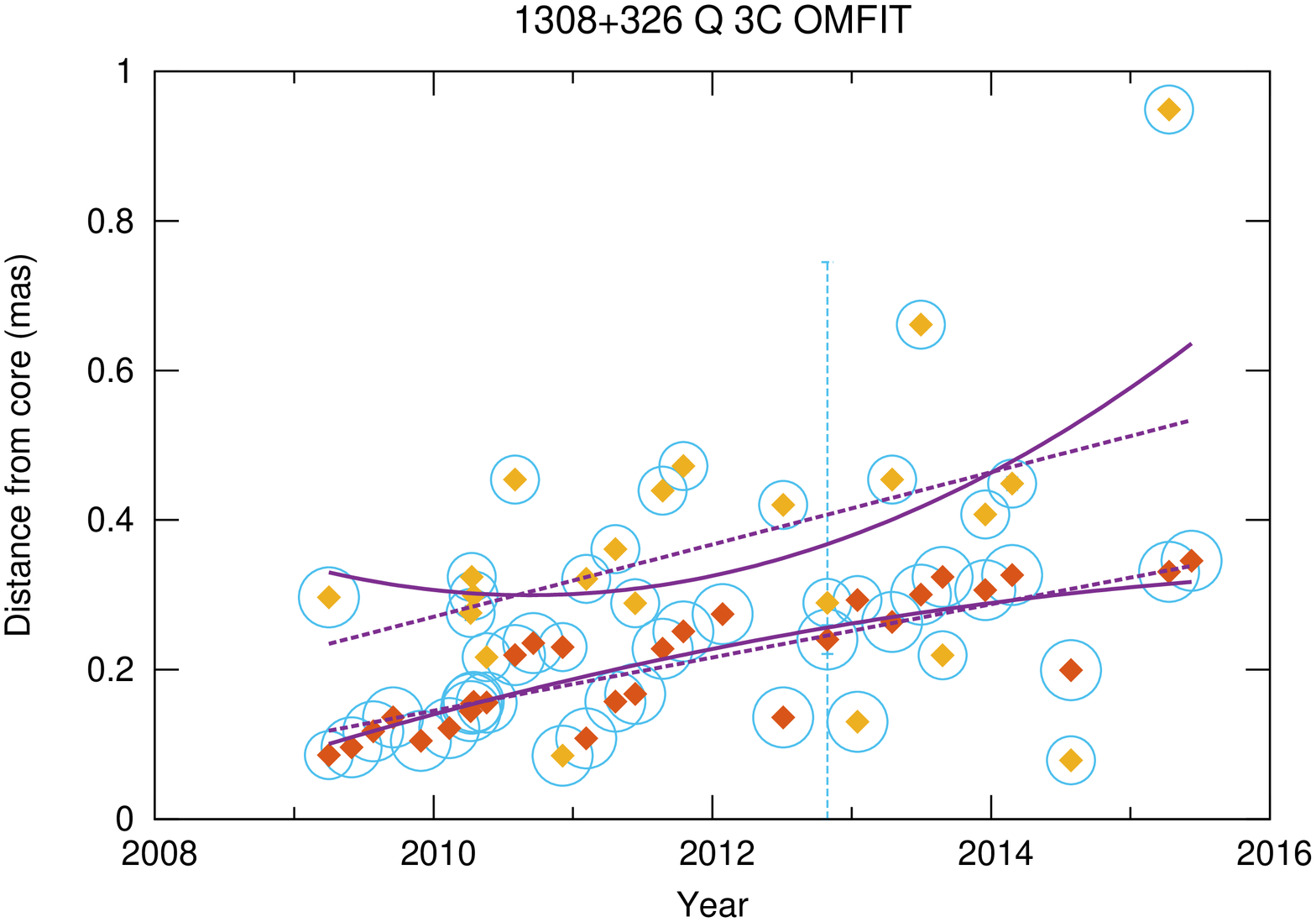}
\includegraphics[width=6cm, angle=-90]{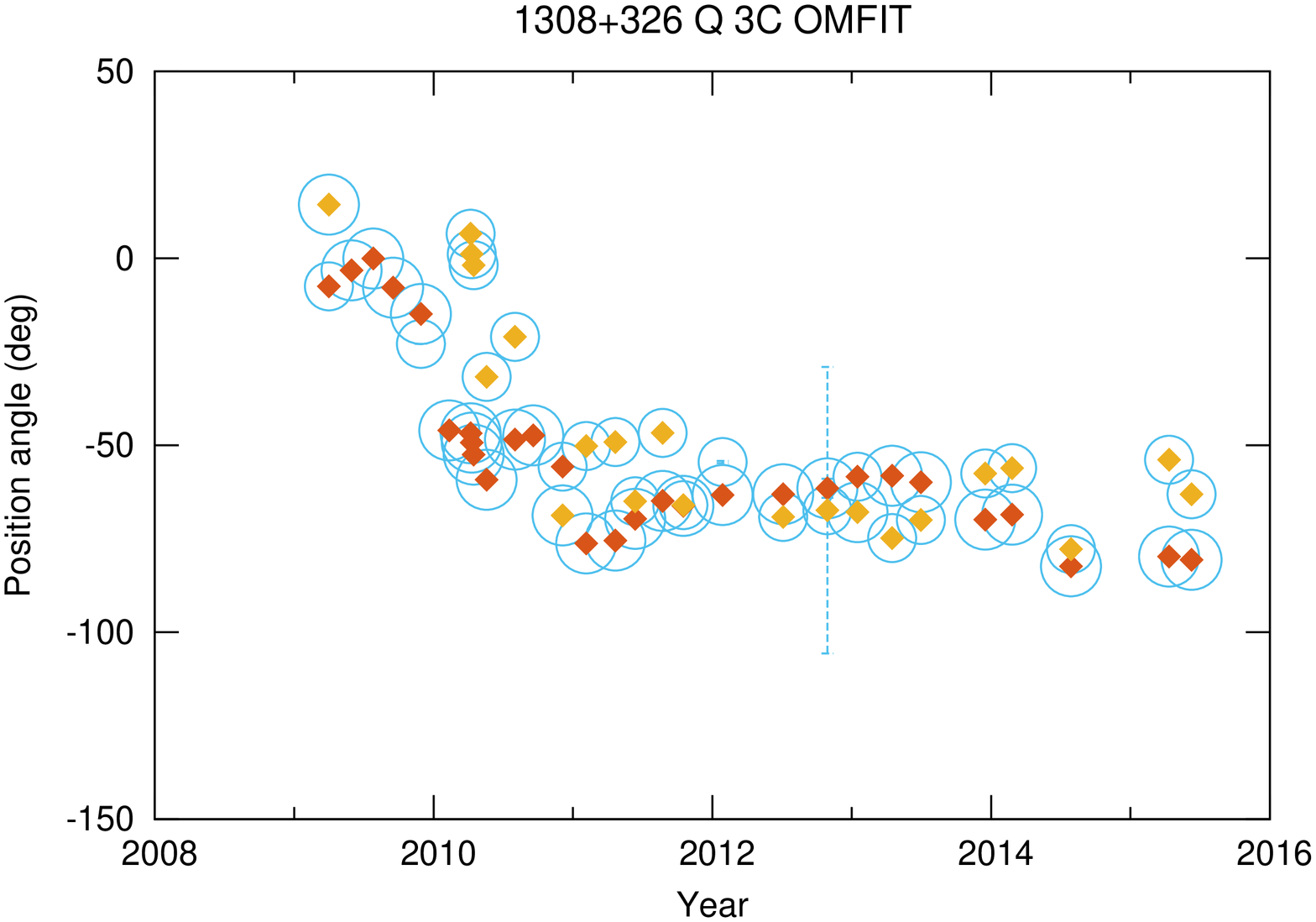}
\caption[Q-band proper motion patterns]{Proper motion patterns for
  1308+326 at the Q band frequency, as inferred from model-fitting in
  the image-plane (upper panels) and $uv$-plane (lower panels). The
  left panels show distances to the core while the right panels show
  radial position angles. Symbols are the same as those in
  Fig.~\ref{fig:psulin}.}
\label{fig:promQ}
\end{figure*}

The Q-band proper motion patterns do not look optimal either, because
the data points are even more scattered, as seen from
Fig.~\ref{fig:promQ}. The $DS$ value of the first two significant
patterns is about 1.6, hence likely to lead to confusion as
illustrated in Fig.~\ref{fig:respconf}. Generally source expansion is
favoured over source contraction, which provides further
constraints. However, contradictory measurements of superluminal
motions when determined by different groups are not unlikely due to
such confusion~\citep{piner.97.apj}.

One potential issue to be pointed out is the registration error in the
alignment of the core component for multi-epoch data. Due to the
regular emergence of jet components, the centroid of the core may
shift with time, hence causing errors in the relative positions of the
outer jet components at different epochs.  Such offsets may be
inferred from the aligned and subtracted multi-epoch images, within
the detection limit of the interferometer, and should be corrected for
the highest accuracy when calculating relative component positions.

\section{Conclusions}

The comprehensive VLBI data reduction pipeline {\sc sand} is designed
to help radio astronomers to improve the efficiency and objectivity of
interferometric data reduction. For multi-epoch multiband
observations, the pipeline provides a complete initial set of source
parameters, derived from imaging and model fitting, which may be
systematically analysed in spatial, temporal and spectral domains at
the post-processing stage. It also offers a means to combine and
analyse heterogeneous data sets from different resources
consistently. With various VLBI observing programmes going on, this
aspect is of high interest and may be viewed as an analogue to what
Virtual Observatory projects pursue.

The {\sc sand} pipeline has a built-in regression {\sc strip}
algorithm which automatically identifies jet component trajectories
and fits proper motions at the same time (see
Sect.~\ref{regstrip}). The algorithm was found to work well when
trajectory patterns between components are well separated.

The pipeline is not a fully-competent artificial-intelligence
substitution to manual VLBI data reduction. However, it provides an
objective approach to cross-examine results from different research
groups. When used for massive data reduction, it also offers a
practical tool to sift out peculiar sources of interest for extended
studies.

\subsection*{Acknowledgements}
MZ is grateful to the Centre National de la Recherche Scientifique
(CNRS) for granting a post-doctoral fellowship at the Laboratoire
d'Astrophysique de Bordeaux (LAB) in France to develop the pipeline
infrastructure reported in this paper. This work was further supported
by the National Basic Research Programme of China (2012CB821804 and
2015CB857100), the National Science Foundation of China (11103055) and
the West Light Foundation of Chinese Academy of Sciences
(RCPY201105). This research has made use of the NASA/IPAC
Extragalactic Database (NED), which is operated by the Jet Propulsion
Laboratory, California Institute of Technology, under contract with
the National Aeronautics and Space Administration (NASA). This
research has also made use of data from the MOJAVE database which is
maintained by the MOJAVE team~\citep{lister.09.aj}, and from the
VLBA-BU-Blazar Monitoring Program at 43~GHz which is funded by NASA
through the Fermi Guest Investigator
Program~\citep{marscher.09.aas}. The VLBA is an instrument of the
National Radio Astronomy Observatory, a facility of the National
Science Foundation operated by Associated Universities, Inc. The
authors also thank the Open Source software packages
Scipy~\citep{jones.01.gnu}, Gnuplot~\citep{william.10.gnu} and
Gnuplot-py~\citep{haggerty.08.gnu}, which made the development of this
pipeline easily achievable and transferable.

\appendix

\section{The SAND pipeline scheme}\label{pipeline}

A major advantage of the {\sc sand} pipeline is that it can benefit
from the robustness of well-tested {\sc aips} tasks and of the
computational functionality of the Python language to develop various
user-customized data reduction algorithms. The pipeline is composed of
structured modules, which facilitates incorporation of initial
calibration and self-calibration procedures, provided that flagging
and calibration tables are available in machine-readable form. This is
generally the case for VLBI post-correlation processing pipelines,
e.g. those implemented at the European VLBI Network (EVN) or Very Long
Baseline Array (VLBA) facilities. Additionally, many databases provide
$uv$ data that are already calibrated. For this reason, we focus on
the imaging and model-fitting procedures behind {\sc sand} in the
following sections.

\subsection{Categorizing the data}

As a general approach, our pipeline has not been designed to deal with
any specific observing programme. It thus has the capability to handle
data from various archives which often have different conventions for
naming and ordering sources. The {\sc sand} pipeline converts between
B1950 and J2000 source names and indexes every data file in a unique
way based on metadata information (source name, observing band,
session number, ...). Heterogeneous data sets are thus easily
imported, and targeted reductions on a certain observing band or
session range, for any particular list of sources, may be
accomplished. All results from processing are stored in specific
repositories with unique identification, allowing multiple pipeline
processes to be run concurrently without confusion in the
outputs. This scheme is especially useful to cross-check the effects
of a certain parameterization or to explore various aspects of the
parameter space in parallel.

\subsection{Imaging}

Deconvolution of the $uv$ data is accomplished using the
\citet{clark.80.a&a} {\sc clean} algorithm with the
\citet{schwab.83.aj} scheme for subtraction in the $uv$-plane, as
implemented in the {\sc aips} task {\sc imagr}. Component search and
extraction from the {\sc clean}ed image is conducted with the task
{\sc sad} which looks through the image plane to find bright peaks
above a certain flux threshold and simultaneously fits those with
elliptical Gaussians. The image quality is closely tied to the $uv$
coverage and depends on the deconvolution algorithm. Gaps in the $uv$
sampling introduce gridding errors which affect the synthetic beam and
cause non-Gaussian residuals during the deconvolution process which
are difficult to eliminate. Conversely, Gaussian noise spikes may be
more easily rejected since the corresponding flux is usually smeared
out below the given flux threshold.

\subsection{Model fitting}\label{modfit}

The on-the-fly model-fitting using the task {\sc sad} is generally
good enough for discrete compact sources. For more complex cases, we
have implemented an extra module based on the specific image-plane
model-fitting task {\sc jmfit}, permitting further verification of the
model parameters. Additionally, we check component separation against
the beam size. If separation is less than half a beam size, the
corresponding component is tagged as confused. Such confused
components are either treated as a single component in a subsequent
processing stage or ``forced'' to have a more significant separation
by constructing a super-resolved map with a smaller beam size.

\subsubsection{Image-plane fitting}

Component extraction in the image plane with task {\sc sad} is based
on the Gaussian fitting subroutine behind task {\sc jmfit}. We have
demonstrated that in most cases the models derived from the two tasks
are identical. However, the task {\sc jmfit} offers a wider range of
options for parameter settings, which is of interest for sources with
subtle structures requiring specific attention. Such sources generally
show extended structures and low-brightness features, requiring
appropriate setting of the window size to get optimum results.

\subsubsection{$uv$-plane fitting}

As an ill-posed inverse problem, a deconvolution has no unique
solution. The main goal of deconvolution algorithms then is to achieve
an optimal convergence. The {\sc clean} algorithm works well in many
cases, especially when source structure is compact. However, compact
radio sources may also show extended features, in particular at low
frequencies. Due to the `{\sc clean} bias'~\citep{condon.98.aj}, a
{\sc clean}ed image might not preserve authentic structures, notably
when the source is extended or weak. Nevertheless, the uncertainties
introduced by the {\sc clean} algorithm should be largely reduced if
fitting models directly in the $uv$~plane.

Model-fitting in the $uv$ plane requires the input structural model of
the source to be well defined beforehand (e.g.~in the form of a single
elliptical Gaussian) to be Fourier-transformed into the $uv$
plane. Having a proper input model is important in this process as
model-fitting in the $uv$~plane does not always converge, in
particular when the actual source structure departs significantly from
the model fed as input or if the source is too weak. It is to be noted
that the task {\sc omfit} used to perform model-fitting in the
$uv$-plane has the same self-calibration capability as {\sc difmap}
during {\sc clean}ing, hence providing consistent schemes. The results
from modelling in the image-plane may thus be used to provide the
required input for modelling in the $uv$-plane.

\subsubsection{Comparison with manual reduction}

\begin{figure*}
\includegraphics[width=\textwidth, angle=-90]{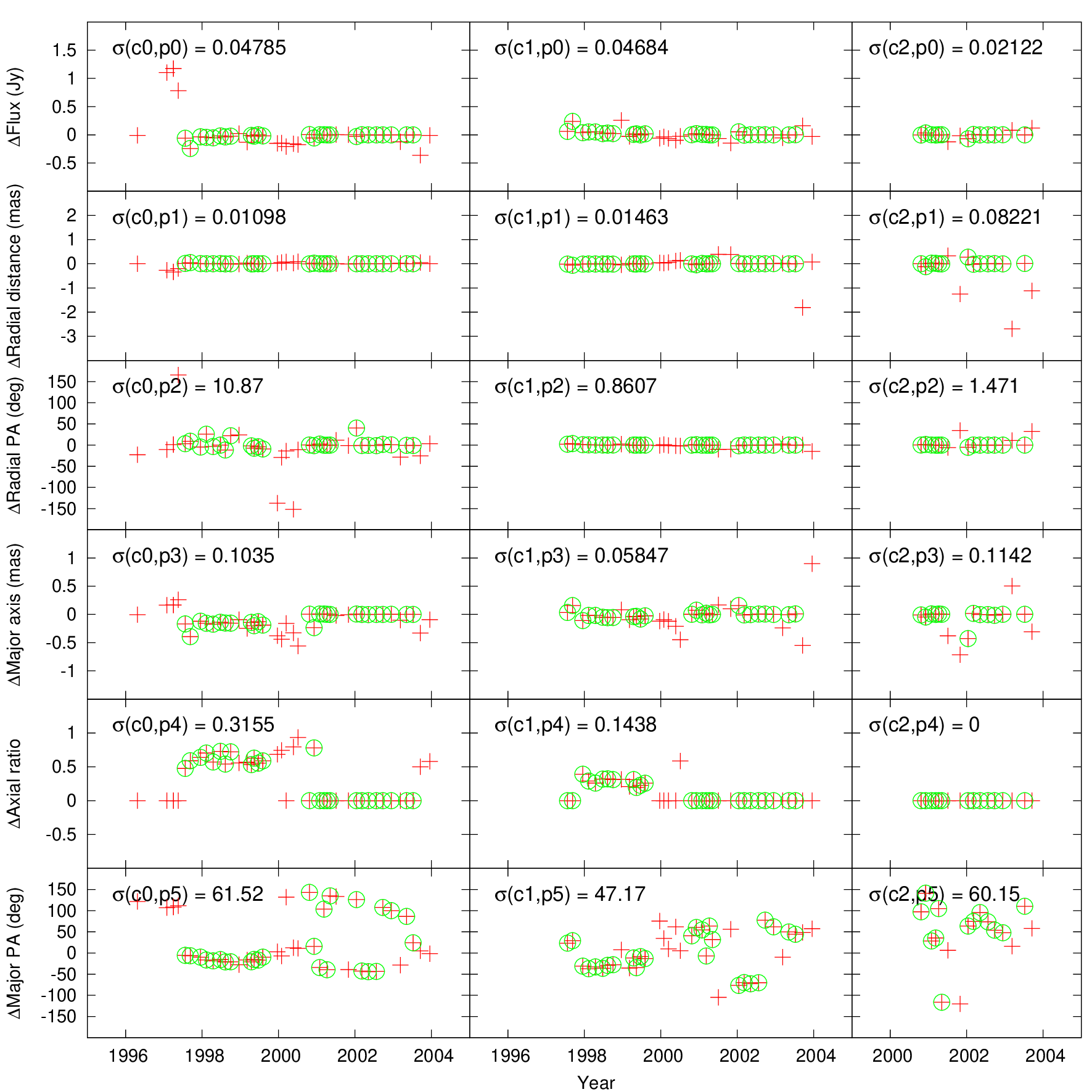}
\caption[Comparison of model parameters]{Differences in fitted
  parameters between the manual reduction of~\citet{piner.12.apj} and
  the {\sc sand} automatic reduction for the X-band RDV observations
  of 1308+326. The component sequence is denoted as \{c0, c1, c2\} in
  decreasing flux order and the parameter sequence is denoted as \{p0,
  p1, p2, p3, p4, p5\}, corresponding to the flux, radial separation,
  radial position angle, major axis, axial ratio and major axis
  position angle in accordance with the {\sc difmap} conventions. The
  circles indicate the epochs for which the number of components in
  the two reductions agrees. }
\label{fig:dfpz}
\end{figure*}

To assess the quality of the models determined by {\sc sand}, we have
compared our pipeline results with those derived from a manual
reduction of the RDV data by~\citet{piner.12.apj}. Due to the
degeneracy between elliptical and circular Gaussians, and to avoid
confusion, we have restricted the shape of our component models in the
$uv$-plane to circular Gaussians. A detailed comparison of the results
from {\sc sand} and from~\citet{piner.12.apj} for the core and the
first two jet components for 1308+326 is shown in
Fig.~\ref{fig:dfpz}. The comparison indicates that the fitted
parameters are quite consistent for the two determinations. The only
noticeable discrepancies are in the position angles of the elliptical
Gaussians, which is not unexpected since we used circular Gaussians
and \citet{piner.12.apj} used elliptical Gaussians.

\subsection{Generation of images and cataloguing}

\subsubsection{Morphological evolution}

\begin{figure*}
\centering
\includegraphics[width=\textwidth, trim=0 0 0 1cm, clip]{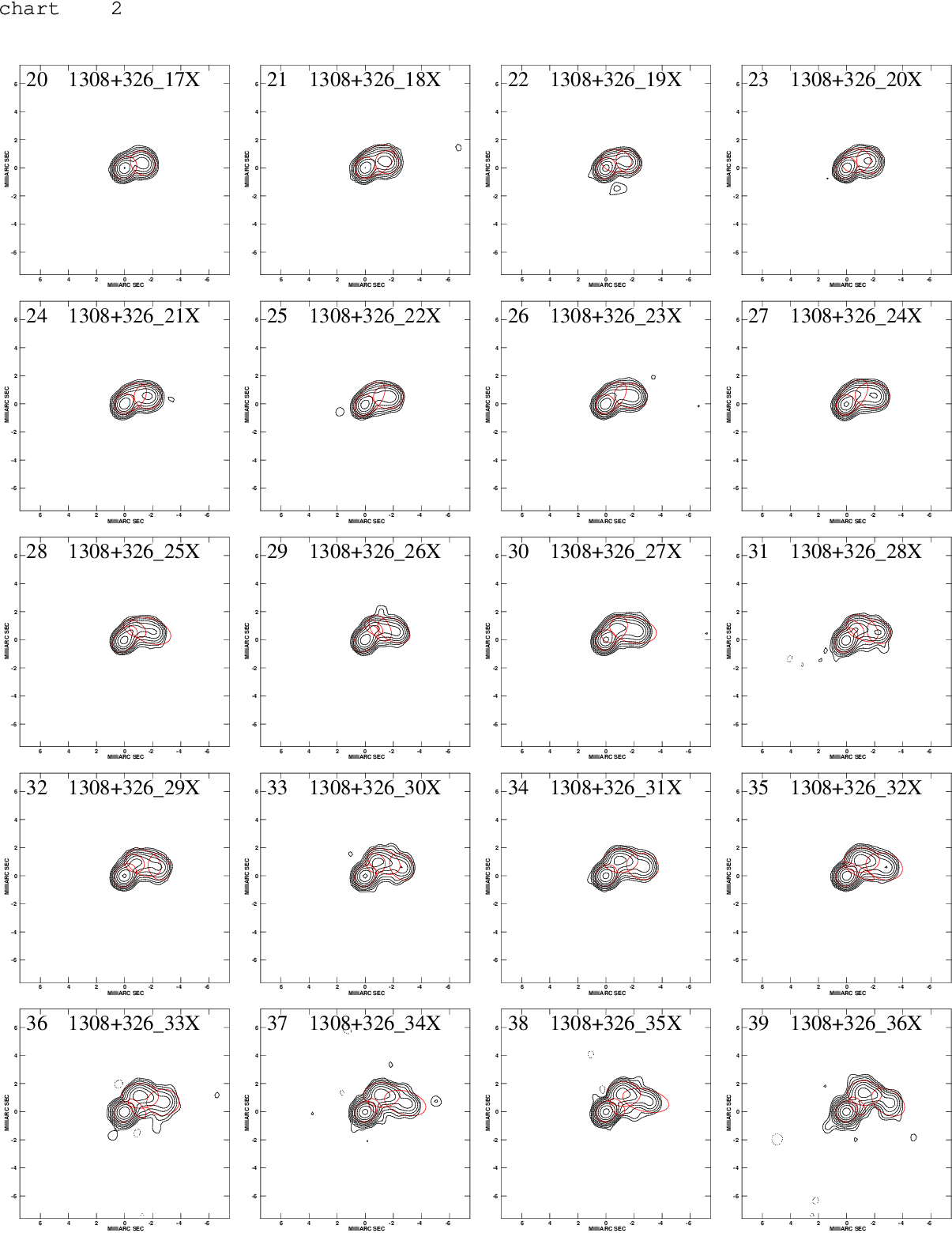}
\caption[{\sc clean}ed images catalogue]{A sample of {\sc clean}ed
  images of 1308+326 at the X band frequency, as derived from RDV
  observations with the {\sc sand} pipeline. Contour levels are
  plotted at (-1, 1, 2, 4, 8, 16, 32, 64, 128, 256, 512, 1024)
  $\times\ 6\sigma$, where $\sigma$ is the off-source noise in the
  map. The red circles denote the modelled components. The axes are
  relative RA and DEC in units of milliarcsecond with the origin of
  the map shifted to the core centroid.}
\label{fig:chtX}
\end{figure*}

\begin{figure*}
\centering
\includegraphics[width=\textwidth,trim=0 0 0 1cm, clip]{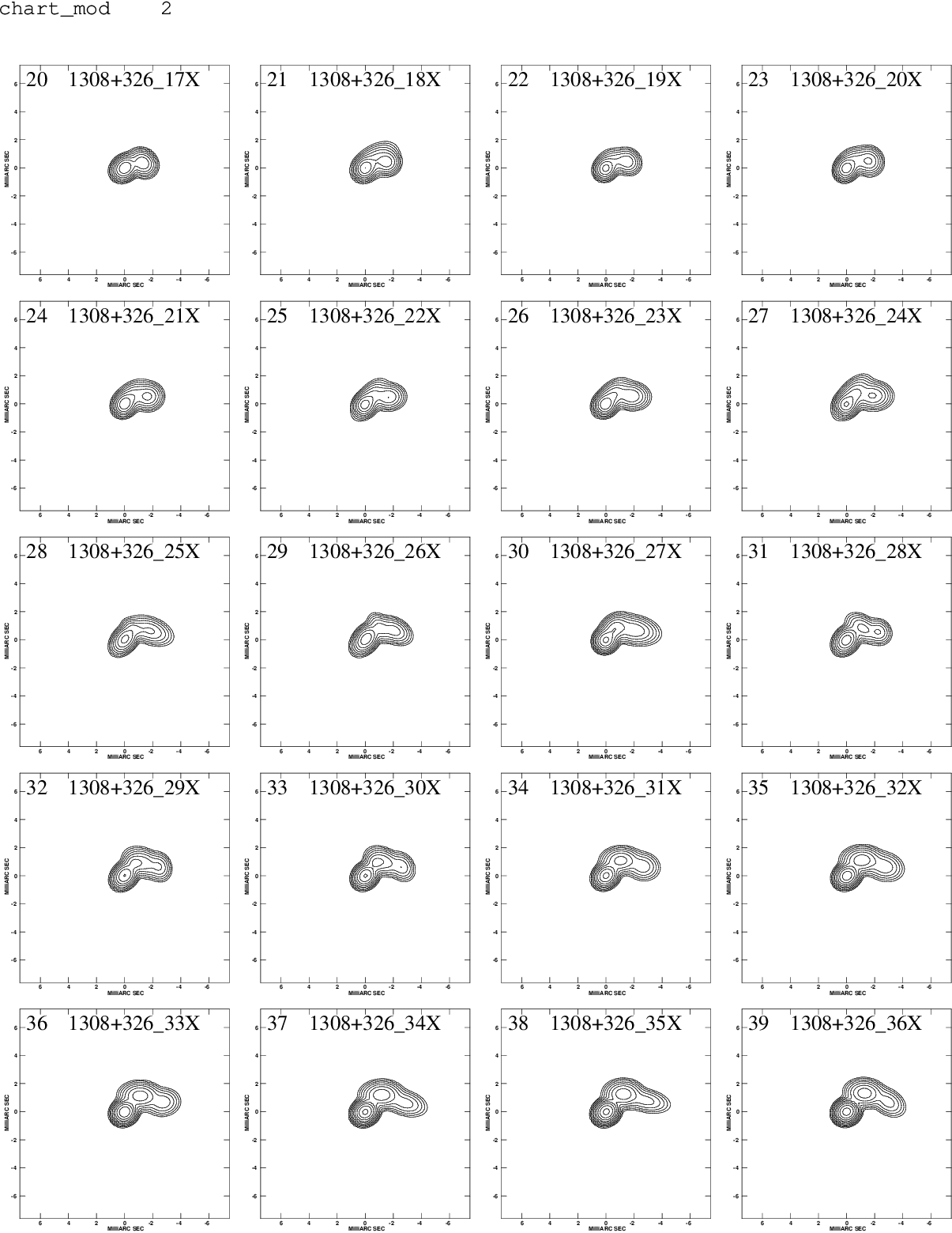}
\caption[Model image catalogue]{A sample of reconstructed models of
  1308+326 at the X band frequency, as derived from RDV observations
  with the {\sc sand} pipeline. The contour levels and units for the
  axes are the same as those in Fig.~\ref{fig:chtX}.}
\label{fig:chtXm}
\end{figure*}

Since our pipeline has been designed to deal with multi-epoch
observations we have built a module that automatically generates
images at every epoch. The images are reconstructed from the fitted
model parameters and are generated in parallel with the {\sc clean}ed
images. See Figs.~\ref{fig:chtX} and~\ref{fig:chtXm} for an
example. By inspecting such images, the evolution of jet structures
can be quickly investigated and any discrepancies from the {\sc
  clean}ed image is easily spotted.  These images are also useful to
identify weak features that sometimes surround major structures.

\subsubsection{Polarization maps}\label{polmap}

\begin{figure*}
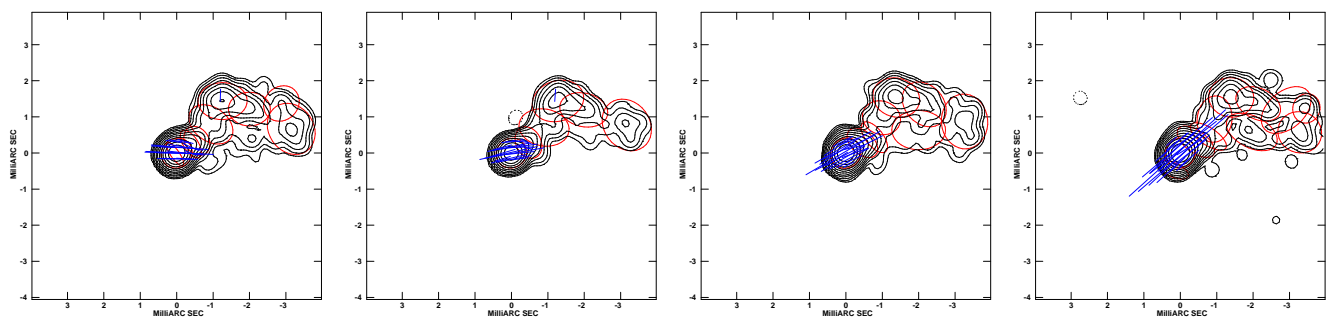

\centering
\includegraphics[width=4.3cm]{lwpla_out.1308+326_0SK.27.ps}
\includegraphics[width=4.3cm]{lwpla_out.1308+326_0TK.28.ps}
\includegraphics[width=4.3cm]{lwpla_out.1308+326_0UK.29.ps}
\includegraphics[width=4.3cm]{lwpla_out.1308+326_0VK.30.ps}
\caption[Polarisation maps]{Illustration of the polarization angle
  rotation in the K-band observations of 1308+326, as determined from
  the {\sc sand} pipeline. The contour levels and units for the axes
  are the same as those in Fig.~\ref{fig:chtX}. The red circles denote
  the modelled components while the blue vectors indicate the
  direction of the polarization emission in the core.}
\label{fig:polK}
\end{figure*}

If the visibility data have full correlated Stokes parameters, i.e.
RR, LL, RL and LR, our pipeline will automatically consider them and
produce polarization maps. As shown from the
series of K-band polarization maps of 1308+326 in Fig.~\ref{fig:polK},
a rotation of the polarisation angle, here in a counter-clockwise
direction, is detected from the maps at the successive
epochs. Based on these maps, one could also infer that the direction
of the jet emission beam from the central engine slewed during this
period, e.g. due to precession.

\subsection{Post-processing}

In addition to the various data reduction steps explained above, some
post-processing procedures have been implemented in the {\sc sand}
pipeline in order to facilitate analysis and interpretation of
multi-epoch multiband results. These procedures comprise the
determination of component trajectories and generation of light curves
and multi-epoch component spectra, as described below.

\subsubsection{Component trajectories}\label{comtraj}

The determination of jet component trajectories is the topic of this
paper and hence is detailed in the main text. See Sect.~\ref{regstrip}
for a description of the algorithm and Sects.~\ref{mock}
and~\ref{trials} for the results of tests with mock and real data.

\subsubsection{Light curves}\label{lcurv}

\begin{figure}
\centering
\includegraphics[width=6cm, angle=-90]{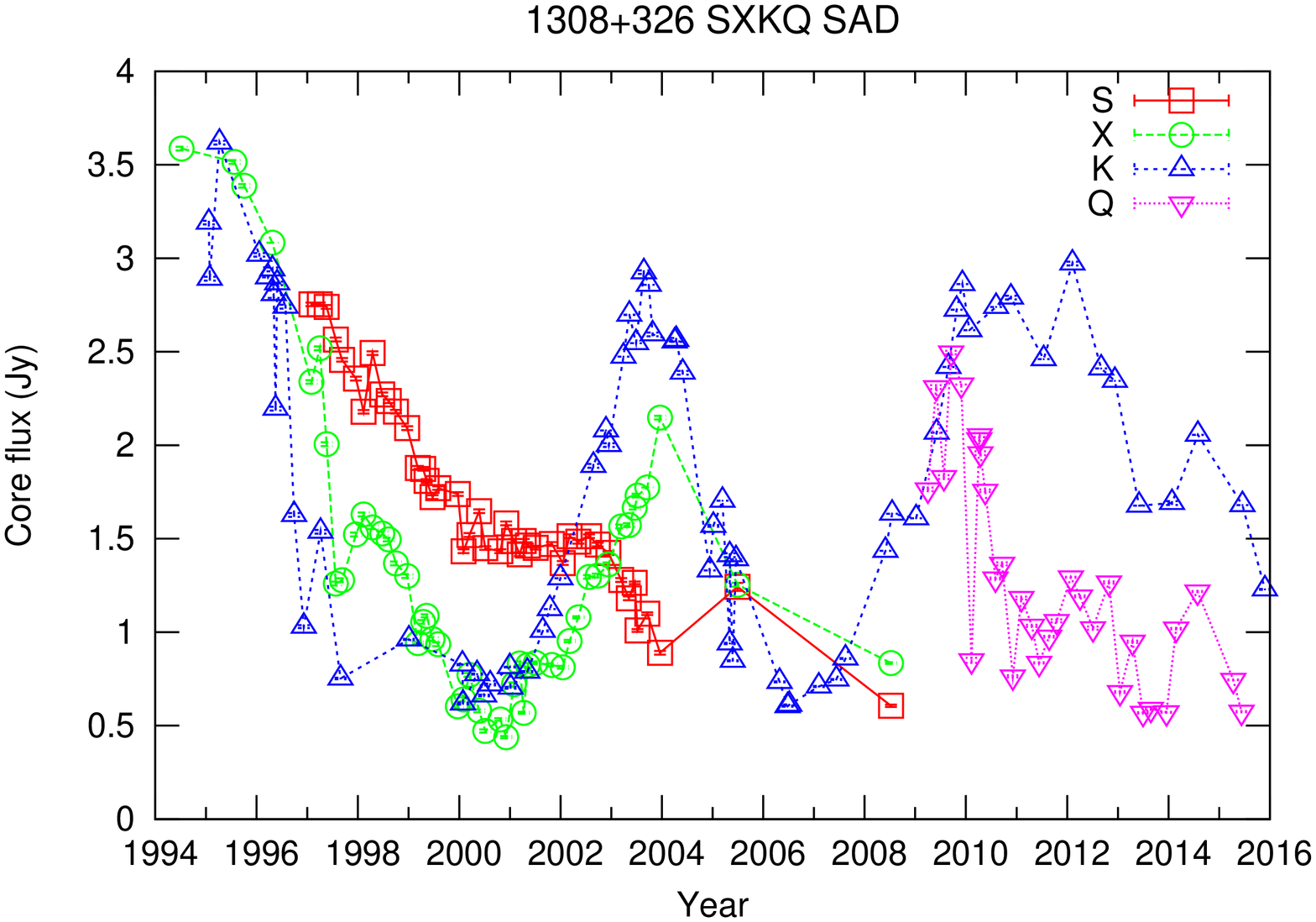}
\caption[Multi-band light curves]{Multi-band variations of the core flux
  density in 1308+326. Symbols in red, green, blue and pink indicate flux
  densities at S band, X band, K band and Q band, respectively.}
\label{fig:lcurv}
\end{figure}

Following model fitting, component flux density is available at every
epoch and for every band. It is thus straightforward to extract the
core flux density from these and derive multi-band light curves. An
example of such a light curve, as determined from {\sc sand}, is shown
in Fig.~\ref{fig:lcurv}. Additionally, a capability to derive
single-side Power Spectral Density (PSD) plots via direct fast Fourier
transform (FFT) was implemented to check the periodicity of light
curves. For simultaneous multi-band VLBI observations, even
from non-coordinated programmes, the pipeline can also automatically
identify and extract observations in overlapping time ranges and
cross-correlate the light curves from the different bands in each range.
This aspect will be discussed in details in a subsequent paper.

\subsubsection{Multi-epoch spectra}\label{spec}

\begin{figure}
\centering
\includegraphics[width=6cm, angle=-90]{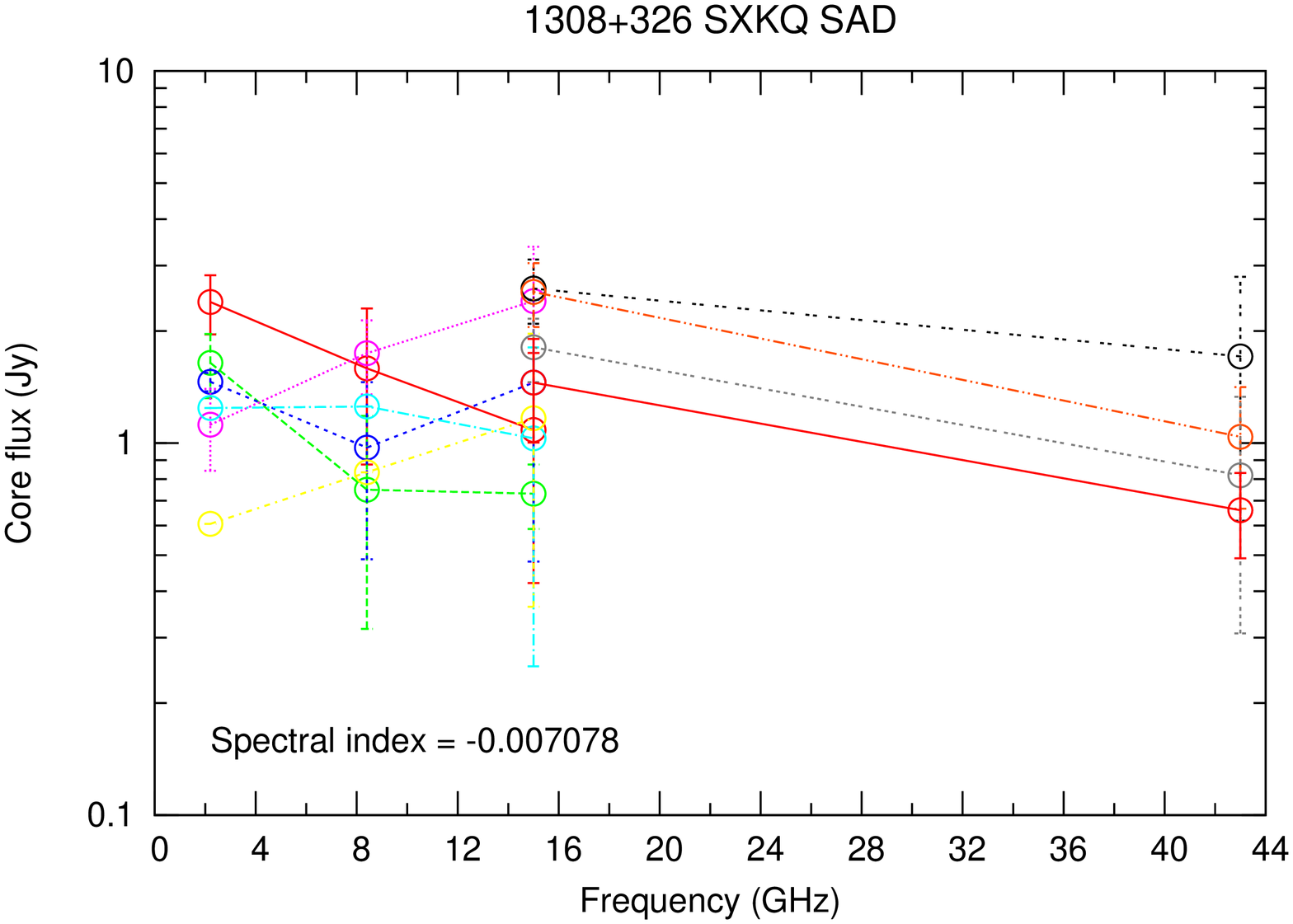}\qquad
\caption[Multi-epoch spectra]{Multi-epoch spectra for the core
  component of 1308+326. Data in the same time bin are plotted with
  symbols of the same colour. Different colours denote different time
  bins. The spectral index value is determined by averaging over
  time.}
\label{fig:spec}
\end{figure}

Using multi-band light curves, one can also check the spectral index
of the radio sources. For this purpose, a procedure that re-bins the
data over the overlapping epoch ranges, calculates the average flux in
each bin, and derives the source spectrum at the different epochs has
been implemented in {\sc sand}.  An example of such multi-epoch
spectra for the flat-spectrum radio source 1308+326 is shown in
Fig.~\ref{fig:spec}.

\bibliography{refs}
\end{document}